\documentclass{article}
\usepackage[margin=1.5in]{geometry}
\makeatletter
\renewcommand\@biblabel[1]{#1.}
\makeatother

\usepackage{graphicx}
\usepackage[super,comma]{natbib}
\usepackage{amssymb}
\usepackage[hypertexnames=false,colorlinks]{hyperref}
\usepackage[utf8]{inputenc}
\usepackage{amsmath}
\usepackage{soul}

\usepackage{url}
\usepackage{mathtools}
\usepackage{physics}
\usepackage[dvipsnames]{xcolor}
\usepackage{caption, subcaption, floatrow}
\usepackage{soul}
\usepackage{lineno}

\title{Spectral index-flux relation for investigating the origins of steep decay in $\gamma$-ray bursts}

\date{}

\begin{document}

\maketitle
\noindent
Samuele Ronchini$^{1,2,3*}$, Gor Oganesyan$^{1,2,3}$, Marica Branchesi$^{1,2,3}$, Stefano Ascenzi$^{4,5,6}$, Maria Grazia Bernardini$^{4}$, Francesco Brighenti$^{1}$, Simone Dall'Osso$^{1,2}$, Paolo D'Avanzo$^{4}$, Giancarlo Ghirlanda$^{4,7}$, Gabriele Ghisellini$^{4}$, Maria Edvige Ravasio$^{4,7}$, Om Sharan Salafia$^{4,8}$\\

\noindent
$^{1}$ Gran Sasso Science Institute, Viale F. Crispi 7, I-67100, L’Aquila (AQ), Italy\\
$^{2}$ INFN - Laboratori Nazionali del Gran Sasso, I-67100, L’Aquila (AQ), Italy\\
$^{3}$ INAF - Osservatorio Astronomico d’Abruzzo, Via M. Maggini snc, I-64100 Teramo, Italy\\
$^{4}$ INAF – Osservatorio Astronomico di Brera, Via E. Bianchi 46, 23807 Merate (LC), Italy\\
$^{5}$ Institute of Space Sciences (ICE, CSIC), Campus UAB, Carrer de Can 
Magrans s/n, 08193, Barcelona, Spain\\
$^{6}$ Institut d’Estudis Espacials de Catalunya (IEEC), Carrer Gran Capita 
2–4, 08034 Barcelona, Spain\\
$^{7}$ Università degli Studi di Milano-Bicocca, Dip. di Fisica “G. Occhialini”, Piazza della Scienza 3, 20126 Milano, Italy\\
$^{8}$ INFN–Sezione di Milano-Bicocca, Piazza della Scienza 3, I-20126 Milano (MI), Italy\\

$^{*}$email: \textbf{samuele.ronchini@gssi.it} \\

\section*{Abstract}

{\noindent \bf $\gamma$-ray bursts (GRBs) are short-lived transients releasing a large amount of energy ($10^{51}-10^{53} \rm erg$) in the keV-MeV energy range. GRBs are thought to originate from internal dissipation of the energy carried by ultra-relativistic jets launched by the remnant of a massive star’s death or a compact binary coalescence. While thousands of GRBs have been observed over the last thirty years, we still have an incomplete understanding of where and how the radiation is generated in the jet. Here we show a relation between the spectral index and the flux found by investigating the X-ray tails of bright GRB pulses via time-resolved spectral analysis. This relation is incompatible with the long standing scenario which invokes the delayed arrival of photons from high-latitude parts of the jet. While the alternative scenarios cannot be firmly excluded, the adiabatic cooling of the emitting particles is the most plausible explanation for the discovered relation, suggesting a proton-synchrotron origin of the GRB emission.
}
\section*{Introduction}
The prompt emission of GRBs is characterized by an erratic superposition of several pulses, whose spectrum typically peaks in the keV-MeV energies. Its physical origin is still matter of discussion and the main open questions concern the composition of the jet (matter\cite{Shemi1990} or magnetic\cite{Usov1992} dominated), the energy dissipation mechanisms (sub-photospheric emission\cite{Pe'er2006}, internal shocks\cite{Rees1994} or magnetic reconnection\cite{Zhang2011}), and the nature of particle radiation. Once the prompt emission ceases, the light curve usually presents a steep decay phase\cite{Nousek2006,Zhang2006,Tagliaferri2005,O'Brien2006} (tail), which can be well monitored in the X-ray band. The duration of the steep decay is around $10^2-10^3$ s and it is characterized by a typical decay power-law slope of $3-5$. After the prompt emission, the jet interacts with the interstellar medium, producing the so called afterglow emission\cite{Paczynski1993,Meszaros1997,Sari1998}. The afterglow models cannot account for such steep slopes and the origin of the steep decay is attributed to the fade-off of the emission mechanism that is responsible of the prompt phase.\\
Considering that the emitting surface of the jet is curved, an on-axis observer first receives photons from the line of sight and later photons from higher latitudes\cite{Fenimore1996,Kumar2000,Liang2006}, which are less Doppler boosted. This gives rise to the so called High Latitude Emission (HLE). Under the assumption of a single power-law spectrum ($F_{\nu}\propto \nu^{-\beta}$), the HLE predicts that the flux decays as $F_{\nu}(t_{\text{obs}})\propto \nu^{-\beta} t_{\text{obs}}^{-(\beta+2)}$. On the other hand, if the spectrum is curved, the HLE can also lead to the transition of the spectral peak across the observing band\cite{Lin2017}, causing a spectral evolution, as often observed in the soft X-rays\cite{Zhang07,Mangano2011}.\\
In this work we find a unique relation between the spectral index and the flux. Here, we systematically analyze the X-ray spectral evolution during the steep decay phase as motivated by fact that temporal and spectral evolution during the tail of prompt pulses can provide clues about emission and cooling processes in GRB jets. Given the same trend followed by all the GRBs of our sample, we search for a common process at the basis of the spectral relation. We find that the standard HLE model cannot account for the observed relation, implying that efficient cooling of particles is disfavoured. We test several assumptions about the dominating cooling mechanisms and we find that the combined action of adiabatic cooling of particles and magnetic field decay robustly reproduces our data. We conclude discussing the implications for the physics of GRB jets, their composition and radiation mechanisms.

\section*{Results}

In order to investigate the spectral evolution during the steep decay phase, we select a sample of GRBs from the archive of the X-ray Telescope (XRT, 0.3-10 keV) on-board the Neil Gehrels Swift Observatory (Swift)\cite{Gehrels2004}. We restrict our study to a sample of GRBs (8 in total) whose brightest pulse in the Burst Alert Telescope (BAT, 15-350 keV) corresponds to the XRT peak preceding the X-ray tail (see as example the Fig.~\ref{peaked_main}a). We perform a time-resolved spectral analysis of the tail in the 0.5-10 keV band assuming a simple power-law model for the photon spectrum $N_{\gamma} \propto E^{-\alpha}$ (see Methods subsections: Sample selection, Time resolved spectral analysis and Spectral modeling). We represent the spectral evolution plotting the photon index $\alpha$ as a function of the flux $F$ integrated in the 0.5-10 keV band, hereafter referred to as the $\alpha - F$ relation. The flux is normalized to the peak value of the X-ray tail. This normalization makes the result independent of the intrinsic brightness of the pulse and of the distance of the GRB.\\
We find a unique $\alpha - F$ relation for the analyzed GRBs as shown in Fig.~\ref{peaked_main}b. This is consistent with a systematic softening of the spectrum; the photon index evolves from a value of $\alpha \sim 0.5-1$ at the peak of the XRT pulse to $\alpha \sim 2-2.5$ at the end of the tail emission, while the flux drops by two orders of magnitude. The initial and final photon indexes are consistent with the typical low- and high-energy values found from the analysis of the prompt emission spectrum of GRBs, namely $\sim 1$ and $\sim 2.3$\cite{Frontera2000,Kaneko2006,Nava2011}, respectively. The $\alpha-F$ relation can be interpreted as being due to a spectral evolution in which the spectral shape does not vary in time, but the whole spectrum is gradually shifted towards lower energies while becoming progressively dimmer (see Fig.~\ref{sketch}). The consistent spectral evolution discovered in our analysis is a clear indication of a common physical mechanism responsible for the tail emission of GRBs and the corresponding spectral softening.\\

\noindent
{\bf Testing High Latitude Emission.} We first compare our results with the expectations from the HLE, which is the widely adopted model for interpreting the X-ray tails of GRBs. When the emission from a curved surface is switched off, an observer receives photons from increasing latitudes with respect to the line of sight. The higher the latitude, the lower the Doppler factor, resulting in a shift towards lower energies of the spectrum in the observer frame. Through an accurate modeling of HLE (as described in Methods subsection: HLE from infinitesimal duration pulse) we derive the predicted $\alpha-F$ relation under the assumption of an abrupt shutdown of the emission, consistent with particles cooling on timescales much smaller than the dynamical timescale. We first consider a smoothly broken power-law (SBPL) comoving spectrum. Regardless of the choice of the peak energy, the bulk Lorentz factor or the radius of the emitting surface, the HLE predicts an $\alpha-F$ relation whose rise is shallower than the observed one (Fig.~\ref{HLE_spe_evo}). We additionally test the Band function, commonly adopted for GRB spectra\cite{Band1993}, and the physically motivated synchrotron spectrum\cite{Rees1994}, obtaining similar results (Supplementary Fig.~12a and Supplementary Fig.~11): the HLE softening is too slow to account for the observed $\alpha-F$ relation. We further relax the assumption of an infinitesimal duration pulse, i.e. considering a shell that is continuously emitting during its expansion and suddenly switches off at radius $R_0$\cite{Genet2009} (see Supplementary Note 1). The contributions from regions $R<R_0$ are sub-dominant with respect to the emission coming from the last emitting surface at $R=R_0$, resulting in a spectral evolution still incompatible with the observations (Supplementary Fig.~1). An interesting alternative is the HLE emission from an accelerating region\cite{Uhm2015} taking place in some Poynting flux dissipation scenarios\cite{Zhang2011}. Even though it can explain the temporal slopes observed in the X-ray tails, also this scenario fails in reproducing the $\alpha-F$ relation (see Supplementary Fig.~2). Our results on HLE are based on the assumption of a common comoving spectrum along the entire jet core. Even changing the curvature (or sharpness) of the spectrum or assuming a latitude dependence of the spectral shape, the disagreement with the data remains, unless we adopt a very fine-tuned structure of the spectrum along the jet core, which is not physically motivated (see Methods subsection: HLE from infinitesimal duration pulse). Alternative models, such as anisotropic jet core\cite{Narayan2009,Duran2016,Geng17} or sub-photospheric dissipation\cite{Pe'er2008}, can hardly reproduce our results (see Supplementary Note 3).\\

\noindent
{\bf Adiabatic cooling.} Since the standard HLE from efficiently cooled particles and its modified versions, as well as alternative scenarios, are not able to robustly interpret the observed $\alpha-F$ relation, we consider a mechanism based on an intrinsic evolution of the comoving spectrum. The most natural process is the adiabatic cooling of the emitting particles\cite{Duran2009}. Here we assume conservation of the entropy of the emitting system $\langle\gamma\rangle^{3} V'$ throughout its dynamical evolution, where $\langle\gamma\rangle$ is the average random Lorentz factor of the emitting particles and $V'\propto R^2 \Delta R'$ the comoving volume\cite{Meszaros1999}. We consider both thick and thin emitting regions, i.e. a comoving thickness of the emitting shell $\Delta R'= \mathrm{const}$ or $\Delta R'\propto R$, respectively. We assume a power law radial decay of the magnetic field $B=B_0(R/R_{0})^{-\lambda}$, with $\lambda>0$, and synchrotron radiation as the dominant emission mechanism. Here $R_0$ is the radius at which adiabatic cooling starts to dominate the evolution of the emitting particles. We compute the observed emission taking also into account the effect of HLE by integrating the comoving intensity along the equal arrival time surfaces (see Methods subsection: Adiabatic cooling). In this scenario, contrary to HLE alone, the emission from the jet is not switched off suddenly, but the drop in flux and the spectral evolution are produced by a gradual fading and softening of the source, driven by adiabatic cooling of particles. The resulting spectral evolution and light curves are shown in Fig.~\ref{ad_int_spe}.\\
Adiabatic cooling produces a much faster softening of $\alpha$ as a function of the flux decay, with respect to HLE alone, in agreement with the data. Assuming a different evolution of the shell thickness, the behavior of the curves changes only marginally (see Supplementary Fig.~3). For large values of $\lambda$ the evolution of $\alpha$ flattens in the late part of the decay (see Fig.~\ref{ad_int_spe}a), indicating that the spectral evolution becomes dominated by the emission at larger angles, rather than by adiabatic cooling in the jet core. Adiabatic cooling can also well reproduce the light curve of X-ray tails (Fig.~\ref{ad_int_spe}b). For comparison, in the same plot we show the light curve given by pure HLE, adopting the same value of $R_0$ and $\Gamma$.\\
In order to fully explore the parameter space of the adiabatic cooling model, we used a Monte Carlo Markov Chain (MCMC) algorithm for the parameter estimation. We consider the joint temporal evolution of flux and photon index and we find agreement of the model with data (see Methods subsection: Parameter estimation via Monte Carlo Markov Chain and Tab.~\ref{tab_AC}). In Supplementary Fig.~9 - 10 we show for each burst the observed temporal evolution of photon index and normalized flux in comparison with the curves produced with 500 random draws from the posterior sample set of the MCMC. We obtain a value of $\lambda$ in the range $0.4-0.7$ (except for 090621A which prefers $\lambda\sim 2$). On average, these values of $\lambda$ are smaller than those expected in an emitting region with a transverse magnetic field ($\lambda=1$ or $\lambda=2$ for a thick or a thin shell, respectively) or magnetic field in pressure equilibrium with the emitting particles ($\lambda=4/3$ or $\lambda=2$ for a thick or a thin shell, respectively\cite{Meszaros1999}).\\
The typical timescale of adiabatic cooling $\tau_{\text{ad}}=R_0/2c\Gamma^2$, i.e. the observed time interval during which the radius doubles, is equal to the HLE timescale\cite{Fenimore1996,Sari1997} and radically affects the slope of X-ray tails. Therefore, the comparison between the model and the observed light curves allows us to constrain the size $R_0$ of the emitting region as in HLE\cite{Lyutikov2006,Lazzati2006}. We find values in the range $0.3\text{ s}\lesssim\tau_{\text{ad}}\lesssim 24 \text{ s}$. These values are quite larger than the typical duration of GRB pulses ($<$1 sec\cite{Walker2000}), which can be due to the following reason. For the spectral analysis to be feasible, we had to choose only tails that are long enough (to be divided down into a sufficient number of temporal bins). Moreover, the prompt emission is usually interpreted as a superposition of several emission episodes: the steep decay observed in XRT is likely dominated by the tails with the slowest decay timescales. Since, in our model, the decay timescale is $\tau_{\text{ad}}=R/2c\Gamma^2$, this could indicate that the emission radius of the pulses that dominate the tail is systematically larger than that of pulses that dominate the prompt emission. If this is the case, a lower magnetic field is also expected, which goes well along with the long radiative timescale and slow (or marginally fast) cooling regime, in agreement with our results. For the range of $\tau_{\text{ad}}$ obtained from the analysis, the corresponding range for the emission radius is $1.8\times 10^{14}(\Gamma/100)^2 \text{ cm}\lesssim R_0\lesssim 1.4\times10^{16}(\Gamma/100)^2 \text{ cm}$.
A different prescription for adiabatic cooling has been suggested in the literature\cite{Duran2009}, in which the particle's momentum gets dynamically oriented transverse to the direction of the local magnetic field. In this case, HLE is the dominant contributor to the X-ray tail emission, which is again incompatible with the observed $\alpha-F$ relation.\\

\noindent
{\bf Extending the sample.} In order to further test the solidity of the $\alpha-F$ relation, we extend our analysis to a second sample of GRBs (composed by 8 elements) which present directly a steep decay at the beginning of the XRT light curve, instead of an X-ray pulse (see Fig.~\ref{spe_evo_extra}a), often observed in early X-ray afterglows\cite{Tagliaferri2005,O'Brien2006}. We require that the XRT steep decay is preceded by a pulse in the BAT light curve (the brightest since its trigger time). We add the data of this second sample to the $\alpha-F$ plot, estimating the peak flux by the extrapolation of the XRT light curve backwards to the peak time of the BAT pulse, under the assumption that BAT peak and XRT peak were simultaneous (see Methods subsection: Extrapolation of $\text{F}_{\text{max}}$). We find that these GRBs follow the overall $\alpha-F$ relation (Fig.~\ref{spe_evo_extra}b), confirming that a common physical process is governing the spectral evolution of X-ray tails. Adiabatic cooling is still capable of reproducing the data of this second sample (Fig.~\ref{spe_evo_extra+HLE}a), provided that we assume a slightly softer high energy intrinsic spectrum ($\alpha \sim 3 $ instead of $\alpha \sim 2.5 $). Alternatively, the introduction of an exponential cutoff in the spectral shape at $\nu=\nu_{\text{c}}\sim \nu_{\text{m}}$ can also reproduce the data (see Fig.~\ref{spe_evo_extra+HLE}b), where $\nu_{\text{c}}$ and $\nu_{\text{m}}$ are the synchrotron characteristic frequencies. The cutoff is formed by a combined action of adiabatic cooling and mild synchrotron cooling (see Supplementary Note 4). We specify that the limited size of our samples is related to the selection requirements, which are necessary for an appropriate time-resolved spectral analysis. Thus our results are proved for X-ray tails firmly connected to prompt emission pulses.

\section*{Discussion}

The $\alpha-F$ relation, found in our analysis, requires a mechanism that produces the X-ray tails of GRBs with a unique law of flux decay and spectral softening. Although other scenarios cannot be ruled out, we find that adiabatic cooling of the emitting particles, together with a slowly decaying magnetic field, is the most plausible scenario able to robustly reproduce this relation. Our results suggest an efficient coupling between a slowly decaying magnetic field and the emitting particles. Our findings are generally in agreement with moderately fast and slow cooling regimes of the synchrotron radiation, which is able to reproduce the overall GRB spectral features\cite{Zhang2020}. In the adiabatic cooling scenario, most of the internal energy is not radiated away before the system substantially expands. If electrons are responsible for the emission, an extremely small magnetic field would be required\cite{Kumar08,Ben13,Ben18}, which is unrealistic for this kind of outflows. Protons radiating through synchrotron emission can solve this problem\cite{Ghisellini2020}. Due to their larger mass, they radiate less efficiently than electrons, explaining why adiabatic cooling dominates the spectral evolution.\\
In conclusion, our results indicate that adiabatic cooling can play a crucial role for the collective evolution of the radiating particles in GRB outflows and consequently for the determination of spectral and temporal properties of prompt emission episodes. The coupling between particles and magnetic field ensures the intrinsic nature and hence the universality of this process, whose effects are independent of the global properties of the system, such as the luminosity of the GRB or the geometry of the jet.
\newpage

\clearpage

\section*{Methods}
\setcounter{figure}{0}
\setcounter{table}{0}

{\noindent \bf  Sample selection.}
We define the steep decay (SD) segment\cite{Nousek2006,Zhang2006,Tagliaferri2005,O'Brien2006} as the portion of the light curve that is well approximated by a power law, $F\propto t^{-\alpha}$ with $\alpha>$3. Such criterion allows us to exclude a decay coming from a forward shock\cite{Paczynski1993,Meszaros1997,Sari1998}. In order to determine the presence of a SD, we analyze the light curve of the integrated flux in the XRT $E=0.3-10$ keV band.\\
From the Swift catalog\cite{Evans2009} as of the end of 2019, we selected all GRBs with an XRT peak flux $F_{\text{p}}^{\text{XRT}}>10^{-8}\text{ erg }\text{cm}^{-2}\text{s}^{-1}$. We selected the brightest pulses in order to have a good enough spectral quality as to perform a time resolved spectral analysis. The peak flux is computed taking the maximum of $F(t_i)$, where $F(t_i)$ are the points of the light curve at each time $t_i$. Among these GRBs, we selected our first sample according to the following criteria:
\begin{enumerate}
\item The XRT light curve shows at least one SD segment that is clean, i.e. without secondary peaks or relevant fluctuations.
\item 
If we call $F_1$ and $F_2$ the fluxes at the beginning and at the end of the SD, respectively, we require that $\frac{F_1}{F_2}>10$. This requirement is necessary to have a sufficient number of temporal bins inside the SD segment and therefore a well sampled spectral evolution.
\item The beginning of the SD phase corresponds to a peak in the XRT light curve, such that we have a reliable reference for the initial time. We stress that the identification of the SD starting time in XRT is limited by the observational window of the instrument. This means that, if the XRT light curve starts directly with a SD phase, with no evidence of a peak, the initial reference time is possibly located before and its value cannot be directly derived.
\item The XRT peak before the SD has a counterpart in BAT, whose peak is the brightest since the trigger time. This requirement is necessary to ensure that XRT is looking at a prompt emission episode, whose typical peak energy is above 100 keV. In a quantitative way, we define two times, $t_{\text{p}}$ and $t_{90}^{\text{stop}}$, where the first indicates the beginning of the peak that generates the SD, while the second is the end time of $T_{90}$\cite{Lien2016}, with respect to the trigger time. We require $t_{90}^{\text{stop}}>t_{\text{p}}$ in order to have an overlap between the last prompt pulses (monitored by BAT) and the XRT peak that precedes the SD phase. Namely, such requirement ensures that a considerable fraction of the energy released by the burst goes into the pulse that generates the X-ray tail.
\end{enumerate}
It is possible that more than one peak is present in the XRT light curve, each with a following SD. In this case we consider only the SD after the brightest peak. If two peaks have a similar flux, we consider the SD with the larger value of $\frac{F_1}{F_2}$.\\
We define then a second sample of GRBs that satisfy the first two points listed before, but have a SD at the beginning of the XRT light curve, namely no initial peak preceding the SD is present. In addition, we require that a BAT pulse precedes the XRT SD and is the brightest since the trigger time. The BAT pulse enables us to constrain the starting time of the SD.\\
The selection criteria limit the size of our sample, but they are unavoidable to perform a well targeted analysis of X-ray tails and to achieve robust conclusions about their origin.\\

{\noindent \bf  Time resolved spectral analysis.}
For each GRB we divided the XRT light curve in several time bins, according to the following criteria:
\begin{enumerate}
\item Each bin contains only data in Windowed Timing (WT) mode or in Photon Counting (PC) mode, since mixed WT+PC data cannot be analyzed as a single spectrum.
\item Each bin contains a total number of counts $N_{\text{bin}}$ in the $E=0.3-10$ keV band larger than a certain threshold $N_0$, which is chosen case by case according to the brightness of the source (see below). The definition of the time bins is obtained by an iterative process, i.e. starting from the first point of the light curve we keep including subsequent points until
\begin{equation}
N_{\text{bin}}=\sum_{t_n=t_i}^{t_f} N(t_n)>N_0
\end{equation}
where $N(t_n)$ are the counts associated to each point of the light curve, while $t_i$ and $t_f$ define the starting and ending time of the bin. Then the process is repeated for the next bins, until $t_f$ is equal to the XRT ending time. Due to the large range of count rates covered during a typical XRT light curve, the choice of only one value for $N_0$ would create an assembly of short bins at the beginning and too long bins toward the end. Therefore we use one value of $N_0$ for bins in WT mode ($N_0^{\text{WT}}$) and a smaller value of $N_0$ for bins in PC mode ($N_0^{\text{PC}}$). In our sample, the SD is usually observed in WT mode, therefore we adjust $N_0^{\text{WT}}$ in order to have at least 4-5 bins inside the SD. A typical value of $N_0^{\text{WT}}$ is around 1500-3000, while $N_0^{\text{PC}}$ is around 500-1000. Using these values, we verified that the relative errors of photon index and normalization resulting from spectral analysis are below $\sim30\%$.

\item For each couple $(N_i,N_j)$ of points inside the bin, the following relation must hold:
\begin{equation}
\frac{\abs{N_i-N_j}}{\sqrt{\sigma_i^2+\sigma_j^2}}<5
\end{equation}
where $\sigma_i$ and $\sigma_j$ are the associated errors. Such requirement avoids large flux variations within the bin itself.

\item The duration of the bin is larger than 5 seconds, in order to avoid pileup in the automatically produced XRT spectra.
\end{enumerate}
It is possible that condition 3 is satisfied only for a duration of the bin $T_{\text{bin}}<T_0$, while condition 2 is satisfied for $T_{\text{bin}}>T_0^{*}$, but $T_0^{*}>T_0$, meaning that they cannot be satisfied at the same time. In this case, we give priority to condition 3, provided that $N_{\text{bin}}$ is not much smaller than $N_0$.\\
Due to the iterative process that defines the duration of the bins, it is possible that the last points in WT and PC mode are grouped in a single bin with a too small $N_{\text{bin}}$, giving a too noisy spectrum. Therefore, they are excluded from the spectral analysis.\\
 
{\noindent \bf  Spectral modeling.} 
The spectrum of each bin is obtained using the automatic online tool provided by Swift for spectral analysis (see Data Availability statement). Each spectrum is analyzed using XSPEC\cite{Xspec}, version 12.10.1, and the Python interface PyXspec. We discard all photons with energy $E<0.5$ keV and $E>10$ keV. The spectra are modeled with an absorbed power law and for the absorption we adopted the Tuebingen-Boulder model \cite{2000ApJ...542..914W}. If the GRB redshift is known, we use two distinct absorbers, one Galactic\cite{Kalberla2005} and one relative to the host galaxy (the XSPEC syntax is tbabs*ztbabs*po). The column density $N_{\text{H}}$ of the second absorber is estimated through the spectral analysis, as explained below. On the other hand, if the GRB redshift is unknown, we model the absorption as a single component located at redshift z=0 (the XSPEC syntax is tbabs*po) and also in this case the value of $N_{\text{H}}$ is derived from spectral analysis.\\
For the estimation of the host $N_{\text{H}}$ we consider only the late part of the XRT light curve following the SD phase. At late time with respect to the trigger we do not expect strong spectral evolution, as verified in several works in the literature\cite{BK2007,Mu16}. Therefore, for each GRB, the spectrum of each bin after the SD is fitted adopting the same $N_{\text{H}}$ which is left free during the fit. Normalization and photon index are also left free, but they have different values for each spectrum. We call $N_{\text{H}}^{\text{late}}$ the value of $N_{\text{H}}$ obtained with this procedure. In principle the burst can affect the ionization state of the surrounding medium, but we assume that such effects are negligible and $N_{\text{H}}$ does not change dramatically across the duration of the burst \cite{Perna2002}. Hence we analyzed separately all the spectra of the SD using a unique value of $N_{\text{H}}=N_{\text{H}}^{\text{late}}$, which is fixed during the fit. Normalization and photon index, instead, are left free.\\
An alternative method for the derivation of $N_{\text{H}}$ is the fitting of all the spectra simultaneously imposing a unique value of $N_{\text{H}}$ that is left free. On the other hand, since $N_{\text{H}}$ and photon index are correlated, an intrinsic spectral evolution can induce an incorrect estimation of $N_{\text{H}}$. For the same reason we do not fit the spectra adopting a free $N_{\text{H}}$, since we would obtain an evolution of photon index strongly affected by the degeneracy with $N_{\text{H}}$.\\
In this regard, we tested how our results about spectral evolution depend on the choice of $N_{\text{H}}$. On average we found that the fits of the SD spectra remain good ($\mathrm{stat/dof\lesssim1}$) for a variation of $N_{\text{H}}$ of about 50$\%$. As a consequence, the photon index derived by the fit would change at most of 30$\%$. Therefore the error bars reported in all the plots $\alpha-F$ are possibly under-estimated, but even considering a systematic error that corresponds to $\sim 30\%$ of the value itself would not undermine the solidity of the results.\\

{\noindent  \bf Extrapolation of $\text{F}_{\text{max}}$.} 
\label{extra_sec}
We explain here how we extrapolated the $F_{\text{max}}$ for the GRBs of the second sample, for which the XRT light curve starts directly with a SD. We consider the peak time $T_{\text{p}}^{\text{BAT}}$ of the BAT pulse that precedes the SD. In the assumption that the SD starts at $T_{\text{p}}^{\text{BAT}}$, we can derive $F_{\text{max}}$ using the following procedure. We consider the 0.5-10 keV flux $F(t_i)$ for each bin time $t_i$ in the SD, derived from spectral analysis. Then we fit these points with a power law
\begin{equation}
F(t_i)=F_{\text{max}}\Big(\frac{t_i}{t_0}\Big)^{-s}
\end{equation}
with $s>0$ and imposing that $t_0=T_{\text{p}}^{\text{BAT}}$. Finally we derive the best fit value of $F_{\text{max}}$ with the associated 1$\sigma$ error. The error of $F_{\text{max}}$ has a contribution coming from the error associated to $s$ and another associated to $t_0$, as well as from the assumption of a power law as fitting function. The value of $T_{\text{p}}^{\text{BAT}}$ is obtained fitting the BAT pulse with a Gaussian profile. Since usually the BAT pulse can have multiple sub-peaks and taking also into account possible lags between XRT and BAT peaks, we adopt a conservative error associated to $T_{\text{p}}^{\text{BAT}}$ equal to 5 seconds.\\

{\noindent  \bf HLE from infinitesimal duration pulse.} We assume that an infinitesimal duration pulse of radiation is emitted on the surface of a spherical shell, at radius $R_0$ from the center of the burst. Such treatment implicitly assumes particles that cool on timescales much smaller than the dynamical timescales. Therefore, all the X-ray tail emission is dominated by photons departed simultaneously from the last emitting surface. The jet has an aperture angle $\vartheta_j$ and it expands with a bulk Lorentz factor $\Gamma$. We assume also that the comoving spectrum is the same on the whole jet surface. The temporal evolution of the observed flux density is given by\cite{Oganesyan2020}:
\begin{equation}\label{f_nu_eq}
F_{\nu}(t_{\text{obs}}) \propto S_{\nu'}(\nu/ \mathcal{D}(\vartheta)) \mathcal{D}^{2}(\vartheta) \cos (\vartheta)
\end{equation} 
with $S_{\nu'}(\nu/ \mathcal{D}(\vartheta))$ the comoving spectral shape, $\mathcal{D}(\vartheta)$ the Doppler factor and $\vartheta$ the angle measured from the line of sight, which is assumed to coincide with the jet symmetry axis. The observer time $t_{\text{obs}}$ is related to the angle $\vartheta$ through this formula:
\begin{equation}\label{t_obs}
t_{\text{obs}}(\vartheta)=t_{\text{em}}(1-\beta \cos \vartheta)
\end{equation}
where $t_{\text{em}}$ is the emission time. Eq.~(\ref{f_nu_eq}) is valid for $\vartheta<\vartheta_j$, while for $\vartheta>\vartheta_j$ the emission drops to zero. This implies that for $t_{\text{obs}}>t_{\text{em}}(1-\beta \cos \vartheta_j)$ the flux drops to zero. At each time $t_{\text{obs}}(\vartheta)$ the observer receives a spectrum that is Doppler shifted by a factor $\mathcal{D}(\vartheta)$ with respect to the comoving spectrum. If the comoving spectrum is curved, i.e. if $\frac{d^2}{d\nu'^2}S_{\nu'}\neq 0$, then also the photon index is a function of time\cite{Lin2017}. The shape of the resulting curve $\alpha-F$ is determined only by the spectral shape and the comoving peak frequency $\nu'_{\text{p}}$, while it is independent on the emission radius $R_0$ and the bulk Lorentz factor $\Gamma$.\\
We notice that the observed photon index goes from $0.5-1.0$ up to $2.0-2.5$, consistent with the slopes of a synchrotron spectrum before and after the peak frequency. Indeed for a population of particles with an injected energy distribution $N(\gamma)\propto \gamma^{-p}$ that has not completely cooled, the expected shape of the spectrum is $F_{\nu}\sim \nu^{1/3}$ $(\alpha=2/3)$ for $\nu<\nu_c$ and $F_{\nu}\sim \nu^{-p/2}$ $(\alpha=p/2+1)$ for $\nu>\nu_{\text{m}}\gtrsim \nu_{\text{c}}$. Hereafter, if not otherwise specified, we assume a spectral shape given by a smoothly broken lower law, which well approximates the synchrotron spectrum below and above the peak frequency. The form of the adopted spectral shape is
\begin{equation}
S_{\nu}\propto \frac{1}{\left(\dfrac{\nu}{\nu_{0}}\right)^{\alpha_{\text{s}}}+\left(\dfrac{\nu}{\nu_{0}}\right)^{\beta_{\text{s}}}}
\end{equation}
with $\alpha_{\text{s}}=-1/3$ and $\beta_{\text{s}}=1.5$. The peak frequency $\nu_{\text{p}}$ of the energy spectrum $\nu S_{\nu}$ is related to $\nu_0$ through the following relation:
\begin{equation}
\nu_{\text{p}}=\left(-\frac{2+\alpha_{\text{s}}}{2+\beta_{\text{s}}}\right)^{\frac{1}{\alpha_{\text{s}}-\beta_{\text{s}}}}\nu_0
\end{equation}
At each arrival time we compute the flux and the photon index in the XRT band using eq.~(\ref{f_nu_eq}). In particular, the XRT flux is given by
\begin{equation}
F_{0.5-10\text{ keV}}(t_{\text{obs}})=\int_{0.5 \text{ keV}/h}^{10 \text{ keV}/h} F_{\nu}(t_{\text{obs}})d \nu
\end{equation}
where $h$ is Planck's constant, while the photon index is computed as\cite{Genet2009,Lin2017}
\begin{equation}
\alpha(t_{\text{obs}})=1-\frac{\log{[F_{\nu=10 \text{ keV}/h}(t_{\text{obs}})/F_{\nu=0.5 \text{ keV}/h}(t_{\text{obs}})]}}{\log{(10 \text{ keV}/0.5 \text{ keV}})}
\end{equation}
This method for the evaluation of photon index is valid in the limit of a spectrum that can be always approximated with a power law as it passes through the XRT band, which is the case for typical prompt emission spectra.\\
In addition to the SBPL, we test HLE also using other spectral shapes. We first adopt a Band function\cite{Band1993} with the following form:
\begin{equation}
B(\epsilon)=
\begin{cases}
\epsilon^{1+\alpha_{\text{s}}}e^{-\epsilon} \quad  \epsilon < \alpha_{\text{s}}-\beta_{\text{s}}\\
(\alpha_{\text{s}}-\beta_{\text{s}})^{\alpha_{\text{s}}-\beta_{\text{s}}}e^{-\alpha_{\text{s}}+\beta_{\text{s}}} \epsilon^{1+\beta_{\text{s}}}\quad  \epsilon > \alpha_{\text{s}}-\beta_{\text{s}}
\end{cases}
\end{equation}
where $\epsilon=\nu/\nu_0$. In this case the peak of the energy spectrum is at $\nu_{\text{p}}=(2+\alpha_{\text{s}})\nu_0$. The resulting spectral evolution is very similar to the case of SBPL, as visible in Supplementary Fig.~12a.\\
As a final attempt, we use synchrotron spectrum emitted by a population of particles with an initial energy distribution $N(\gamma)\propto \gamma^{-p}$. Synchrotron is considered the dominant radiative process in prompt emission of GRBs\cite{Rees1994,Zhang2020}. In the fast cooling regime, the particle distribution becomes
\begin{equation}
N(\gamma)\propto
\begin{cases}
\gamma^{-2} & \gamma_c<\gamma<\gamma_{\text{m}}\\
\gamma^{-(p+1)} & \gamma>\gamma_{\text{m}}
\end{cases}
\end{equation}
The only three parameters that define the shape of the synchrotron spectrum are $\nu_{\text{m}}\propto \gamma_{\text{m}}^2$, $\nu_{\text{c}}\propto \gamma_{\text{c}}^2$ and $p$. For the computation of the spectrum we use\cite{Rybicky}:
\begin{equation}
F_{\nu}\propto \int_{\gamma_{\text{c}}}^{\infty}P(\nu, \gamma)N(\gamma)d\gamma
\end{equation}
with
\begin{equation}
P(\nu, \gamma)\propto B \left[\left(\frac{\nu}{\nu_{\text{ch}}}\right)\int_{\frac{\nu}{\nu_{\text{ch}}}}^{\infty}K_{5/3}(x)dx\right], \quad \nu_{\text{ch}}\propto \gamma^2 B
\end{equation}
where $B$ is the magnetic field and $K_{5/3}(x)$ is the modified Bessel function of order 5/3. The resulting spectral evolution for values of $\nu_{\text{m}}/\nu_{\text{c}}=1$ and $\nu_{\text{m}}/\nu_{\text{c}}=10$ is reported in Supplementary Fig.~11. A value of $\nu_{\text{m}}/\nu_{\text{c}}\sim1$ is expected in the marginally fast cooling regime\cite{Kumar08,Ben13,Ben18}, which is favored by broad-band observations of GRB prompt spectra\cite{Oga17,Oga18,Oga19,Rav18,Rav19,Bur2020}.
Finally we test how the sharpness of the spectral peak can affect our results. In particular we consider again a SBPL and we generalize the formula adding a sharpness parameter $n$:
\begin{equation}
S_{\nu}^{(n)}\left(\frac{\nu}{\nu_{0}}\right)\propto \left[\frac{1}{\left(\dfrac{\nu}{\nu_{0}}\right)^{n\alpha_{\text{s}}}+\left(\dfrac{\nu}{\nu_{0}}\right)^{n\beta_{\text{s}}}}\right]^{1/n}
\end{equation}
where larger values of $n$ correspond to sharper spectral peaks. As visible in Supplementary Fig.~12b, where we have adopted $n=4$, the shape of the curves becomes flatter at the beginning and at the end of the decay, but with no substantial steepening of the intermediate part. This is attributable to HLE that imposes an evolution of the observed peak frequency like $t_{\text{obs}}^{-1}$. Thus, while the initial and final values of photon index are dictated by the spectral shape, the steepness of the transition from the initial to the final value is governed by HLE and is independent on the spectral shape. In conclusion, no one of the alternative spectral shapes that we tested is able to reconcile HLE with the observed spectral evolution.\\
We finally test how the $\alpha-F$ relation changes if we assume a structured jet with an angle-dependent comoving spectrum. In particular, we consider a spectral peak energy that is nearly constant inside an angle $\vartheta_{c}$ (measured with respect to the line of sight) and starts to decrease outside it. Regardless of the choice of the specific law for the angular dependence (e.g. Gaussian or power law), the HLE can reproduce the $\alpha-F$ relation only if all the analyzed GRBs have a fine-tuned value of $\vartheta_{c}<1^{\circ}$. Such a small value of $\vartheta_{c}$, on the other hand, would imply a very short steep decay, in contradiction with observations.\\

{\noindent  \bf Adiabatic Cooling.} 
In this section we derive the effect of adiabatic cooling of the emitting particles\cite{Pan19} on the light curve and the spectral evolution of X-ray tails. We assume that the emission is dominated by a single species of particles that can be treated as a relativistic gas in adiabatic expansion. We assume also that there is no interaction with other species of particles. If the particles are embedded in a region of comoving volume $V'$, an adiabatic expansion satisfies the equation
\begin{equation}
\expval{\gamma}^3V'=\mathrm{const}
\label{eq:adiabatic_expansion}
\end{equation}
where $\expval{\gamma}$ is the average Lorentz factor of the emitting particles in the comoving frame. The last equation is valid in the limit in which the adiabatic cooling timescale is smaller than the cooling time of other radiative processes, such as synchrotron or inverse Compton. Namely, particles radiate only a negligible fraction of their internal energy during the expansion of the system. Regarding the radial dependence of the volume $V'$, we distinguish two cases:\\
1) thick shell, with a comoving width $\Delta R'$ that does not evolve with time, hence $V'\propto R^2 \Delta R' \propto R^2$
\\
2) thin shell, with a comoving width $\Delta R'$ that evolves linearly with $R$, hence $V'\propto R^2 \Delta R' \propto R^3$.\\
We assume that the dominant radiative process is synchrotron. The evolution of the spectrum in the observer frame is therefore fully determined once we know how the spectrum normalization $F_{\nu_{\text{p}}}$ and the peak frequency $\nu_{\text{p}}$ evolve in time. These two quantities, under the assumption of constant total number of emitting particles and constant bulk Lorentz factor $\Gamma$, take the following form:
\begin{equation}
\label{ad_eq}
F_{\nu_{\text{p}}}\propto B, \quad \nu_{\text{p}}\propto \expval{\gamma}^2B
\end{equation}
where $B$ is the magnetic field (assumed tangled) as measured by a comoving observer. As described in the main text, we adopt the following parametrization for the magnetic field:
\begin{equation}
B=B_0\left(\frac{R}{R_{0}}\right)^{-\lambda}
\end{equation} 
where $\lambda\geq0$, under the reasonable assumption that magnetic field has to decrease or at most remain constant during the expansion. The value of $R_0$ corresponds to the radius where particles are injected, namely when adiabatic cooling starts to dominate. We use the integration along the Equal Arrival Time Surfaces (EATS) to compute the evolution of flux, as done for HLE from finite-duration pulse (see Supplementary Note 1), with the only difference that in this case the emission never switches off. The final form of the integral is
\begin{equation}\label{F_EATS}
F_{\nu}(t_{\text{obs}}) \propto \int_{0}^{\vartheta_j} S_{\nu^\prime}(\nu/\mathcal{D}(\vartheta)) \left(\frac{R(\vartheta,t_{\text{obs}})}{R_{0}}\right)^{-\lambda} \mathcal{D}^{3}(\vartheta) \sin \vartheta \cos \vartheta d \vartheta
\end{equation}
where the factor $(R/R_0)^{-\lambda}$ comes from  $I_{\nu_{\text{p}}^{\prime}}^{\prime} \propto B$, while $\nu_{\text{p}}'$ evolves in time according to eq.~(\ref{ad_eq}).\\

{\noindent  \bf Parameter estimation via Monte Carlo Markov Chain (MCMC).} 
In order to fully explore the parameter space of the adiabatic cooling model, we performed a MCMC, using the emcee algorithm \cite{Foreman-Mackey}. The setup of our analysis is described in the following points:\\
1) The model contains as free parameters $E_{\text{p}}$, $\lambda$ and $\tau_{\text{ad}}=R/2c\Gamma^2$, which are the peak energy at the beginning of the steep decay, the decay index of the magnetic field, and the adiabatic time scale. The inclusion of $\Gamma$ as free parameter returns a flat posterior distribution, indicating that the model is insensitive to it. Therefore we performed the analysis fixing $\Gamma=100$.\\
2) The MCMC in performed jointly for flux and photon index evolution. The adopted likelihood is:
\begin{equation}
\log(\mathcal{L})=-\frac{1}{2} \sum_{n}\left[\frac{\left(\phi_{n}-\bar{\phi}(t_n)\right)^{2}}{s_{\phi,n}^{2}}+\ln \left(2 \pi s_{\phi,n}^{2}\right)\right]-\frac{1}{2} \sum_{n}\left[\frac{\left(\alpha_{n}-\bar{\alpha}(t_n)\right)^{2}}{s_{\alpha,n}^{2}}+\ln \left(2 \pi s_{\alpha,n}^{2}\right)\right]
\end{equation}
where $\phi=F/F_{\text{max}}$, $\alpha $ is the photon index and with $\bar{\phi}$, $\bar{\alpha} $ we indicate the value predicted by the model at each time $t_n$. Moreover,
\begin{equation}
s_{\phi,n}^{2}=\sigma_{\phi,n}^2+f_{\phi}\cdot \bar{\phi}(t_n), \quad s_{\alpha,n}^{2}=\sigma_{\alpha,n}^2+f_{\alpha}\cdot \bar{\alpha}(t_n)
\end{equation}
where $\sigma_{\phi}$ and $\sigma_{\alpha}$ are the errors, while $f_{\phi}$ and $f_{\alpha}$ are introduced to take into account possible underestimation of the errors. Since keeping $f_{\alpha}\neq0$ leads to a posterior distribution of $f_{\alpha}$ peaked around $\sim 10^{-7}$, the parameter estimation was performed fixing $f_{\alpha}=0$. Instead, we keep $f_{\phi}\neq0$ taking into account that the error on the flux resulting from the fitting of time-averaged spectrum may not represent the true flux error over the time bin.\\
3) The MCMC runs until the number of steps exceeds 100 times the autocorrelation time (its maximum) and the averaged autocorrelation time (over 100 steps) becomes constant within 1$\%$ accuracy. The burn-in is chosen as twice of autocorrelation time. As an example, we show in Supplementary Fig.~7 the evolution of the autocorrelation time as a function of the steps.\\
The resulting parameter estimation is summarized in Tab.~\ref{tab_AC}. The uncertainties are reported based on the 16th, 50th, and 84th percentiles of the samples in the marginalized distributions ($1\sigma$ level of confidence). An example of corner plot obtained via MCMC is shown in Supplementary Fig.~8. In Supplementary Fig.~9 - 10, we show for each burst the observed temporal evolution of photon index and normalized flux in comparison with the curves produced with 500 random draws from the posterior sample set of the MCMC.\\
We performed an analogous MCMC analysis adopting the model of HLE from an instantaneous emission. However, the algorithm is unable to converge, demonstrating that the model cannot successfully match with the observations. The only way to obtain converged chains by this model is to admit extreme and unrealistic values of $f_{\phi}$, of the order $10^4-10^8$. The only exception is GRB 090621, which can be fitted by HLE alone. This is the only case where it is meaningful to compute the Bayes factor between HLE and AC, which results to be $\sim200$. Thus we prove that the adiabatic cooling model is strongly preferred for all the analyzed cases.\\
The model comparison (adiabatic cooling + HLE against HLE-only) is also done assuming different spectral shapes: SBPL, Band and synchrotron. The spectral parameters are the same as those adopted before. For synchrotron, we use $\nu_{\text{c}}=\nu_{\text{m}}$ (the case $\nu_{\text{c}}\neq\nu_{\text{m}}$ does not improve the goodness of fit). In order to compare the goodness of fit of the two models, we used the Akaike Information Criterion (AIC), which is defined as $\mathrm{AIC}=2k-2\ln(\mathcal{L})$, where $k$ in the number of parameters of the model, and $\mathcal{L}$ is the best fit likelihood, that is, $2\ln(\mathcal{L})=-\chi^2$. In Tab.~\ref{Tab:fit_comp} we show the value of $\Delta_\mathrm{AIC}=\mathrm{AIC_{HLE}-AIC_{AC}}$ for each spectral shape. For all cases, the adiabatic cooling is significantly favoured with respect to HLE.

\section*{Data availability}
Raw data are public and available in the UK Swift Science Data Centre at the University of Leicester. The light curve data are taken at this link:\\
\url{https://www.swift.ac.uk/xrt\_curves/GRB\_ID/flux.qdp}\\
where \url{GRB\_ID} is the GRB observation ID. The spectra are obtained at this link:\\
\url{https://www.swift.ac.uk/xrt\_spectra/addspec.php?targ=GRB\_ID}\\
where \url{GRB\_ID} is the ID number of the GRB. The details of the automatic spectral analysis can be found here:\\
\url{https://www.swift.ac.uk/xrt\_spectra/docs.php}\\
Derived data are available from the corresponding author on request.

\section*{Code availability}
Codes used to produce the plots in this paper are available in this public repository:\\
\url{https://github.com/samueleronchini/Nature\_communications}\\
XSPEC and PyXspec are freely available online at the following links:\\
\url{https://heasarc.gsfc.nasa.gov/xanadu/xspec/}\\
\url{https://heasarc.gsfc.nasa.gov/docs/xanadu/xspec/python/html/index.html}

{\noindent  \bf Acknowledgements}\\
The research leading to these results has received funding from the European Union’s Horizon 2020 Programme under the AHEAD2020 project (Grant agreement n. 871158). G. Ghir. acknowledges the support from the ASI-Nustar Grant (1.05.04.95). M.B., P.D., and G.G. acknowledge support from PRIN-MIUR 2017 (Grant 20179ZF5KS). G.O. acknowledges financial contribution from the agreement ASI-INAF n.2017-14-H.0. SA acknowledges the PRIN-INAF "Towards the SKA and CTA era: discovery, localization and physics of transient sources" and the ERC Consolidator Grant “MAGNESIA” (nr.817661). MGB and PDA acknowledge ASI grant I/004/11/3. OS acknowledges the INAF-Prin 2017 (1.05.01.88.06) and the Italian Ministry for University and Research grant “FIGARO” (1.05.06.13) for support. GO and SR are thankful to INAF – Osservatorio Astronomico di Brera for kind hospitality during the completion of this work. This work made use of data supplied by the UK Swift Science Data Centre at the University of Leicester.\\

{\noindent  \bf Author Contributions}\\
S.R., G.O. and M.B. conceived the idea behind the paper. S.R. performed the sample selection and the spectral analysis, with the help of M.G.B. and P.D.. S.R. and G.O. implemented the models and related numerical calculations. S.R. and G.O. developed the codes to compare models and data. S.R., G.O., M.B., S.A., M.G.B., F.B., S.D., P.D., G.Ghir., G.Ghis., M.E.R. and O.S.S. contributed in the discussion and in the interpretation of the results, as well as in writing and revising the manuscript.\\

{\noindent  \bf Competing Interests} The authors declare no competing
interests

\clearpage
\begin{figure}
	\centering
	\includegraphics[width=0.65\textwidth]{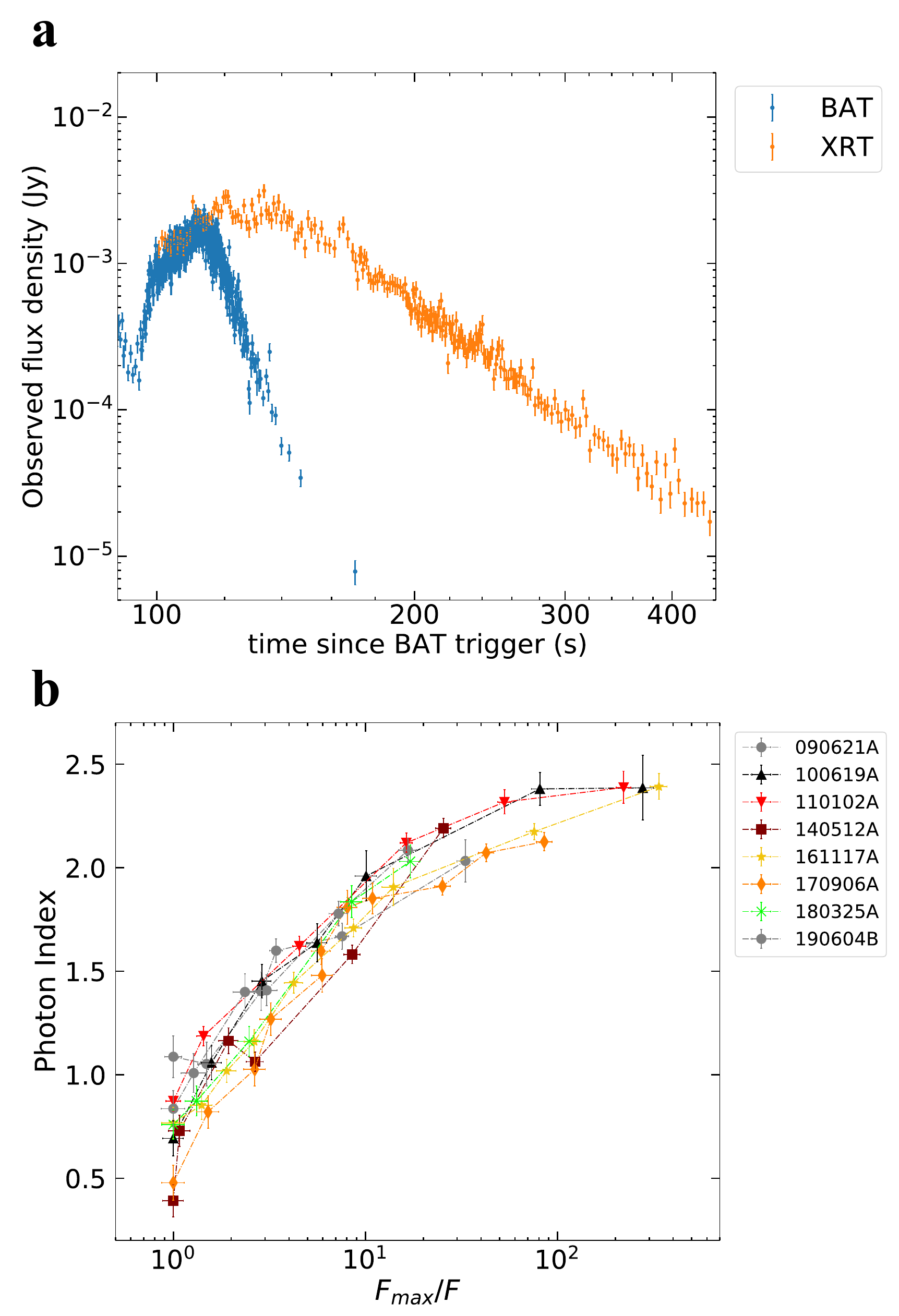}
	\caption{{\bf The steep decay phase and the correspondent spectral evolution.} In panel \textbf{a} we show an example of a light curve of an X-ray tail selected from our sample, taken from the GRB 161117A. We show on the same plot the XRT (orange) and the BAT (blue) flux density at 1 keV and 50 keV, respectively. The XRT light curve decays less steeply than BAT because of the evolution of the peak energy. The error bars represent $1\sigma$ uncertainties and they are derived from the Swift archive. In panel \textbf{b} we report the spectral evolution of the X-ray tail for all the GRBs in the first sample (shown with different colors). The photon index $\alpha$ is represented as a function of the reciprocal of the normalized flux $F_{\text{max}}/F$. Time flows from left to right. The error bars represent $1\sigma$ uncertainties, calculated via spectral fitting in XSPEC. In the legend we report the name of each GRB.}
	\label{peaked_main}
\end{figure}

\begin{figure}
	\centering
	\includegraphics[width=0.75\textwidth]{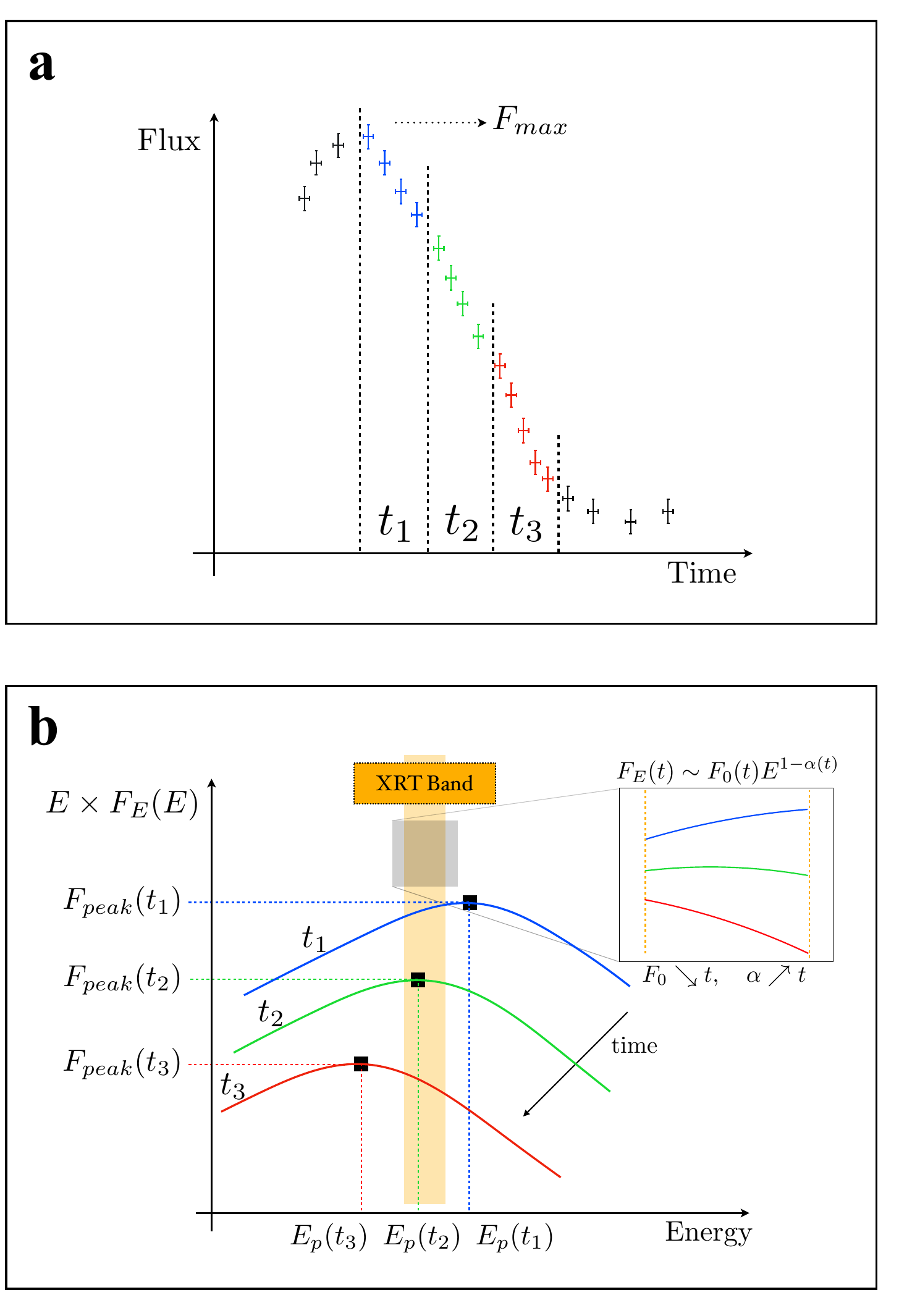}
	\caption{{\bf Illustration of the spectral evolution caused by a shift of the spectrum towards lower energies.} The transition of the spectral peak through the XRT band explains the observed spectral softening. The spectra in panel {\bf b} coloured in blue, green and red correspond to the three temporal bins shown in panel {\bf a} with the same colours. The inset in panel {\bf b} shows how the local spectral slope evolves as observed in the XRT band. Since in the panel {\bf b} we plot the flux density, the local slope in the XRT band is given by $1-\alpha$, where $\alpha$ is the photon index. Both the x and y axes in panels {\bf a} and {\bf b} have arbitrary units.}
	\label{sketch}
\end{figure}

\begin{figure}
	\centering
	\includegraphics[width=0.7\textwidth]{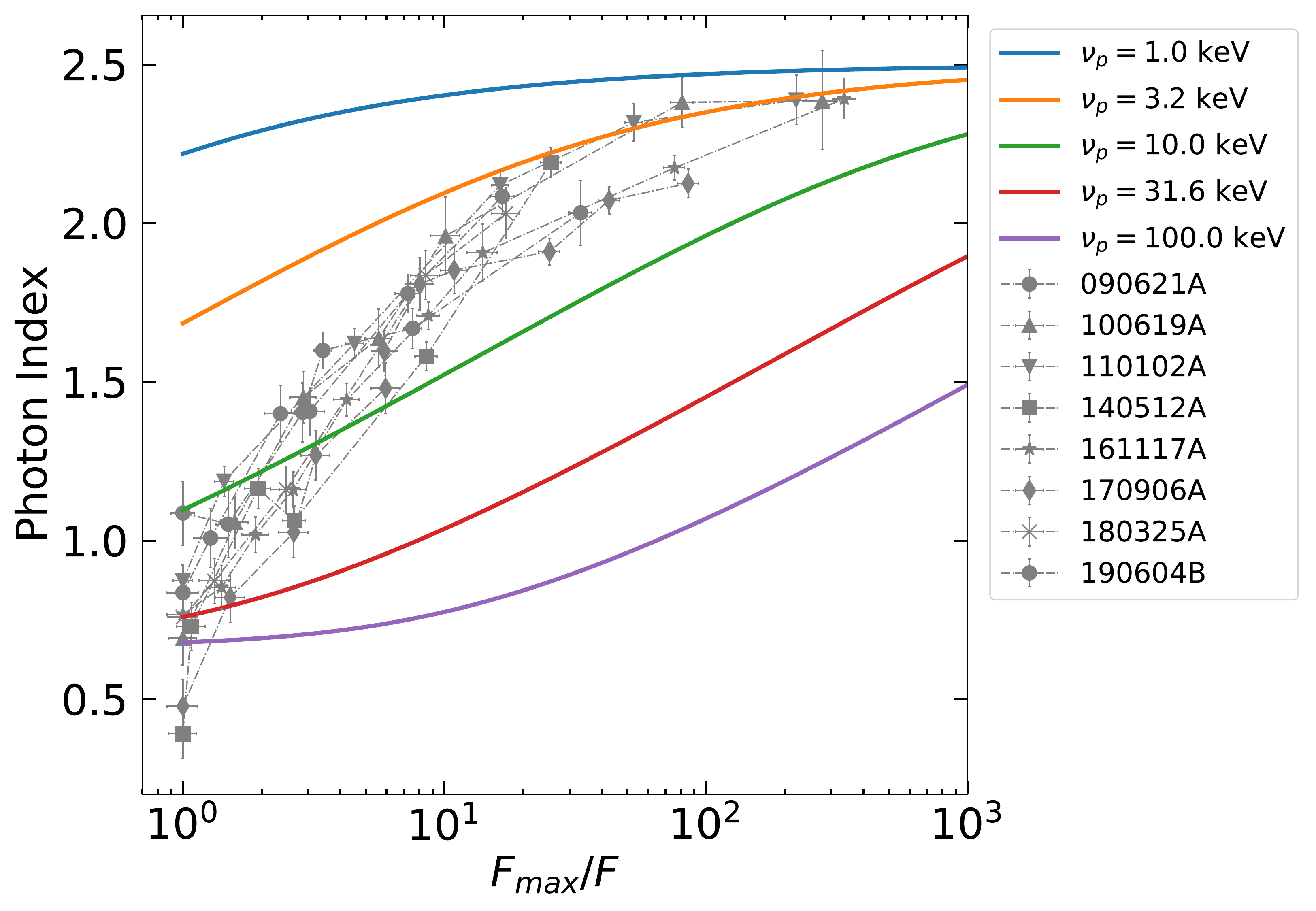}
	\caption{{\bf Spectral evolution expected for HLE from an infinitesimal duration pulse}. The comoving spectrum is assumed to be a SBPL. The several colors indicate the observed peak frequency at the beginning of the decay. The error bars represent $1\sigma$ uncertainties, calculated via spectral fitting in XSPEC. In the legend we report the name of each GRB.}
	\label{HLE_spe_evo}
\end{figure}

\begin{figure}
	\centering
	\includegraphics[width=0.6\textwidth]{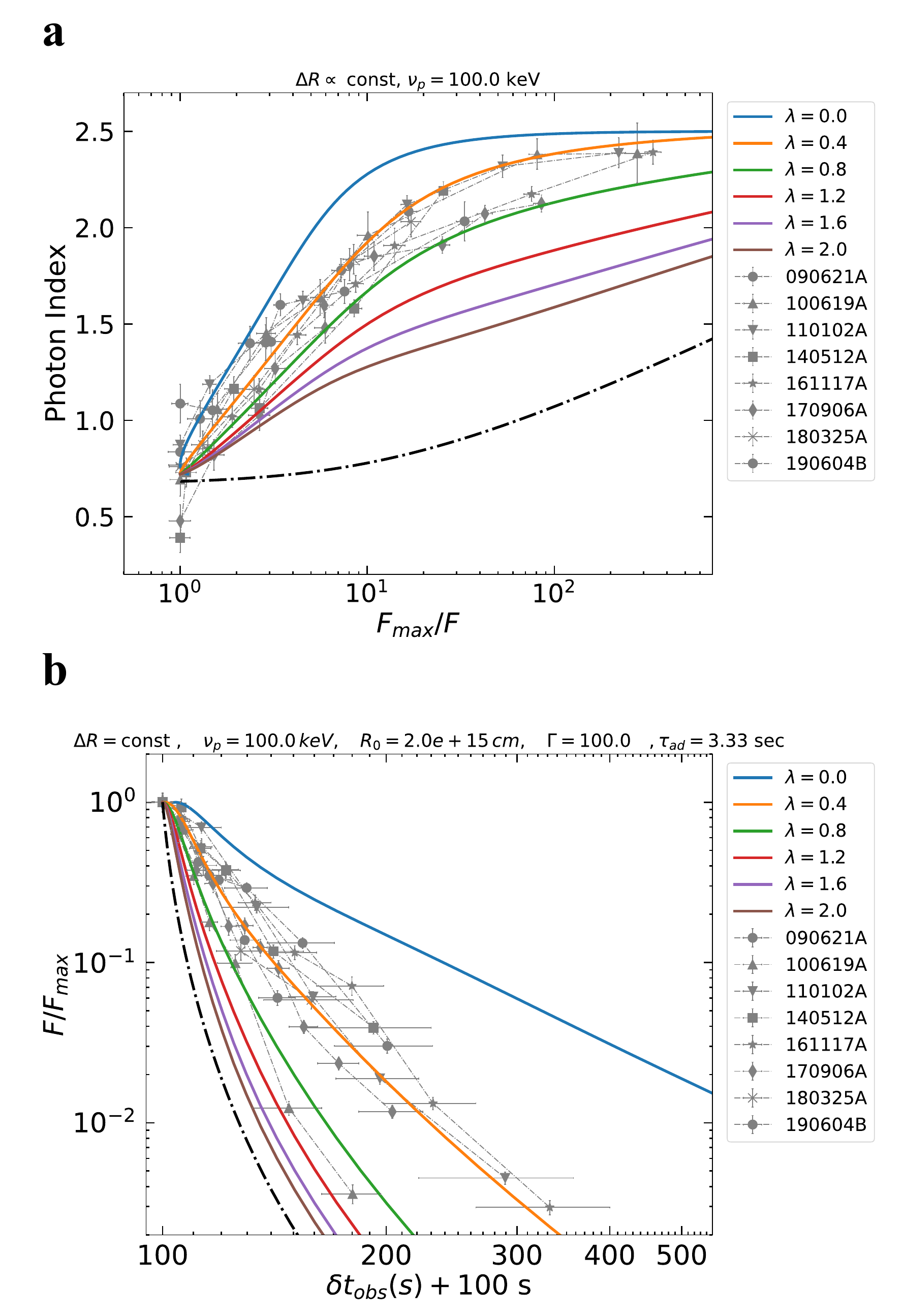}
	\caption{{\bf Spectral and temporal evolution in case of adiabatic cooling.} In panel {\bf{a}} we show the $\alpha-F$ relation expected in the case of adiabatic cooling (solid lines). The theoretical curves are computed taking also into account the effect of HLE. The value of $\lambda$ specifies the evolution of the magnetic field. We adopt a SBPL as spectral shape with $\alpha_\text{s}=-1/3$ and $\beta_\text{s}=1.5$, an initial observed peak frequency of 100 keV and a thickness of the expanding shell that is constant in time. The dot-dashed line is the evolution expected in case of HLE without adiabatic cooling, assuming the same spectral shape and initial observed peak frequency. The error bars represent $1\sigma$ uncertainties, calculated via spectral fitting in XSPEC. In panel {\bf{b}} we show the temporal evolution of normalized flux expected in case of adiabatic cooling. $\delta t_{\text{obs}}+100 $ s is the time measured from the peak of the decay shifted at 100 s, the typical starting time of the tail emission detected by XRT. We adopt the same parameters as in {\bf{a}}, assuming $R_0=2\times 10^{15}$ cm and $\Gamma=100$. The dot-dashed line is the corresponding HLE model without accounting for adiabatic cooling. $\tau_{\text{ad}}=R_0/2c\Gamma^2$ indicates the timescale of adiabatic cooling, which is the same of HLE. The vertical error bars represent $1\sigma$ uncertainties and they are calculated via spectral fitting in XSPEC, while horizontal error bars represent the width of the time bin. In the legend we report the name of each GRB.}
	\label{ad_int_spe}
\end{figure}

\begin{figure}
	\centering
	\includegraphics[width=0.65\textwidth]{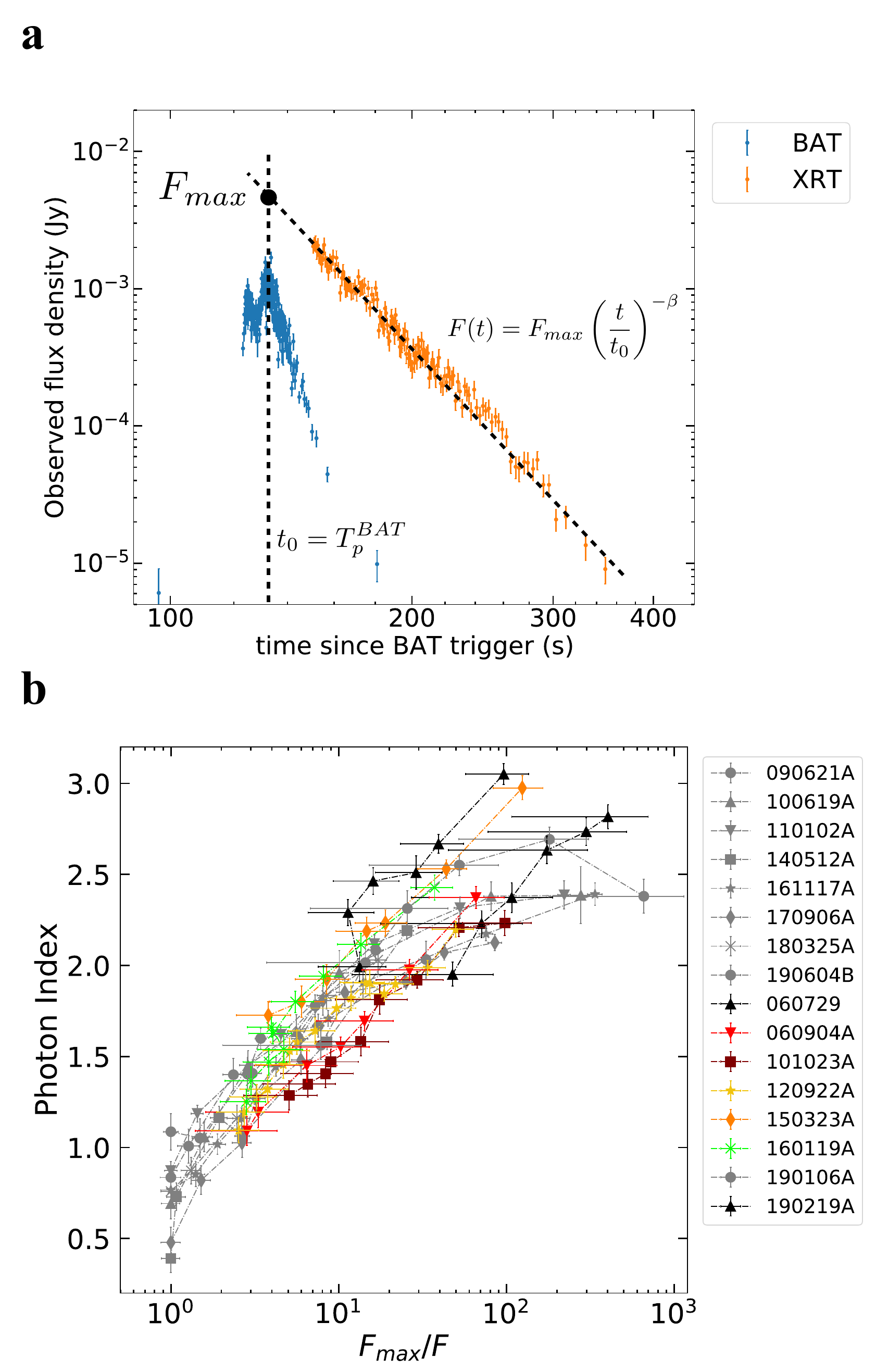}
	\caption{{\bf The steep decay phase and the correspondent spectral evolution for the extended sample.} In panel {\bf a} we show an example of a light curve of an X-ray tail selected for our extended sample, taken from GRB 150323A. We report on the same plot the XRT (orange) and the BAT (blue) flux density at 1 keV and 50 keV, respectively. The peak flux $F_{\text{max}}$ is estimated extrapolating the X-ray tail back to the BAT peak. The error bars represent $1\sigma$ uncertainties and they are derived from the Swift archive. In panel {\bf b} we show the spectral evolution of our extended sample of GRBs, which present a steep decay at the beginning of the XRT light curve, preceded by the brightest BAT pulse since the trigger time. The evolution of $\alpha$ lies on the same region of the plane occupied by the original sample, indicated in grey. The error bars represent $1\sigma$ uncertainties, calculated via spectral fitting in XSPEC. In the legend we report the name of each GRB.}
	\label{spe_evo_extra}
\end{figure}

\begin{figure}
	\centering
	\includegraphics[width=0.6\textwidth]{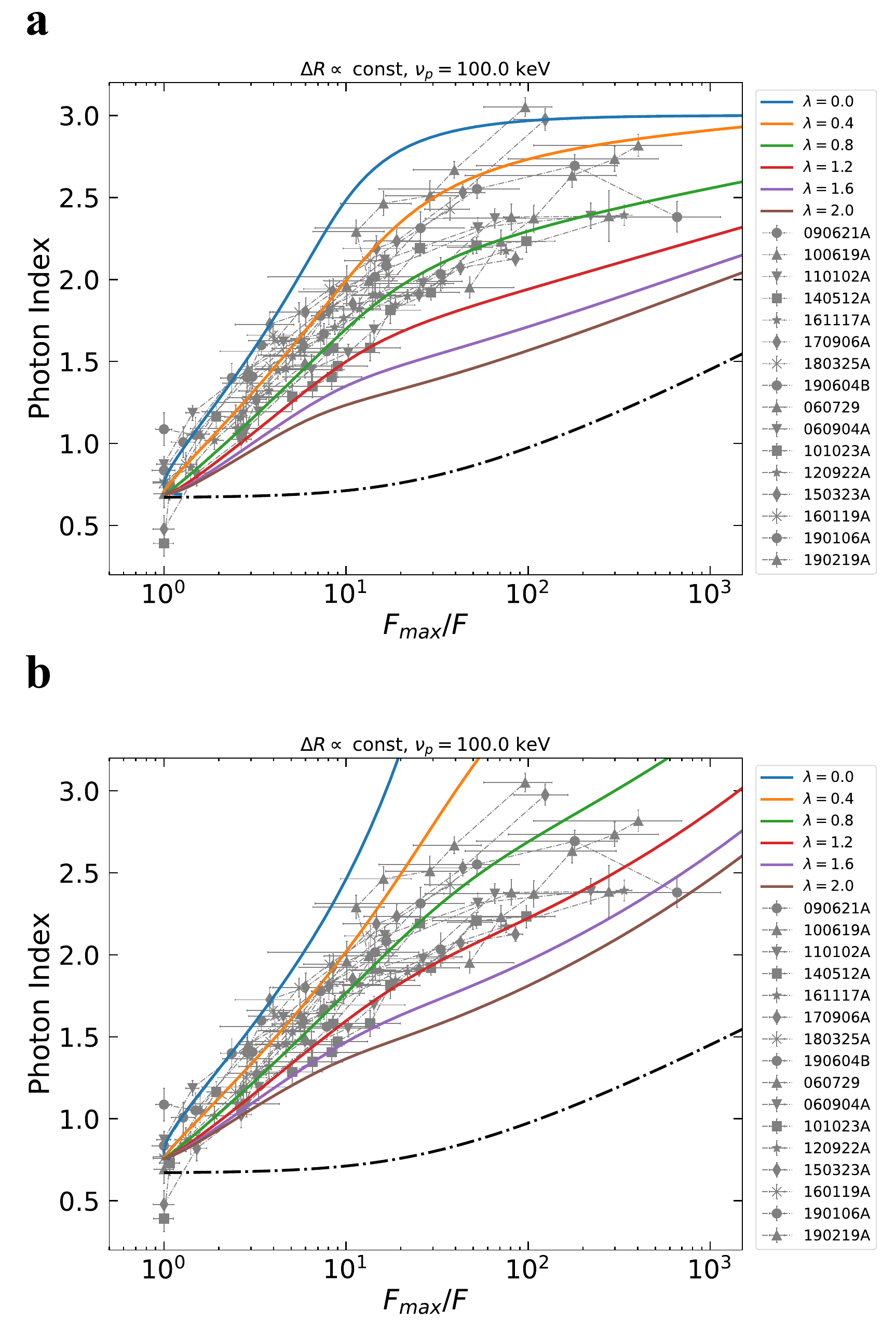}\\
	\caption{{\bf Spectral evolution expected in case of adiabatic cooling (solid lines) superimposed to the extended sample.} In panel {\bf a} the theoretical curves are computed considering adiabatic cooling and inefficient synchrotron cooling, taking also into account the effect of HLE. The value of $\lambda$ specifies the evolution of the magnetic field. We adopt a SBPL as spectral shape with $\alpha_\text{s}=-1/3$ and $\beta_\text{s}=2.0$, an initial observed peak frequency of 100 keV and a thickness of the expanding shell that is constant in time. In panel {\bf b} we show the spectral evolution expected in case of combined adiabatic cooling and mild synchrotron cooling. The adopted spectral shape is a SBPL plus an exponential cutoff. The initial peak frequency is 100 keV. The theoretical curves are computed taking also into account the effect of HLE. In both panels, the error bars represent $1\sigma$ uncertainties, calculated via spectral fitting in XSPEC, and the dot-dashed line is the evolution expected considering only HLE, assuming the same spectral shape and initial observed peak frequency. In the legend we report the name of each GRB.}
	\label{spe_evo_extra+HLE}
\end{figure}

\newpage

\renewcommand{\arraystretch}{1.5}
\begin{table}
\begin{tabular}{|c|c|c|c|}
\hline
GRB & E$_{\rm peak}$ {\rm (keV)}& $\lambda$ & $\tau_{\text{ad}}${\rm (s)}  \\ \hline

090621A & $18_{-2}^{+3}$ & $2.11_{-0.54}^{+0.56}$ & $24.4_{-3.0}^{+4.7}$ \\ \hline
100619A & $>129$ & $0.47_{-0.07}^{+0.11}$ & $0.3_{-0.2}^{+1.0}$ \\ \hline
110102A & $46_{-9}^{+15}$ & $0.61_{-0.10}^{+0.10}$ & $5.8_{-1.1}^{+1.9}$ \\ \hline
140512A & $>323$ & $0.48_{-0.03}^{+0.04}$ & $0.9_{-0.4}^{+0.9}$ \\ \hline
161117A & $80_{-21}^{+55}$ & $0.69_{-0.10}^{+0.10}$ & $6.2_{-2.3}^{+2.0}$ \\ \hline
170906A & $135_{-53}^{+204}$ & $0.66_{-0.09}^{+0.10}$ & $3.0_{-1.5}^{+1.6}$ \\ \hline
180325A & $>122$ & $0.39_{-0.05}^{+0.06}$ & $0.8_{-0.5}^{+1.3}$ \\ \hline
190604B & $54_{-20}^{+227}$ &  $0.45_{-0.15}^{+0.25}$ & $3.5_{-2.8}^{+2.6}$ \\

\hline
\end{tabular}
\caption{{\bf Results of the parameter estimation via MCMC, adopting the adiabatic cooling model}. The confidence intervals and the lower limits represent the 16th, 50th, and 84th percentiles of the samples in the marginalized distributions (i.e. $1\sigma$ level of confidence).}
\label{tab_AC}

\end{table}

\begin{table}

\begin{tabular}{|c| c| c| c|}
\hline
GRB & $(\Delta_{\text{AIC}})|_{\text{SBPL}}$ & $(\Delta_{\text{AIC}})|_{\text{Band}}$ & $(\Delta_{\text{AIC}})|_{\text{Sync}}$\\\hline
090621A &8     &6      &9\\\hline
100619A &69    &66     &67\\  \hline
110102A &145   &141    &145\\ \hline
140512A &193   &190    &43\\\hline
161117A &132   &124    &133\\ \hline
170906A &148   &135    &145\\ \hline
180325A &80    &76     &91\\\hline
190604B &61    &59     &65\\\hline
\end{tabular}
\caption{{\bf Comparison of best fit statistics between Adiabatic Cooling (AC) and HLE, adopting a SBPL, a Band function or a synchrotron spectrum, using the Akaike Information Criterion (AIC)}. The large values of $(\Delta_{\text{AIC}})|_{\text{spectrum}}=(\text{AIC}_{\text{HLE}}-\text{AIC}_{\text{AC}})|_{\text{spectrum}}$ indicate that, regardless of the assumed spectral shape, the HLE from efficiently cooled particles is strongly disfavoured with respect to the adiabatic cooling model.}
\label{Tab:fit_comp}

\end{table}

\clearpage

\section*{Supplementary information}

{\noindent  \bf Supplementary Note 1. HLE from finite-duration pulse. }

If we relax the assumption of infinitesimal duration of the pulse (in the jet comoving frame), we can assume that the jet continuously emits until it switches off at a radius $R_0$. For the computation of the flux as a function of time we therefore integrate the comoving intensity along the equal-arrival-time surfaces (EATS)\cite{Fenimore1996,Dermer2004,Genet2009,Salafia2016}. Photons emitted at different times along the EATS arrive simultaneously to the observer. Knowing that $t_{\text{obs}}(\vartheta)=t_{\text{em}}(1-\beta \cos \vartheta)$ and imposing that $t_{\text{obs}}(\vartheta,R)=\mathrm{const}$, the polar equation which describes the EATS is given by:
\begin{equation}\label{eats}
R(\vartheta,t_{\text{obs}})=\frac{\beta c t_{\text{obs}}}{1-\beta \cos{\vartheta}}
\end{equation}
where we have expressed the emission time as $t_{\text{em}}=R/\beta c$, in the assumption of constant expansion velocity. From the above equation, we see that our assumption that the emission switches off when the radius crosses $R_0$ translates to a $\vartheta$-dependent switching off in the observer frame. At any time $t_{\text{obs}}> (1-\beta)R_0/\beta c$, the observer receives the photons emitted along a surface given by the intersection of the EATS and the jet cone, defined by $R<R_0$, $\vartheta<\vartheta_j$ and $0<\phi<2\pi$, where $\phi$ is the azimuth angle. The resulting surface extends from a minimum angle $\vartheta_{\text{min}}(t_{\text{obs}})$ out to $\vartheta=\vartheta_j$, where the former is given by
\begin{equation}
\vartheta_{\text{min}}(t_{\text{obs}})=\arccos\left(\frac{1}{\beta}-\frac{ct_{\text{obs}}}{R_0}\right)    
\end{equation}
The flux density is given by
\begin{equation}
F_{\nu}(t_{\text{obs}})=\int_\mathrm{EATS} I_{\nu}(\vartheta_{\text{obs}}) \cos \left(\vartheta_{\text{obs}}\right) d \Omega_{\text{obs}}
\end{equation}
where $I_{\nu}$ is the specific intensity and $d \Omega_{\text{obs}}$ is the solid angle in the observer frame. Transforming to the comoving frame we have $I_{\nu}(\nu)=\mathcal{D}^3I'_{\nu^{\prime}}(\nu/\mathcal{D})$. We decompose the comoving intensity as
\begin{equation}
I_{\nu^{\prime}}^{\prime}=I_{\nu_{p}^{\prime}}^{\prime} \cdot S_{\nu^{\prime}} \end{equation}
where $I^\prime_{\nu_p^\prime}$ is the comoving intensity at the peak frequency $\nu_p^\prime$ and $ S_{\nu^{\prime}}$ is the comoving spectral shape, normalized so that $S_{\nu^\prime}(\nu_p^\prime)=1$. In general, $I_{\nu_{p}^{\prime}}^{\prime} \propto N_{\text{tot}}/R^{2}$, where $N_{\text{tot}}$ is the number of emitting particles. If the emission process is synchrotron,  $I_{\nu_{p}^{\prime}}^{\prime}$ is also proportional to $B$, the magnetic field as measured by an observer comoving with the jet, which is assumed to evolve as $B=B_0(R/R_0)^{-\lambda}$, with $\lambda\geq0$ is a free parameter.
If we assume $N_{\text{tot}}$ to be constant in time, then $I_{\nu_{p}^{\prime}}^{\prime} \propto R^{-2}$ and, since $d\Omega_{\text{obs}}\propto R^2\sin\vartheta$, the final form of the integral is
\begin{equation}\label{f_fin_dur}
F_{\nu} (t_{\text{obs}})\propto \int_{\vartheta_{\text{min}}(t_{\text{obs}})}^{\vartheta_j} S_{\nu^\prime}(\nu/\mathcal{D}(\vartheta)) \left(\frac{R(\vartheta,t_{\text{obs}})}{R_0}\right)^{-\lambda} \mathcal{D}^{3}(\vartheta) \sin \vartheta \cos \vartheta d \vartheta
\end{equation} 
The $\alpha-F$ relation for several values of $\lambda$ is plotted in Supplementary Fig.~\ref{HLE_fin_dur}.\\

{\noindent  \bf Supplementary Note 2. HLE from an accelerating shell. }

In this section we test the effect of relaxing the assumption that the shell which generates HLE expands with a constant bulk Lorentz factor $\Gamma$\cite{Uhm2015,Uhm2016,Uhm2016b,Uhm2018}. For our treatment we consider that the emission starts at $R=R_{\text{in}}$ and finishes at $R=R_{\text{off}}$. We assume also that $\Gamma$ evolves as a power law with the radius, namely
\begin{equation}\label{gamma_evo}
\Gamma(R)=\Gamma_0\left(\frac{R}{R_{\text{in}}}\right)^k
\end{equation}
with $k>0$ if the shell accelerates or $k<0$ if the shell decelerates. We consider the emission of a photon at radius $R_{\text{em}}$ and an angle $\vartheta=\vartheta_{\text{em}}$, then we define $\Delta t_{\text{em}}$ the time necessary to expand from $R_{\text{in}}$ to $R_{\text{em}}$. During the same interval of time, a photon emitted at radius $R_{\text{in}}$ and an angle $\vartheta=0$ travels a distance $c \Delta t_{\text{em}}$. Therefore the delay between these two photons is $\Delta t_{\text{obs}}=(R_{\text{in}}+c \Delta t_{\text{em}}-R_{\text{em}}\cos{\vartheta})/c$. From eq.~(\ref{gamma_evo}) we can write
\begin{equation}
    \frac{1}{\sqrt{1-\frac{1}{c^2}\left(\frac{dR}{dt}\right)^2}}=\Gamma_0\left(\frac{R}{R_{\text{in}}}\right)^k
\end{equation}
from which we derive
\begin{equation}
\frac{d R}{\sqrt{1-\frac{1}{\Gamma_{0}^{2}}\left(\frac{R}{R_{\text{in}}}\right)^{-2k}}}=c\, dt\end{equation}
In the limit of $\Gamma_0\gg\left(\frac{R}{R_{\text{in}}}\right)^{-k}$, we can write
\begin{equation}
    \int_{R_{\text{in}}}^{R_{\text{em}}} \left[1+\frac{1}{2 \Gamma_{0}^{2}}\left(\frac{R}{R_{\text{in}}}\right)^{-2k}\right]dR\simeq c \Delta t_{\text{em}}
\end{equation}
Thus, the delay time is
\begin{equation}
\Delta t_{\text{obs}}=\frac{R_{\text{em}}}{c}(1-\cos \theta)+\frac{1}{2 c} \int_{R_{\text{in}}}^{R_{\text{em}}}\frac{1}{\Gamma^2}dR
\end{equation}
Given an arrival time $\Delta t_{\text{obs}}$, this equation allow us to associate a radius $R_{\text{em}}$ to each angle $\vartheta_{\text{em}}$ through the following expression:
\begin{equation}
\cos \vartheta_{\text{em}}=1-\frac{c \Delta t_{\text{obs}}}{R_{\text{em}}}+\frac{{R_{\text{in}}}}{2 {R_{\text{em}}}} \frac{1}{\Gamma_{0}^{2}} \frac{1}{1-2 k}\left[\left(\frac{R_{\text{em}}}{R_{\text{in}}}\right)^{1-2 k}-1\right]
\end{equation}
Inverting this equation, we obtain the polar equation $R_{\text{em}}(\vartheta_{\text{em}},\Delta t_{\text{obs}})$ which defines the EATS, namely all the photons emitted from this locus of points arrive to the observer with a time delay $\Delta t_{\text{obs}}$ with respect to the first photon coming from $R=R_{\text{in}}$ and $\vartheta=0$. The computation of the flux as a function of time is again done using eq.~(\ref{f_fin_dur}), with the only difference that now $\beta$ and $\Gamma$, which appear in the Doppler factor $\mathcal{D}(\vartheta)$, depend on $R(\vartheta)$. The light curve and the spectral evolution for values of $k$ in the range $-0.4\leq k\leq 0.4$ are shown in Supplementary Fig.~\ref{HLE_acc}.\\

{\noindent  \bf Supplementary Note 3. Alternative scenarios shaping the X-ray tails.} 

In this section we explore other possible models of prompt emission which can drive the evolution during the X-ray tails. We can consider, for instance, an anisotropic emission from the jet core, made of mini-jets\cite{Narayan2009,Duran2016,Geng17} with angular sizes $< 1/\Gamma$. In order to model such anisotropy we adopt an angular distribution of the emission in the form $P(\theta') \propto (\sin \theta')^{n}$, where $n$ is the degree of anisotropy and $\theta'$ is the angle between the direction of the emitted photons and the local radial direction, as measured in the comoving frame. We evaluate the resulting HLE flux received by the observer as
\begin{equation}
F_{\nu} \propto P(\theta')\mathcal{D}^2(\theta)S\left(\frac{\nu}{P(\theta')\mathcal{D}(\theta)\nu_c'}\right)
\end{equation}
where the dependence on time is implicit in $\theta$. The resulting $\alpha-F$ relation for $n=2$, $n=5$ and $n=10$ is shown in Supplementary Fig.~\ref{mini_jets}. The figure demonstrates that the larger the value of $n$, the more the predicted curves move away from the data. Since for $n=0$ we are in the limit of standard HLE, which is already disfavoured by our study, we conclude that also mini-jets are not able to successfully reproduce the $\alpha$ - F relation.\\
Within the HLE scenario, only models which assume a dissipation occurring above the jet photosphere, such as in internal shocks\cite{Rees1994} or in magnetic reconnection scenarios\cite{Drenkhahn2002,Lyutikov2003}, are able to reproduce the typical duration of X-ray tails ($\sim100$ s).
Photospheric models\cite{Pe'er2008}, where dissipation occurs at radii $R_{\text{ph}} \sim 10^{12}$ cm\cite{Piran1999}, give smaller times scales of $\sim 10^{-2}$ s, incompatible with observations. Only a common declining activity of the central engine \cite{Hascoet2012,Kumar2008} and a fine-tuned intrinsic spectral softening\cite{Beloborodov2011} would be required to account for the $\alpha-F$ relation.\\
In slow heating/reacceleration scenarios\cite{Asano2009}, as soon as the shock crosses the shell, particle acceleration is halted along with the generation of magnetic field (as both rely on the presence of shock-generated turbulence), leading to an abrupt switch-off of the emission. This leads again to HLE being the dominant effect in determining the tail flux and spectral evolution, which is clearly disfavoured by our analysis. A slower decay of the magnetic field after the shock crossing\cite{Asano2015}, along with a decaying particle acceleration, could be compatible with our results, but we still need adiabatic cooling (which is anyway unavoidable) to play the leading role in the spectral evolution, as discussed in the next section. We therefore conclude that, while the slow-heating scenario is not \textit{per se} rejected by our results, it cannot be invoked as the main mechanism behind the $\alpha$ - F relation.\\

{\noindent  \bf Supplementary Note 4. Possible temporal evolution of the comoving spectral shape.}

In the derivation of spectral evolution from adiabatic cooling, we implicitly assumed that we are in the early post-prompt phase, namely where no more particles are injected/accelerated and the particle distribution only evolves according to the cooling processes. If adiabatic cooling is dominant, the energy of all particles evolve at the same way, according to the following equation:
\begin{equation}
\label{ad_eq}
\gamma^3V'=\text{const}
\end{equation}
where $\gamma$ is the Lorentz factor of the particle and $V'$ is the comoving volume. Therefore the shape of the particle distribution, and hence of the spectrum, does not change in time, but is only rigidly shifted at lower energies.\\
If adiabatic and radiative cooling are competing on comparable timescales, in the post-prompt phase an exponential cutoff appears above the cooling energy $\gamma_c$. If also the magnetic field decays, the adiabatic cooling tends to dominate with time and also the cutoff energy would eventually evolve according to eq.~\ref{ad_eq}, going again in the limit of rigidly shifted spectrum.\\
The assumption of rigidly shifted spectrum might not hold if the particle injection gradually decreases in time, instead of ceasing abruptly. The temporal evolution of particle distribution in case of decreasing injection of particles is studied solving numerically the cooling equation
\begin{equation}
\frac{\partial{N}}{\partial{t}}=\dot{N}_{\text{inj}}-\frac{\partial}{\partial{\gamma}}(N\dot{\gamma})
\end{equation}
where $N(\gamma)=dN_\mathrm{e}/d\gamma$ ($N_\mathrm{e}$ being the number of emitting particles). The resulting evolution of the synchrotron spectrum is shown in Supplementary Fig.~\ref{N_dec_syn}a, where we assumed an injection term of the form $\dot{N}_{\text{inj}}\propto (t/t_{\text{inj}})^{-y}$, with $y>0$, and a constant magnetic field. Imposing that, at the beginning, the $F_{\nu}$ peak is just above the observing band, the resulting spectral evolution would be an initial softening followed by a hardening, as shown in Supplementary Fig.~\ref{N_dec_syn}b. In the case of a decay of both the magnetic field and $\dot{N}_{\text{inj}}$, the corresponding effects on the spectral shape tend to compensate each other\cite{Pan19}, giving a bare modification of the spectral shape or even a hardening (see Supplementary Fig.~\ref{N_dec_B_dec}). In conclusion, intrinsic modifications of the spectral shape can hardly give an agreement with data comparable to the case of a rigidly shifting spectrum.\\

{\noindent  \bf Supplementary Note 5. Supplementary discussion.} 

As stated in the main text, HLE and adiabatic cooling have the same timescale, but the relevance of one process with respect to the other is determined by the decay of the magnetic field, which governs the drop of the spectrum normalization. The expected value of $\lambda$ can be derived in several scenarios, according to the process that rules the magnetic field evolution. In case of conservation of magnetic flux, the perpendicular and parallel component of $B$ evolve as $B_{\perp} \sim 1/(\Delta R' \cdot r)$ and $B_{/ /} \sim r^{-2}$, where $r$ is the transverse radial dimension of the jet in a cylindrical reference system ($r,\phi,z$). If the jet is conical then $r\propto R$, leading to $B\sim R^{-1}$ ($\lambda=1$) for $\Delta R'=\mathrm{const}$ and $B\sim R^{-2}$ ($\lambda=2$) for $\Delta R'\propto R$. Another possibility predicts equipartition between magnetic energy density and particle energy density, giving $B^2 \sim \expval{\gamma}/V \sim V^{-4/3}$, where in the last step we used eq.~(\ref{ad_eq}). In this case $B\sim R^{-4/3}$ ($\lambda=4/3$) for $\Delta R'=\mathrm{const}$ and $B\sim R^{-2}$ ($\lambda=2$) for $\Delta R'\propto R$. All these predicted values of $\lambda$ are larger than the range found from our analysis. Such tension can be solved, for instance, if the shell thickness decreases as the jet expands, or if the jet is not conical (e.g. paraboloidal, with $r\propto \sqrt{R})$.

\clearpage
\section*{Supplementary Figures}

\begin{figure}[ht!]
  \centering
  \includegraphics[width=0.8\textwidth]{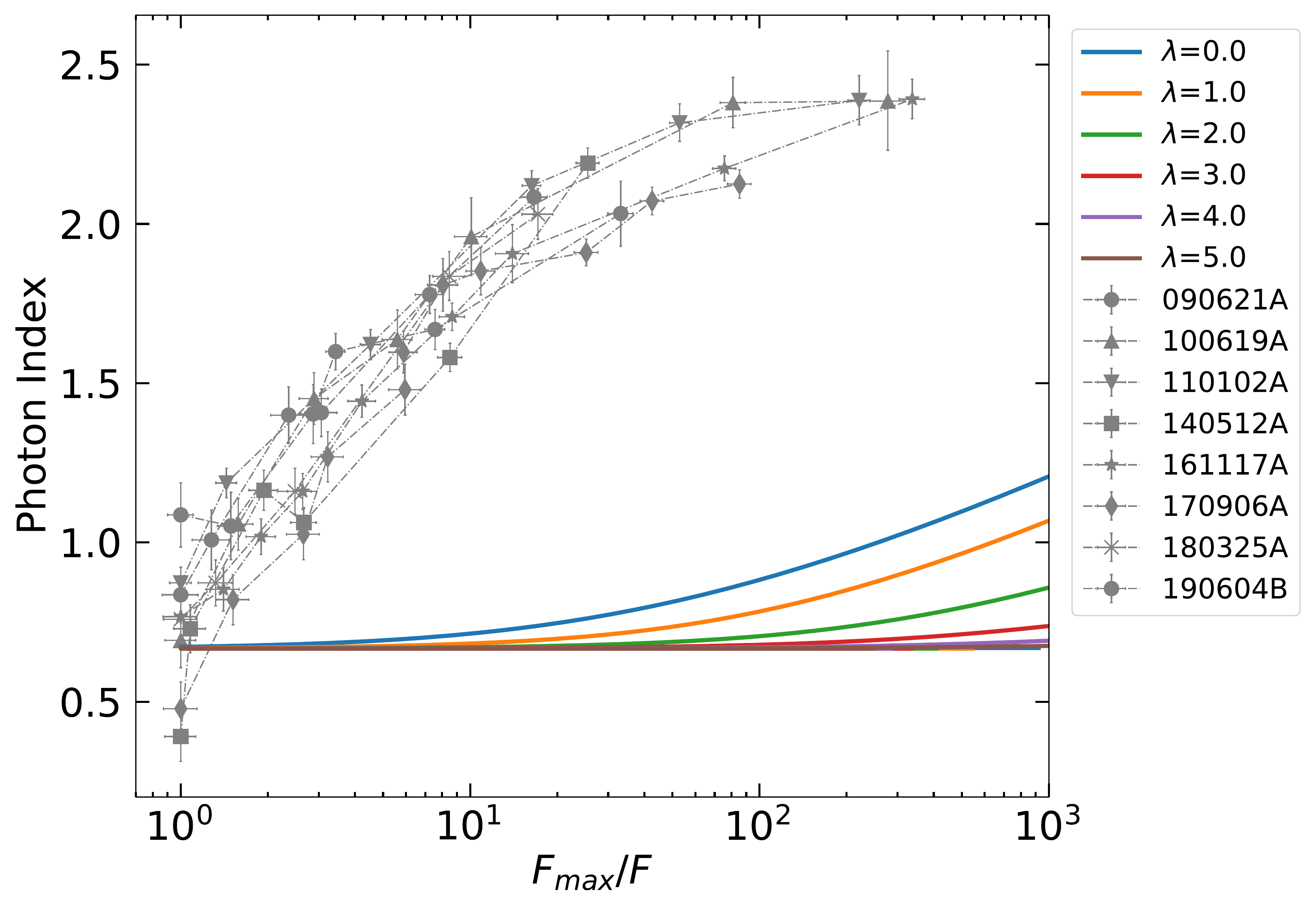}

  \caption{{\bf Spectral evolution in case of HLE from a finite-duration pulse.} The adopted parameters are $R_{\text{in}}=3\times10^{15}$ cm, $R_{\text{off}}=9\times10^{15}$ cm, $\Gamma_0=100$ and $\nu_p=100$ keV. The adopted spectral shape is a SBPL. The value of $\lambda$ specifies the evolution of the magnetic field. The error bars represent $1\sigma$ uncertainties, calculated via spectral fitting in XSPEC. In the legend we report the name of each GRB.}
  \label{HLE_fin_dur}
\end{figure}

\begin{figure}[ht!]
  \centering
  \includegraphics[width=0.7\textwidth]{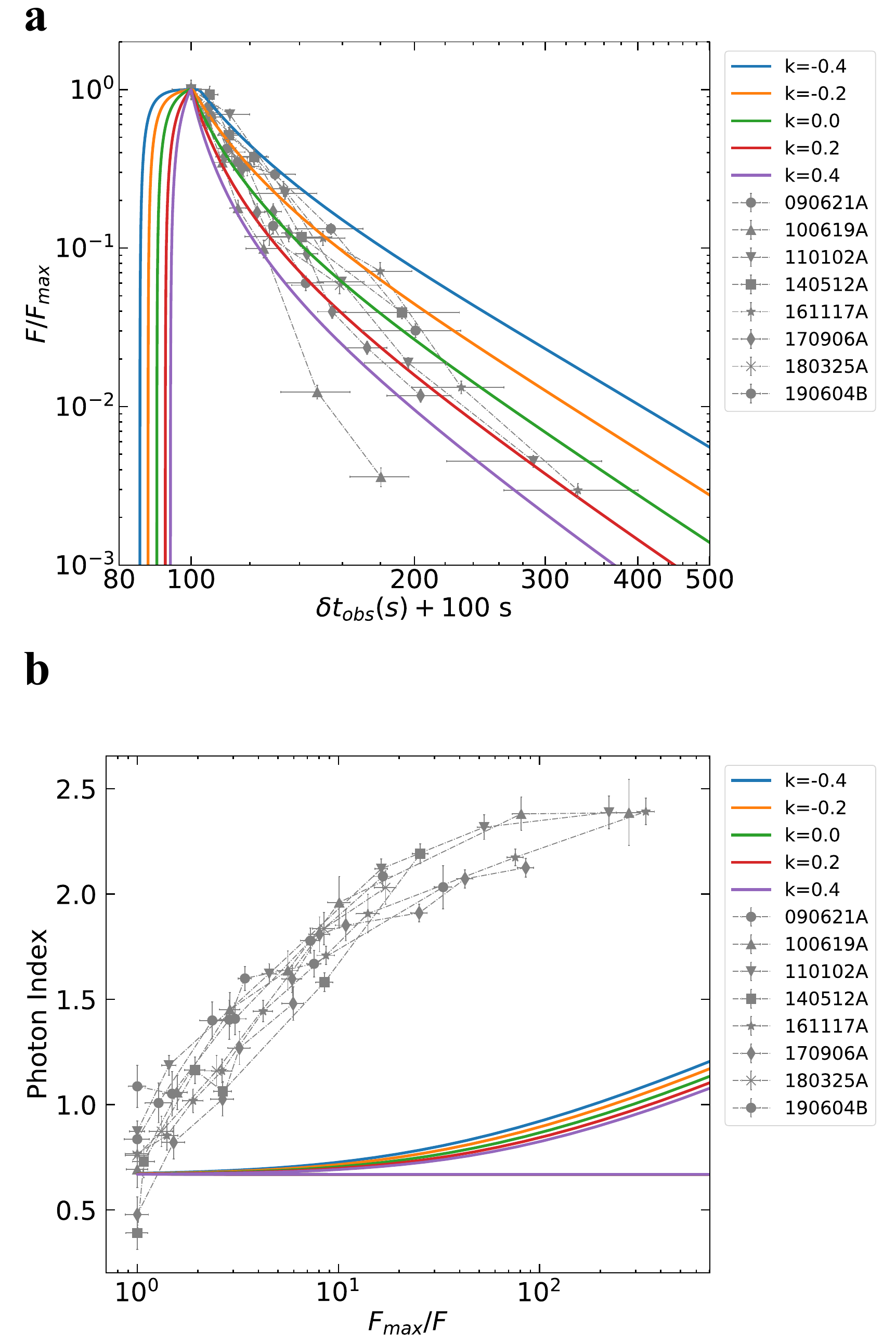}\\

  \caption{{\bf Temporal (a) and spectral (b) evolution for HLE from a finite-duration pulse, in case of not constant $\Gamma$.} The magnetic field does not evolve with radius, i.e $\lambda=0$. The adopted parameters are $R_{\text{in}}=3\times10^{15}$ cm, $R_{\text{off}}=9\times10^{15}$ cm, $\Gamma_0=100$ and $\nu_p=100$ keV. The value of $k$ specifies the evolution of $\Gamma$. The adopted spectral shape is a SBPL. The peak of each curve is shifted at 100 s. In {\bf a} the vertical error bars represent $1\sigma$ uncertainties, calculated via spectral fitting in XSPEC, while horizontal error bars represent the width of the time bin. In {\bf b} the error bars represent $1\sigma$ uncertainties, calculated via spectral fitting in XSPEC. In the legend we report the name of each GRB.}
  \label{HLE_acc}
\end{figure}

\begin{figure}[ht!]
  \centering
  \includegraphics[width=0.8\textwidth]{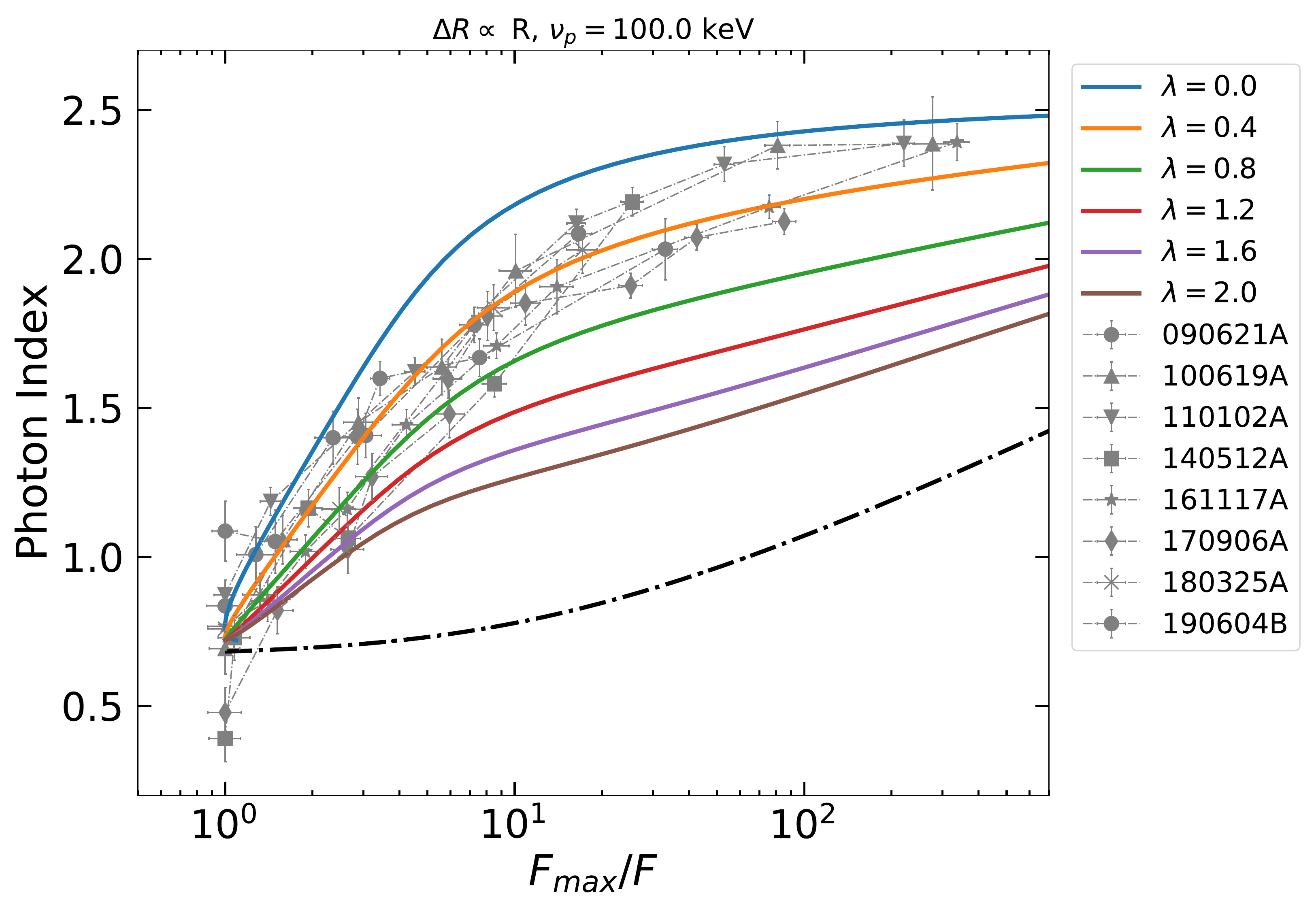}
  \caption{{\bf$\alpha-F$ plot in case of adiabatic cooling, but with the shell thickness $\Delta R \propto R$, instead of $\Delta R =$ const.} The theoretical curves are computed taking also into account the effect of HLE. The value of $\lambda$ specifies the evolution of the magnetic field. We adopt a SBPL as spectral shape with $\alpha_{\text{s}}=-1/3$ and $\beta_{\text{s}}=1.5$, an initial observed peak frequency of 100 keV. The dot-dashed line is the evolution expected in case of HLE without adiabatic cooling, assuming the same spectral shape and initial observed peak frequency. The error bars represent $1\sigma$ uncertainties, calculated via spectral fitting in XSPEC. In the legend we report the name of each GRB.}
  \label{dr_propto_r}
\end{figure}

\begin{figure}[ht!]
  \centering
  \includegraphics[width=0.65\textwidth]{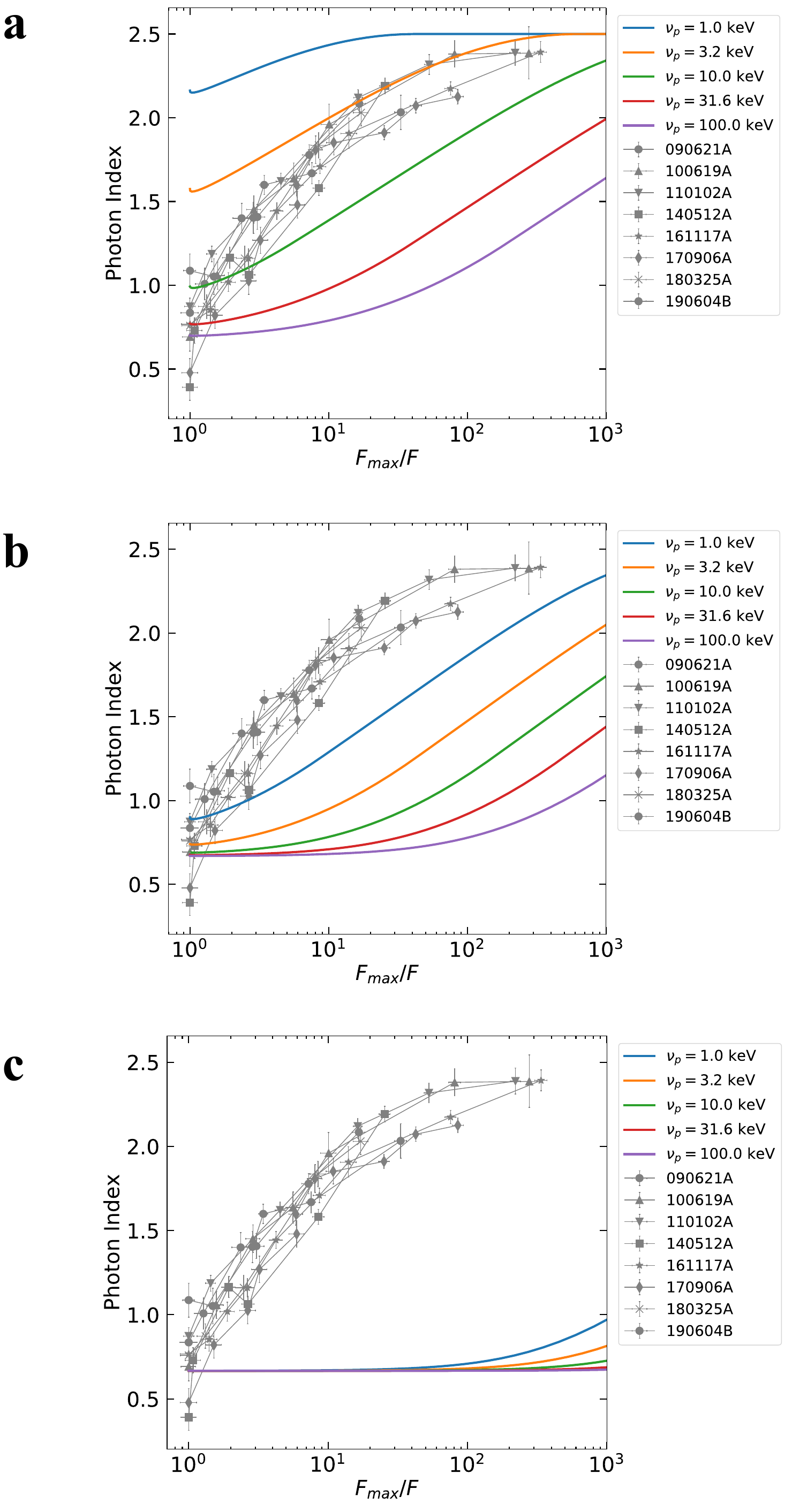}\\

  \caption{{\bf Predicted $\alpha-F$ relation in case of mini-jets model}. We show the evolution for $n=2$ ({\bf a}), $n=5$ ({\bf b}) and $n=10$ ({\bf c}). The error bars represent $1\sigma$ uncertainties, calculated via spectral fitting in XSPEC. In the legend we report the name of each GRB.}
  \label{mini_jets}
\end{figure}

\begin{figure}[ht!]
  \centering
  \includegraphics[width=0.7\textwidth]{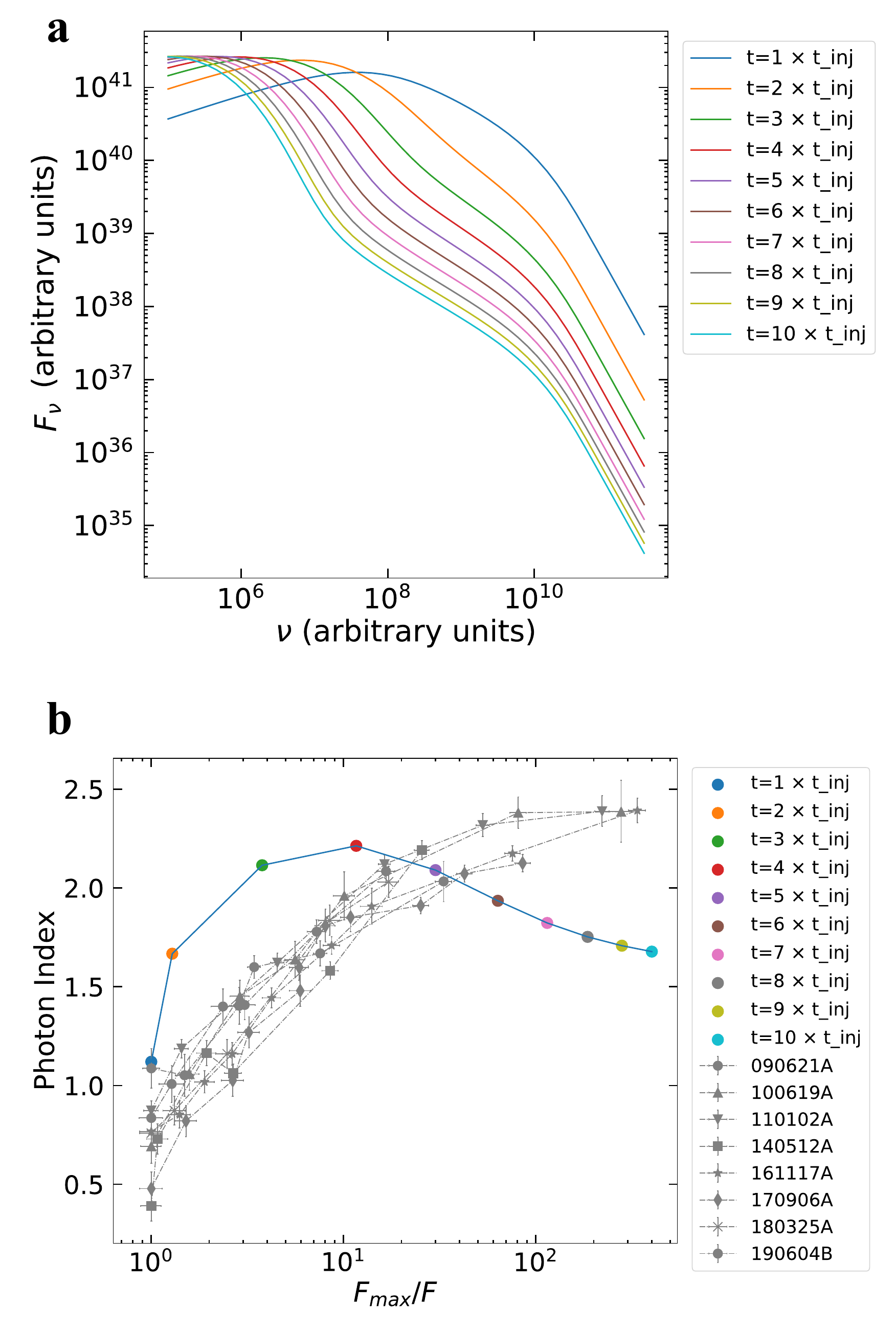}\\
  \caption{{\bf Spectral evolution of the synchrotron spectral shape for a decaying particle injection}. {\bf a} shows how the spectral shape evolves, adopting a decaying index for the injection rate $y=3$ and a constant magnetic field. The time goes from $t=t_{\text{inj}}$ (blue line) up to $t=10 \times t_{\text{inj}}$ (cyan line), with steps of $t_{\text{inj}}$. In panel {\bf b} the blue line shows the corresponding $\alpha-F$ relation imposing that the observing band is below the initial spectral peak, which ensures that the initial photon index is $\sim 2/3$. With different coloured points we indicate the evolution in the $\alpha-F$ plane as a function of time, from $t=t_{\text{inj}}$ (blue point) up to $t=10 \times t_{\text{inj}}$ (cyan point), with steps of $t_{\text{inj}}$. The error bars represent $1\sigma$ uncertainties, calculated via spectral fitting in XSPEC.}
  \label{N_dec_syn}
\end{figure}

\begin{figure}[ht!]
  \centering
  \includegraphics[width=0.8\textwidth]{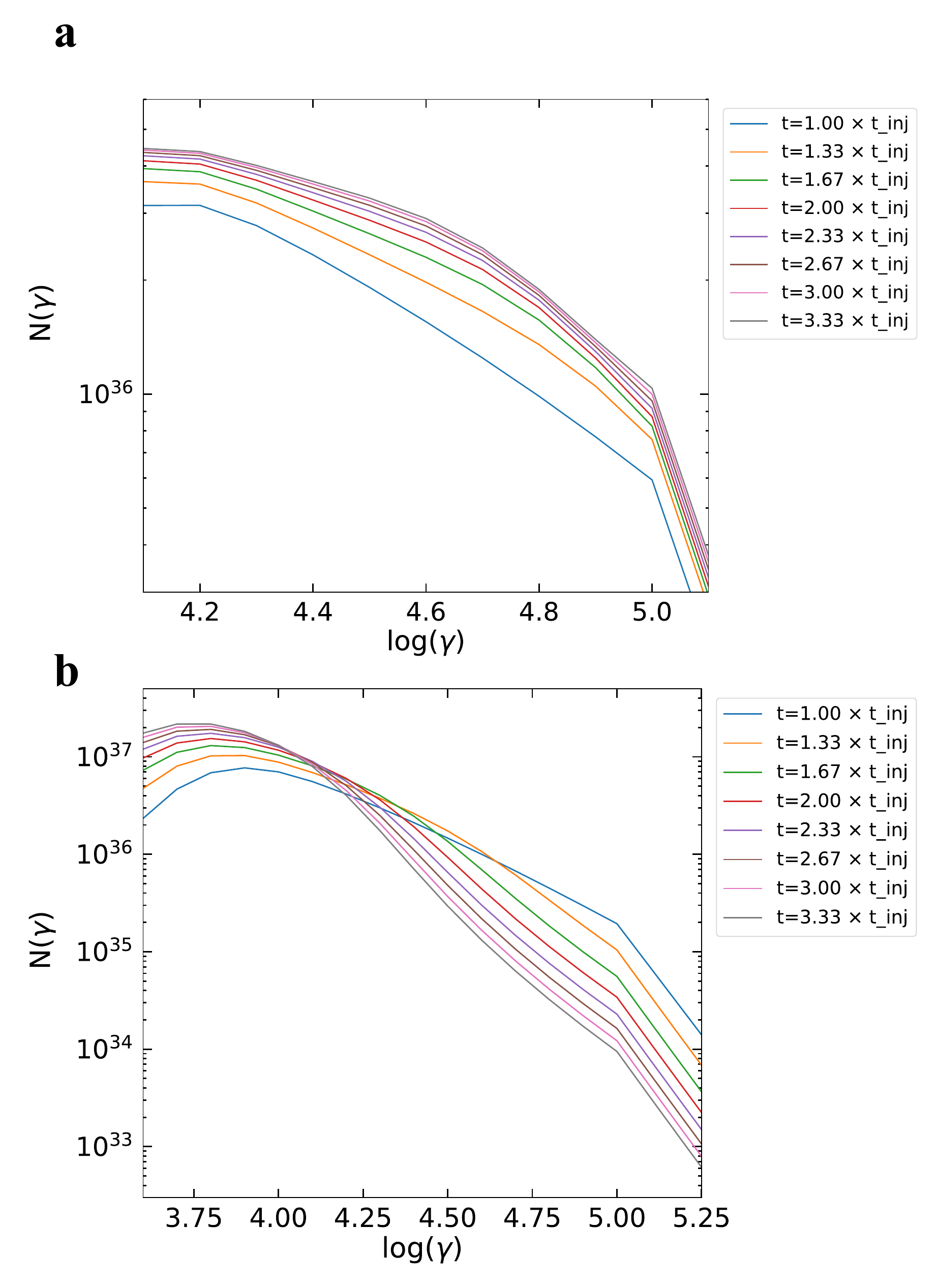}\\

  \caption{{\bf Temporal evolution of the particle distribution for a decay of both $\dot{N}_{\text{inj}}$ and magnetic field}. The adopted parameters are $y=2$ and $\lambda=2$ for {\bf a}, $y=4$ and $\lambda=1$ for {\bf b}. The evolution is followed from $t=t_{\text{inj}}$ (blue line), which is the standard $\gamma^{-2}$ cooling branch of the distribution) up to $t=3.33 \times t_{\text{inj}}$ (grey line), with steps of $\sim 4/3 \, t_{\text{inj}}$.}
  \label{N_dec_B_dec}
\end{figure}

\begin{figure*}[ht!]
\begin{center}
   {
  \includegraphics[width=0.80\textwidth]{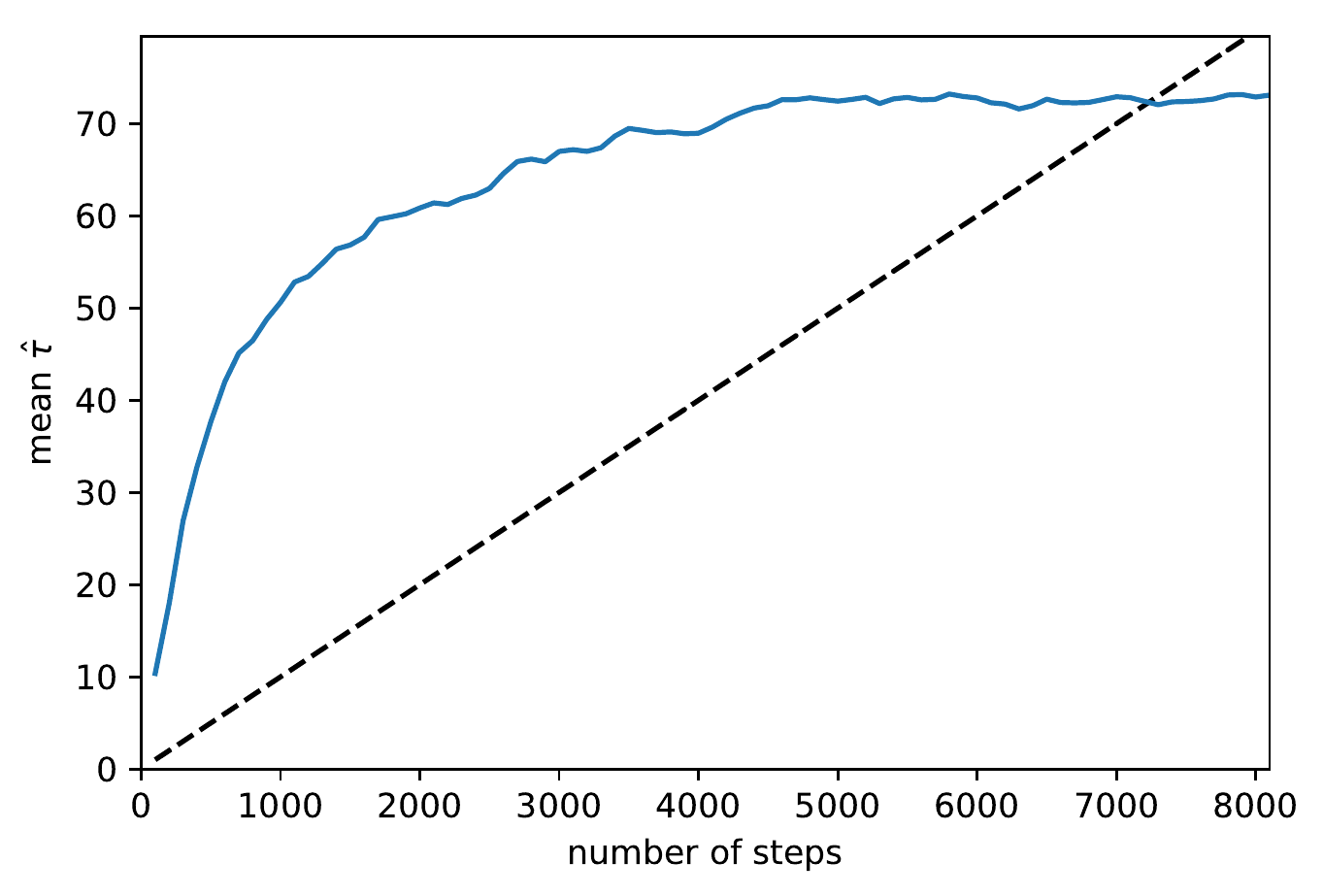}
  }
\caption{{\bf An example of autocorrelation time-MCMC steps plot (GRB 161117A)}. The blue line indicates how $\hat{\tau}$, the autocorrelation time (adimensional quantity), evolves as a function of the number of MCMC steps. The dashed line corresponds to number of steps $=100\times\hat{\tau}$.}
\label{corr_plot}
\end{center}
\end{figure*}

\begin{figure*}[ht!]
\begin{center}
   {
  \includegraphics[width=0.95\textwidth]{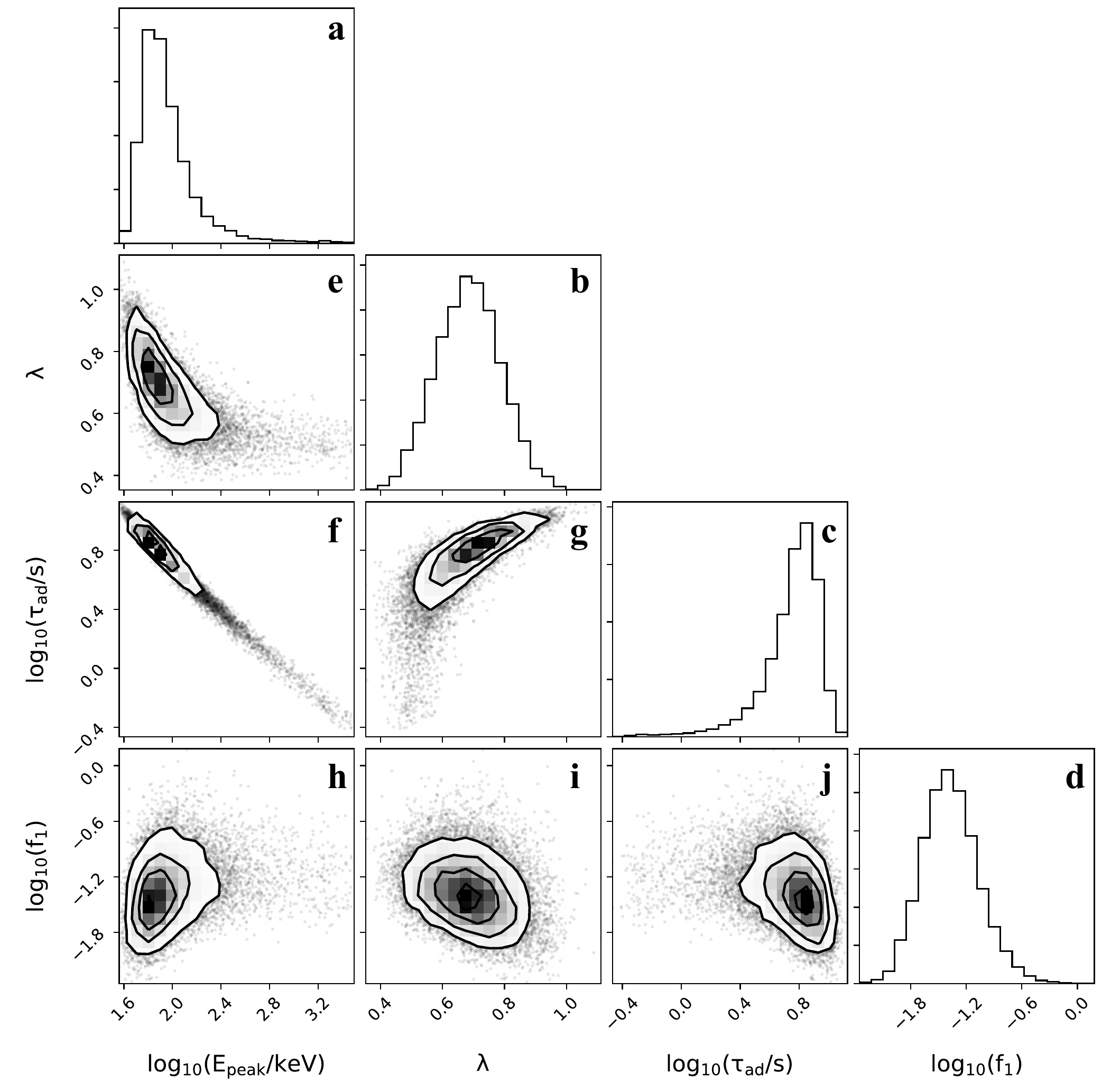}

  }
\caption{{\bf An example of corner plot from the MCMC (GRB 161117A).} $E_{\text{peak}}$ is the peak energy at the beginning of the steep decay, $\lambda$ is the decaying index of magnetic field (adimensional parameter), $\tau_{\text{ad}}$ is the adiabatic timescale, $f_1$ is a parameter used in the definition of the likelihood (see the methods section in the main text for further details). The panels {\bf a-d} show the 1D posterior probability distribution of each parameter; since the y axis is a measure of probability density, it has an arbitrary scale. The panels {\bf e-j} show the 2D posterior distribution for each couple of parameters and the contour lines represent the confidence regions at $0.5\sigma$, $1\sigma$, $1.5\sigma$ and $2\sigma$ level of confidence (if only 3 contours are visible, this means that the inner one, corresponding to $0.5\sigma$, is so small that it is reduced to a point and is not shown).}
\label{fig:banana_170906}
\end{center}
\end{figure*}

\begin{figure*}[ht!]
   {
  \includegraphics[width=0.45\textwidth]{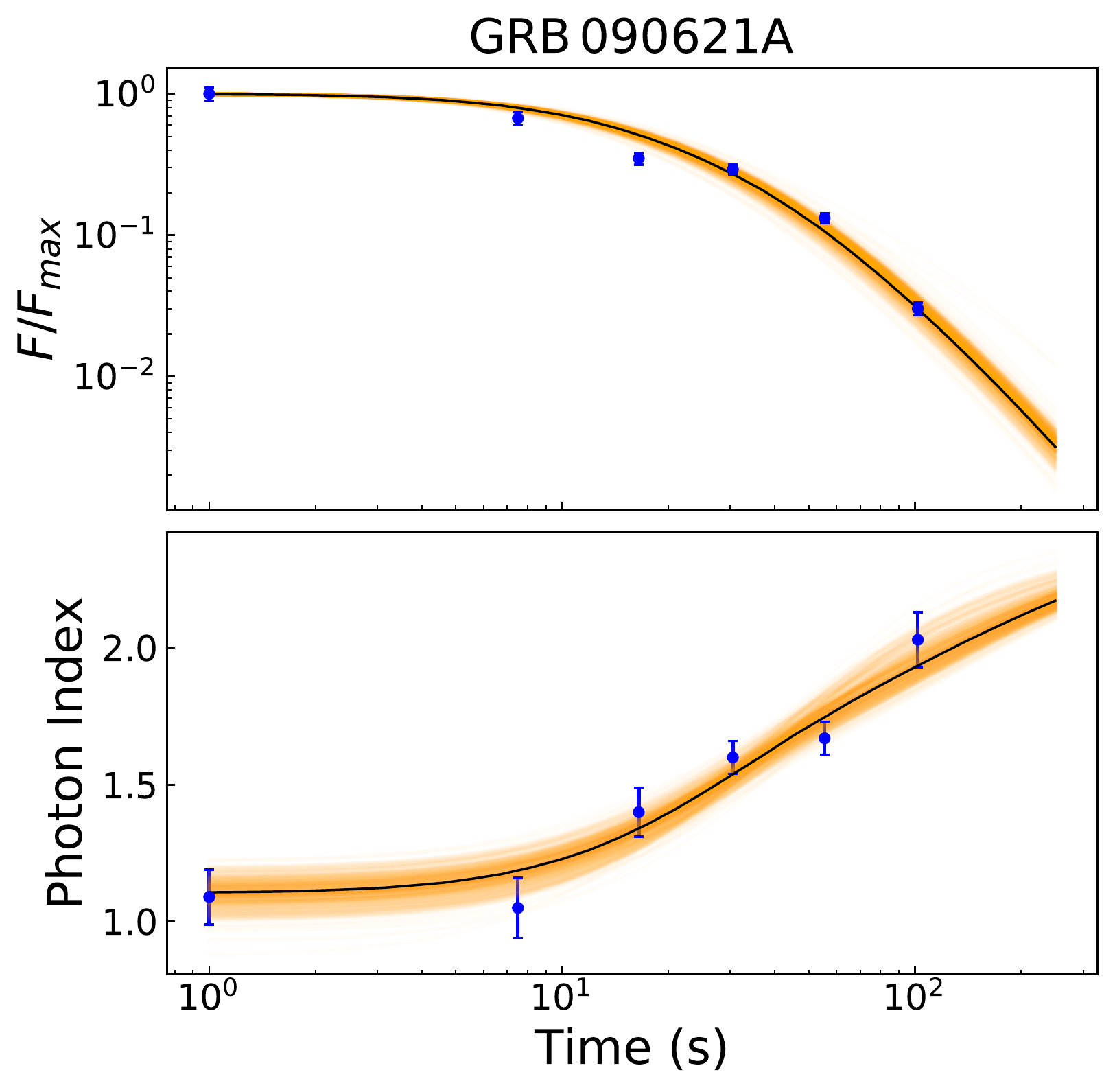}
  \includegraphics[width=0.45\textwidth]{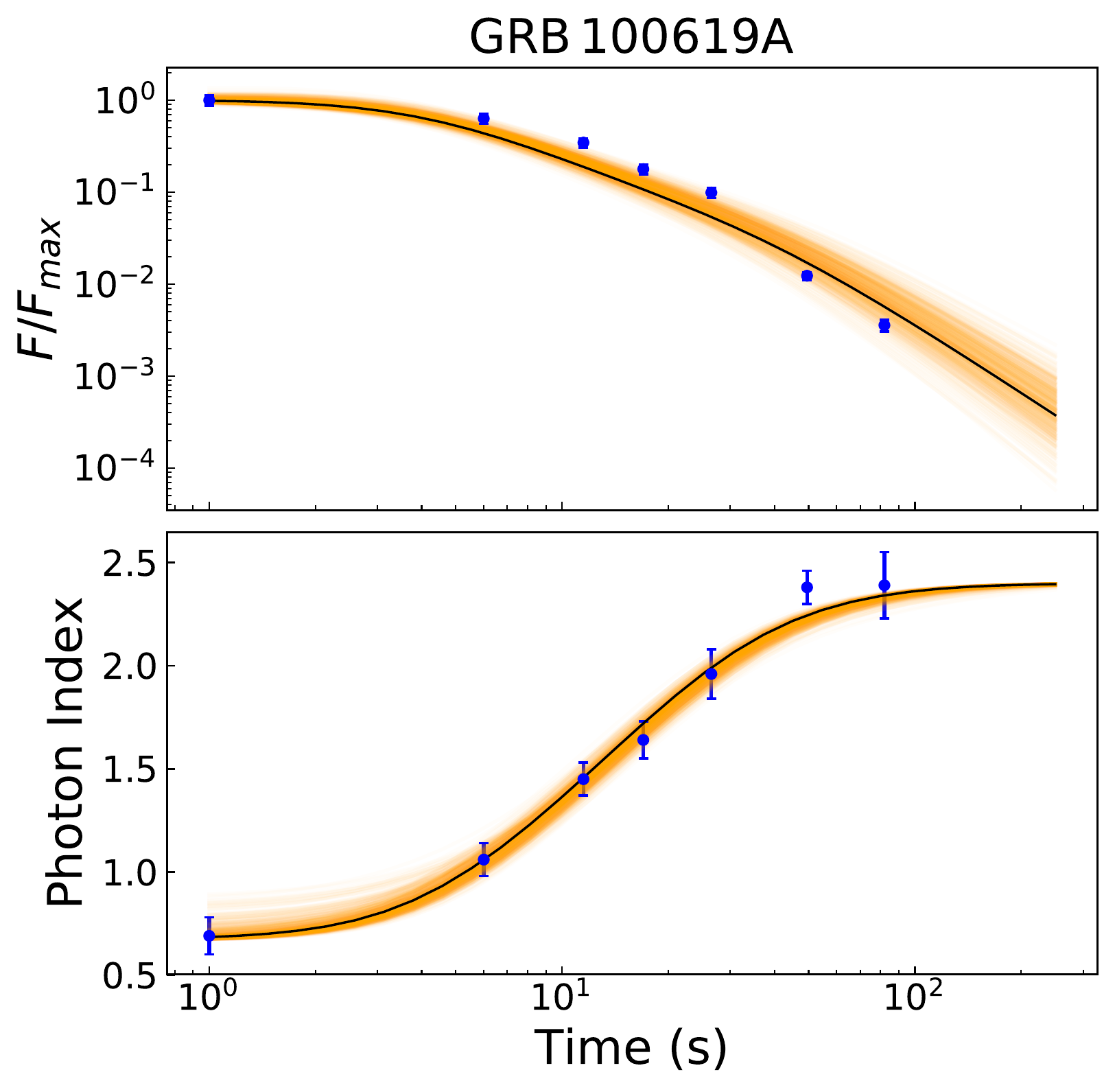}
  \includegraphics[width=0.45\textwidth]{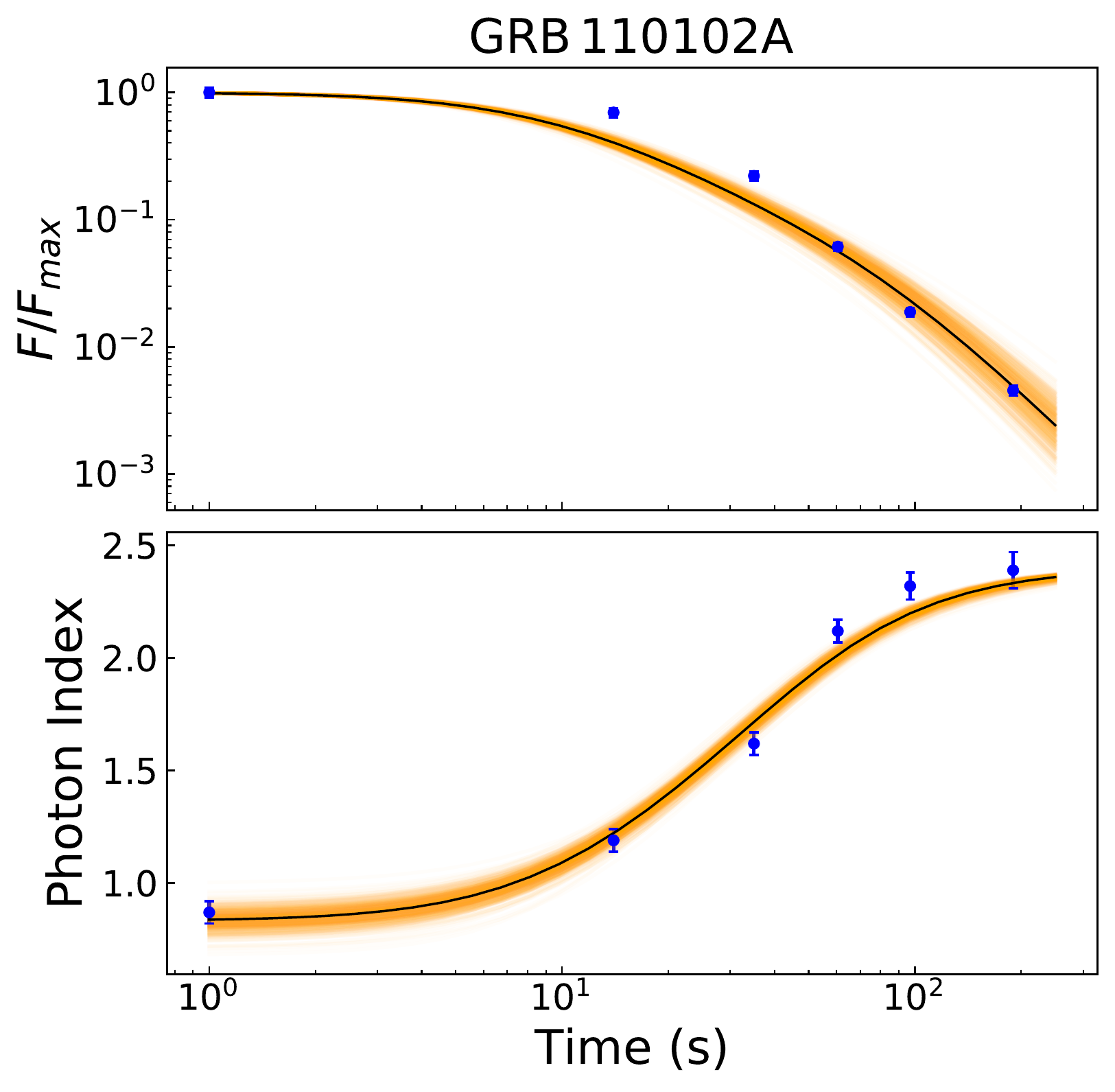}
  \includegraphics[width=0.45\textwidth]{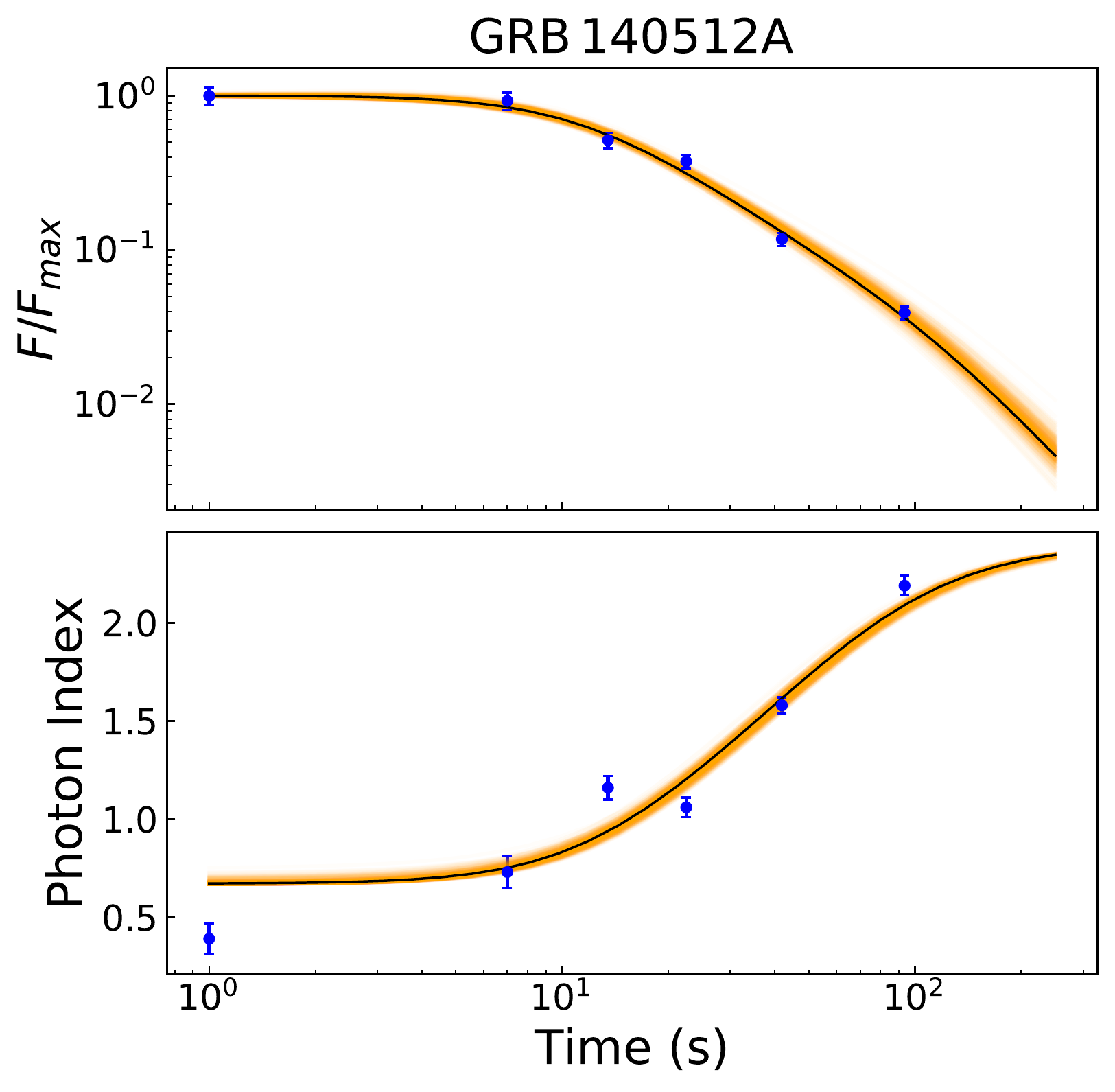}

  }
 
\caption{{\bf Joint temporal evolution of normalized flux and photon index.} For each GRB we compare the data (blue points) with the best fit curve of the adiabatic cooling model (black line). The orange lines are curves produced extracting randomly the model parameters from the posterior distribution obtained from the MCMC. 500 lines are plotted together and their superposition creates a confidence band of the model. In some regions of the plot the band appears narrower because the parameters uncertainty produces a smaller scatter of the lines. The error bars represent $1\sigma$ uncertainties and they are derived from spectral analysis.} 

\end{figure*}

\begin{figure*}[ht!]
   {

  \includegraphics[width=0.45\textwidth]{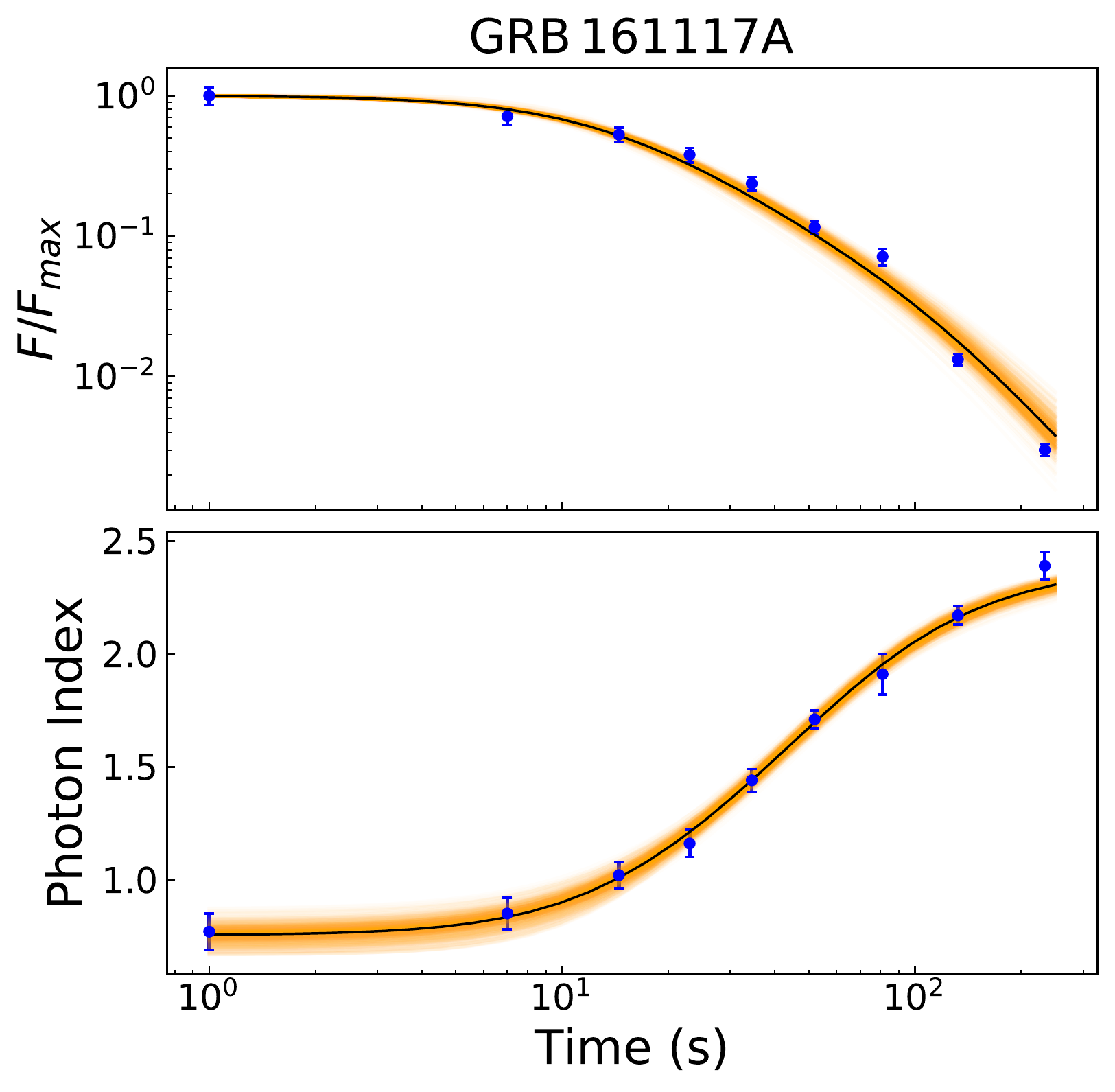}
  \includegraphics[width=0.45\textwidth]{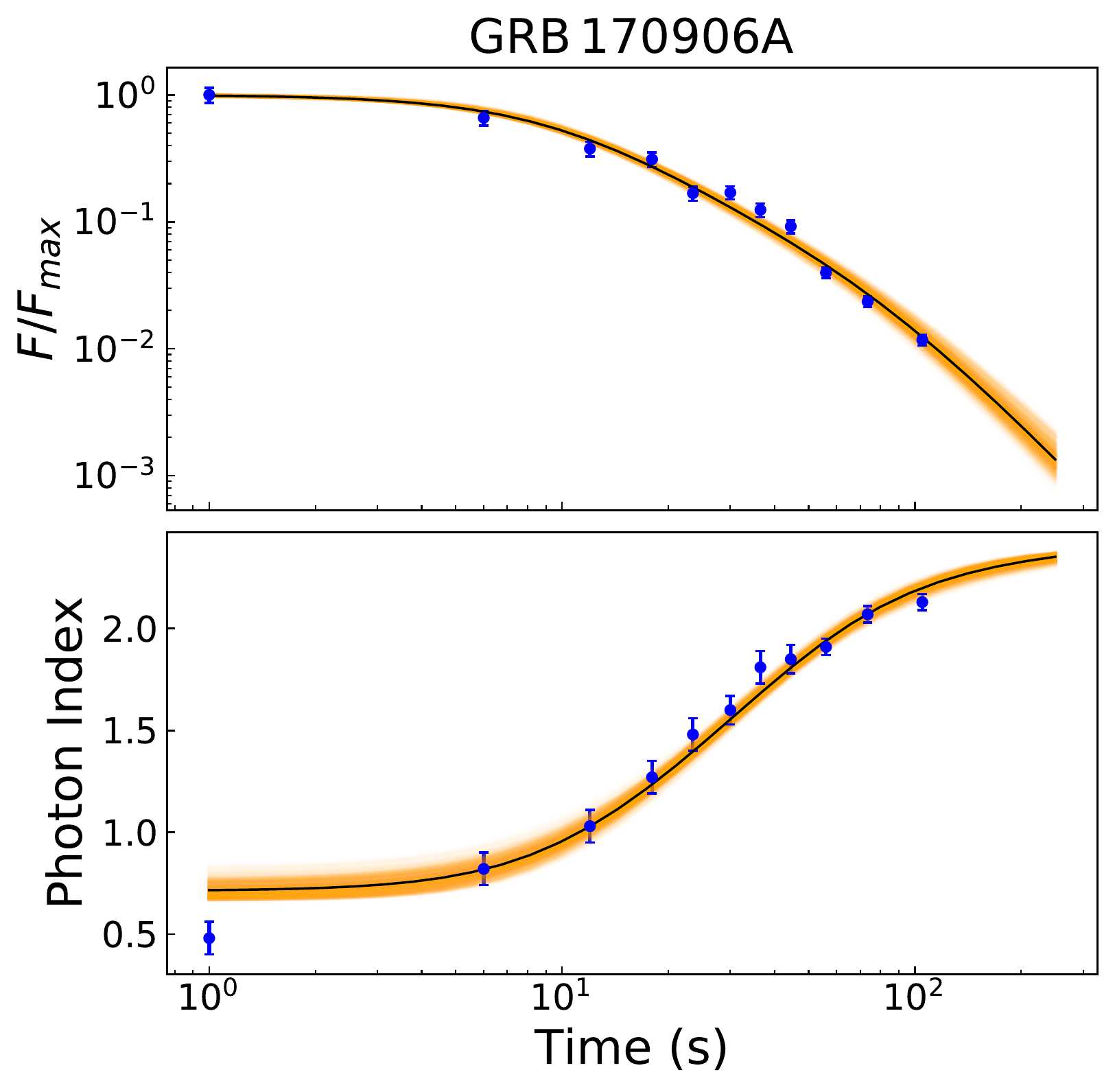}
  \includegraphics[width=0.45\textwidth]{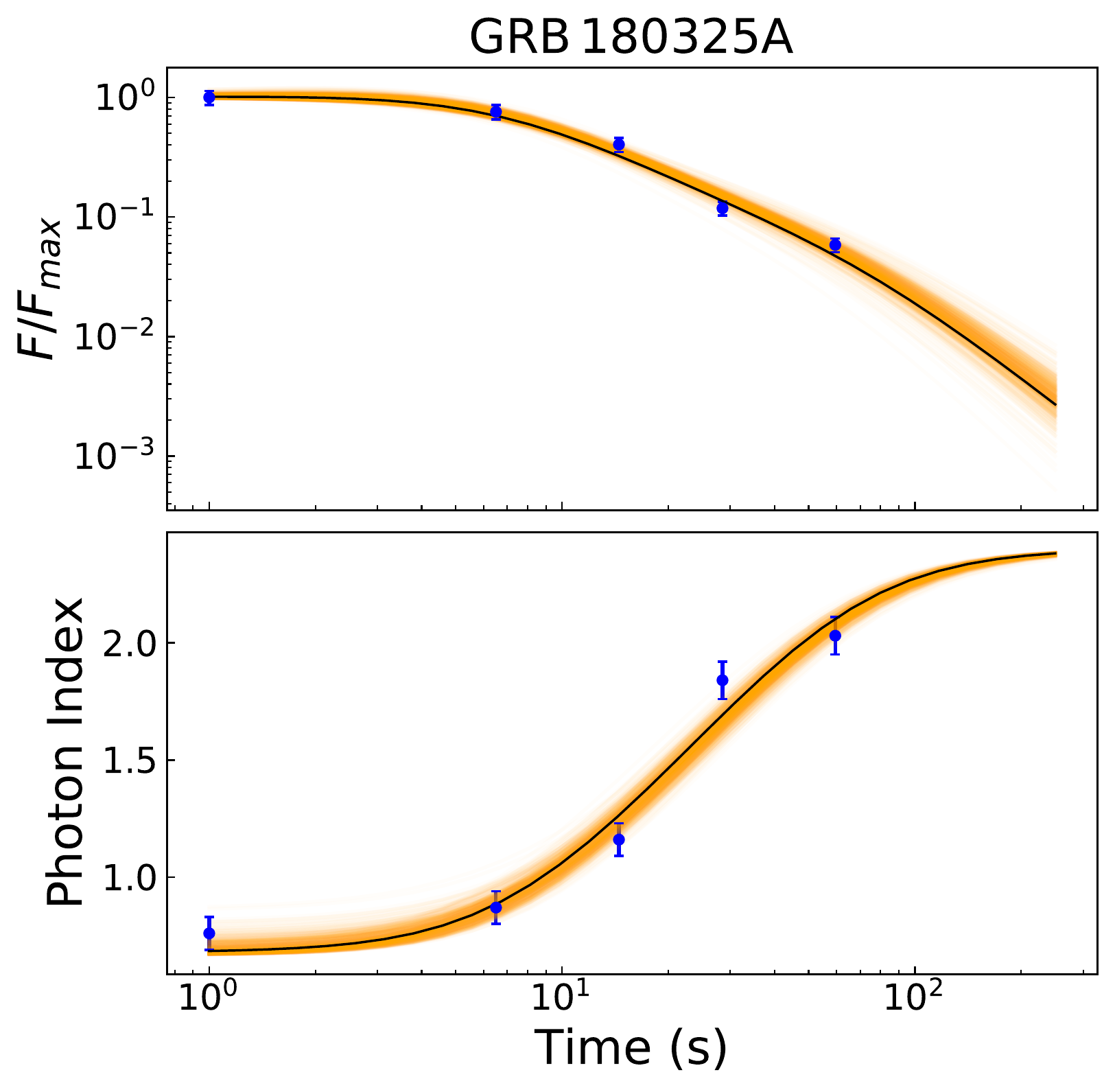}
  \includegraphics[width=0.45\textwidth]{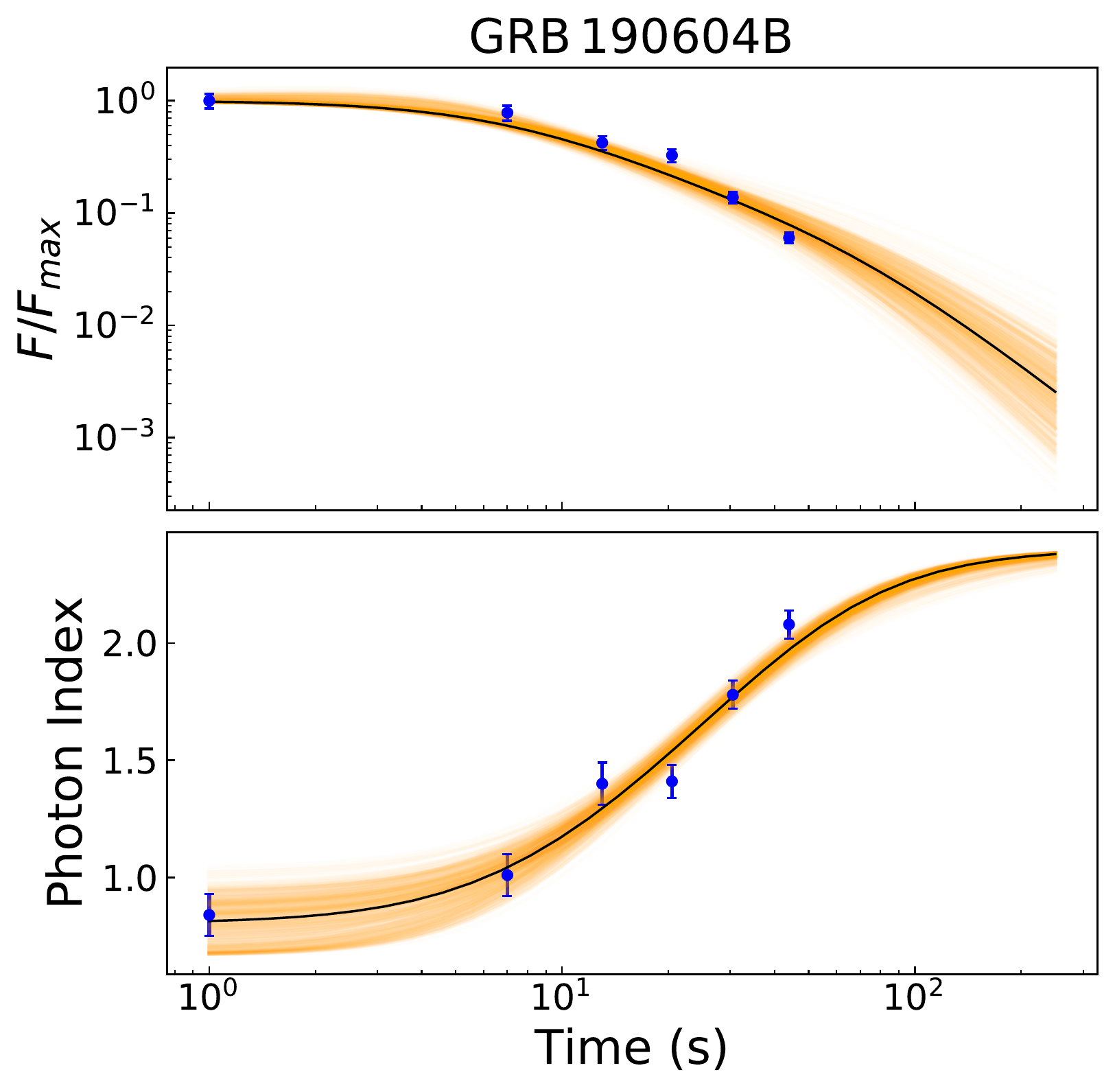}

  }
\caption{{\bf Joint temporal evolution of normalized flux and photon index-continued.} For each GRB we compare the data (blue points) with the best fit curve of the adiabatic cooling model (black line). The orange lines are curves produced extracting randomly the model parameters from the posterior distribution obtained from the MCMC. 500 lines are plotted together and their superposition creates a confidence band of the model. In some regions of the plot the band appears narrower because the parameters uncertainty produces a smaller scatter of the lines. The error bars represent $1\sigma$ uncertainties and they are derived from spectral analysis.} 

\end{figure*}

\begin{figure}[ht!]
  \centering
  \includegraphics[width=0.75\textwidth]{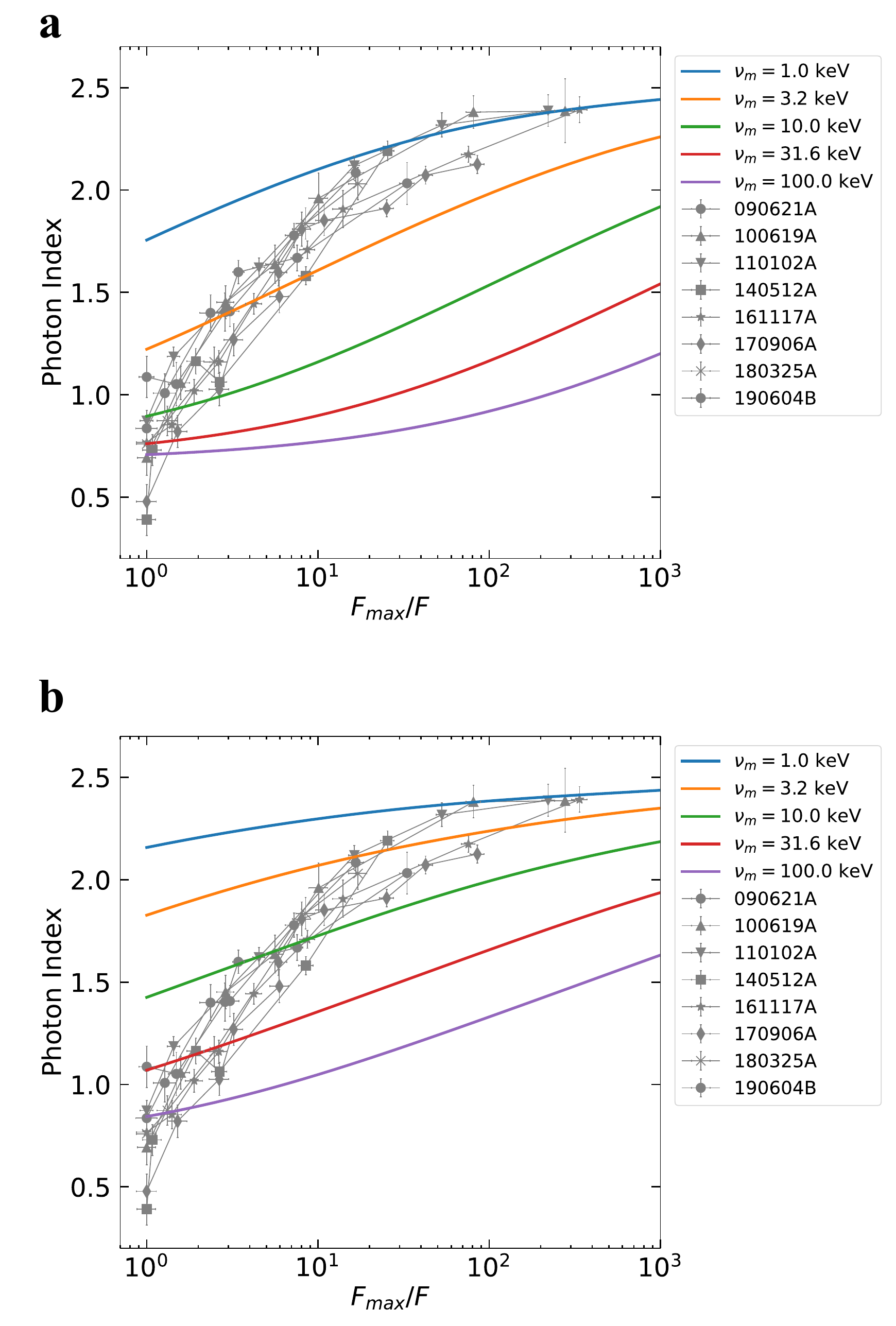}\\
  \caption{{\bf Spectral evolution expected for HLE from a infinitesimal duration pulse, assuming a synchrotron spectrum as spectral shape.} In {\bf a} we adopt $\nu_\text{m}/\nu_\text{c}=1$, while in {\bf b} $\nu_\text{m}/\nu_\text{c}=10$. The several colors indicate the observed peak frequency at the beginning of the decay. In {\bf b} the spectral evolution appears slightly steeper with respect to the case $\nu_m/\nu_c=1$ because for $\nu_\text{c}<\nu<\nu_\text{m}$ the spectrum goes like $F_{\nu}\sim \nu^{-p/2}$. The error bars represent $1\sigma$ uncertainties, calculated via spectral fitting in XSPEC. In the legend we report the name of each GRB.}
  \label{HLE_sync}
\end{figure}

\begin{figure}[ht!]
  \centering
  \includegraphics[width=0.75\textwidth]{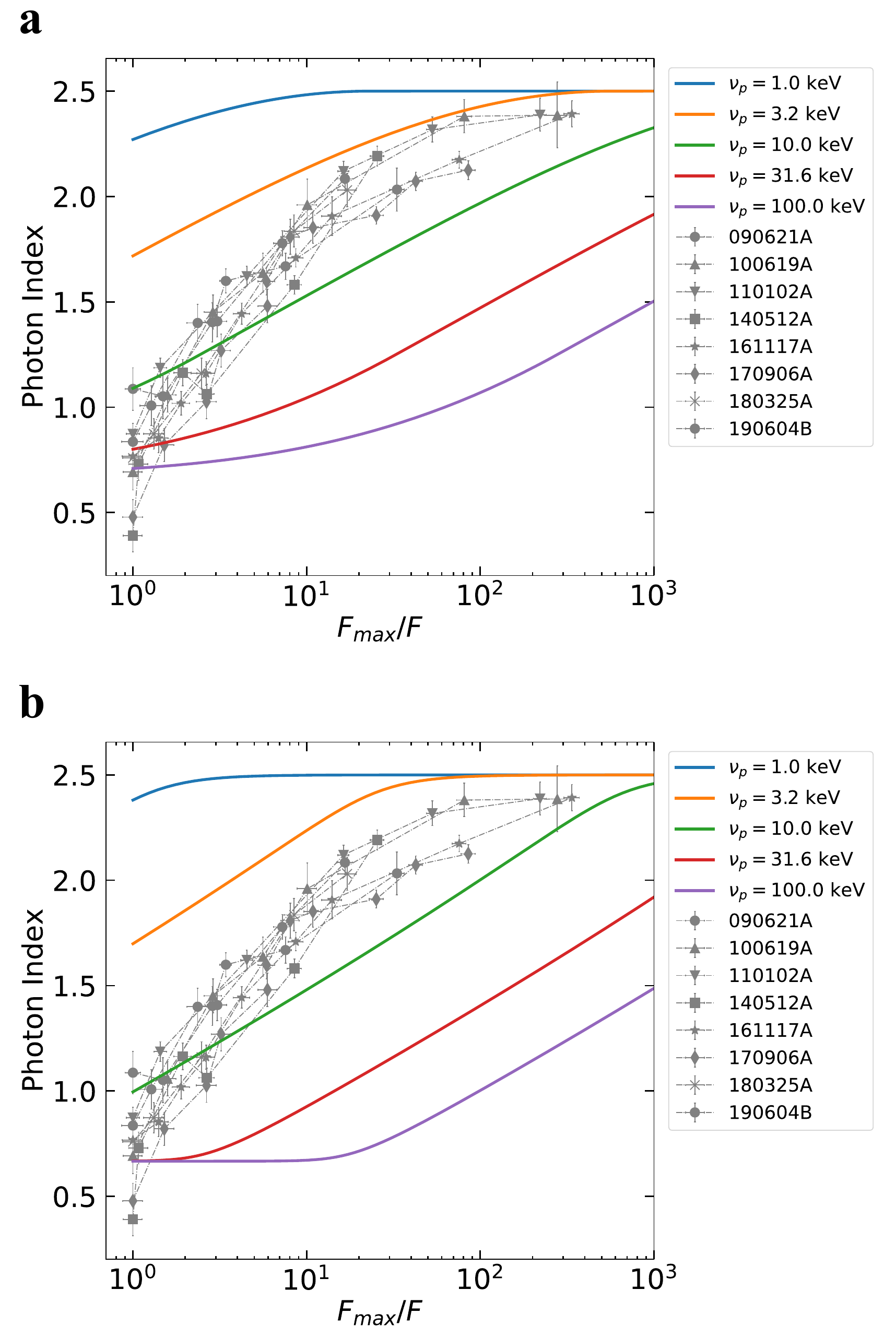}\\

  \caption{{\bf Spectral evolution expected for HLE from a infinitesimal duration pulse, for alternative spectral shapes.} In {\bf a} we adopt a Band function, while in {\bf b} we adopt a SBPL with sharpness parameter $n=4$. The error bars represent $1\sigma$ uncertainties, calculated via spectral fitting in XSPEC. In the legend we report the name of each GRB.}
  \label{HLE_Band+SBPL}
\end{figure}

\clearpage

{\noindent  \bf Supplementary Tables}

\renewcommand{\arraystretch}{1.3}

\begin{table}[h!]
\centering
\begin{tabular}{|c|c|c|c|c|}
\hline
GRB   & z     & $N_H(10^{22}\rm{cm}^{-2})$  & $T_i(\rm{s})$   & $T_f(\rm{s})$ \\
\hline  
090621A & -   & 1.53  & 268   & 369 \\\hline
100619A & -   & 0.46  & 88    & 168 \\\hline
110102A & -   & 0.14  & 266   & 455 \\\hline
140512A & 0.725 & 0.12  & 124   & 216 \\\hline
161117A & 1.549 & 0.58  & 121   & 353 \\\hline
170906A & -   & 0.23  & 90    & 194 \\\hline
180325A & 2.25  & 0.71  & 87    & 146 \\\hline      
190604B & -   & 0.21  & 219   & 262 \\\hline

\end{tabular}
\caption{{\bf Main information about the GRBs of the first sample.} z is the redshift, when available. $N_H$ is the column density adopted in the spectral analysis of the X-ray tail. $T_i$ and $T_f$ are the central times of the initial and final bins of the spectral analysis, respectively. }
\end{table}

\begin{table}[h!]
\centering
\begin{tabular}{|c|c|c|c|c|c|}
\hline
GRB   & z     & $N_H(10^{22}\rm{cm}^{-2})$  & $T_i(\rm{s})$   & $T_f(\rm{s})$   & $T_p^{\text{BAT}}(\rm{s})$  \\\hline

060729  & 0.54  & 0.03  & 133   & 163   & 93    \\\hline
060904A & -   & 0.13  & 74    & 148   & 56    \\\hline
101023A & -     & 0.17  & 91    & 180   & 63    \\\hline
120922A & -   & 0.09  & 128   & 296   & 100   \\\hline
150323A & 0.593 & 0.44  & 153   & 241   & 136   \\\hline
160119A & -   & 0.14  & 166   & 267   & 147   \\\hline
190106A & 1.86  & 0.68  & 99    & 227   & 76    \\\hline
190219A & -   & 0.14  & 114   & 186   & 66    \\\hline

\end{tabular}
\caption{{\bf Main information about the GRBs of the second sample.} z is the redshift, when available. $N_H$ is the column density adopted in the spectral analysis of the X-ray tail. $T_i$ and $T_f$ are the central times of the initial and final bins of the spectral analysis, respectively. $T_p^{\text{BAT}}$ is the peak time of the BAT pulse preceding the X-ray tail, used for the extrapolation of $F_{\text{max}}$.}
\end{table}

\footnotesize
\begin{table}
\centering
\begin{tabular}{|c|c|c|c|c|}
\hline
GRB & time (s) & $\alpha$ & $F_{(0.5-10)\text{ keV}}$ $(10^{-10}\text{erg cm}^{-2}\text{s}^{-1})$ & cstat/dof\\\hline
090621A&&&&\\\hline
& $265-271$ &$  1.09_{-0.1}^{+0.1}$ & $397.6_{-20.5}^{+19.7}$ & $321.3/377$\\\hline
& $271-278$ &$  1.05_{-0.11}^{+0.11}$ & $267.0_{-14.6}^{+13.9}$ & $282.7/351$\\\hline
& $278-289$ &$  1.4_{-0.09}^{+0.09}$  & $138.8_{-6.2}^{+6.0}$ & $274.8/381$\\\hline
& $289-306$ &$  1.6_{-0.06}^{+0.06}$  & $116.0_{-3.1}^{+3.0}$ & $509.3/548$\\\hline
& $306-339$ &$  1.67_{-0.06}^{+0.06}$ & $52.6_{-1.5}^{+1.5}$  & $452.0/527$\\\hline
& $339-399$ &$  2.03_{-0.1}^{+0.1}$ & $12.0_{-0.6}^{+0.6}$  & $253.7/346$\\\hline

100619A&&&&\\\hline
& $85-90$ &$  0.69_{-0.08}^{+0.09}$ & $1311.0_{-82.1}^{+77.8}$  & $289.5/363$\\\hline
& $90-95$ &$  1.06_{-0.08}^{+0.08}$ & $829.3_{-49.1}^{+46.6}$ & $259.4/347$\\\hline
& $95-101$  &$  1.45_{-0.08}^{+0.08}$ & $454.1_{-25.1}^{+23.8}$ & $253.1/325$\\\hline
& $101-106$ &$  1.64_{-0.09}^{+0.09}$ & $234.0_{-13.7}^{+13.1}$ & $233.5/277$\\\hline
& $106-120$ &$  1.96_{-0.12}^{+0.12}$ & $130.0_{-8.9}^{+8.4}$ & $151.4/170$\\\hline
& $120-152$ &$  2.38_{-0.08}^{+0.08}$ & $16.2_{-0.6}^{+0.6}$  & $267.6/320$\\\hline
& $152-185$ &$  2.39_{-0.16}^{+0.15}$ & $4.7_{-0.4}^{+0.4}$ & $130.9/164$\\\hline

110102A&&&&\\\hline
& $260-272$ &$  0.87_{-0.05}^{+0.05}$ & $484.3_{-21.3}^{+20.5}$ & $462.9/524$\\\hline
& $272-286$ &$  1.19_{-0.05}^{+0.05}$ & $337.0_{-13.6}^{+13.1}$ & $463.2/485$\\\hline
& $286-314$ &$  1.62_{-0.05}^{+0.05}$ & $107.0_{-4.0}^{+3.8}$ & $451.9/424$\\\hline
& $314-337$ &$  2.12_{-0.05}^{+0.05}$ & $29.7_{-0.9}^{+0.9}$  & $324.3/386$\\\hline
& $337-387$ &$  2.32_{-0.06}^{+0.06}$ & $9.1_{-0.3}^{+0.3}$ & $265.6/354$\\\hline
& $387-523$ &$  2.39_{-0.08}^{+0.08}$ & $2.2_{-0.1}^{+0.1}$ & $272.9/299$\\
\hline

\end{tabular}
\caption{{\bf Results of time resolved spectral analysis for the first sample of GRBs.} For each bin we report the time window, the photon index $\alpha$, the un-absorbed flux $F_{(0.5-10)\text{ keV}}$ and the statistics over the degrees of freedom (dof). The uncertainties are reported with $1\sigma$ level of confidence and they are calculated via spectral fitting in XSPEC.}
\end{table}

\begin{table}
\centering
\begin{tabular}{|c|c|c|c|c|}
\hline
GRB & time (s) & $\alpha$ & $F_{(0.5-10)\text{ keV}}$ $(10^{-10}\text{erg cm}^{-2}\text{s}^{-1})$ & cstat/dof\\\hline
140512A&&&&\\\hline
& $121-127$ &$  0.39_{-0.08}^{+0.08}$ & $347.3_{-22.1}^{+21.1}$ & $279.3/414$\\\hline
& $127-133$ &$  0.73_{-0.08}^{+0.08}$ & $322.2_{-21.2}^{+20.1}$ & $271.6/359$\\\hline
& $133-140$ &$  1.16_{-0.06}^{+0.06}$ & $179.3_{-9.5}^{+9.1}$ & $388.3/416$\\\hline
& $140-151$ &$  1.06_{-0.05}^{+0.05}$ & $130.4_{-5.1}^{+5.0}$ & $482.6/530$\\\hline
& $151-179$ &$  1.58_{-0.04}^{+0.04}$ & $40.8_{-1.4}^{+1.4}$  & $459.8/480$\\\hline
& $179-254$ &$  2.19_{-0.05}^{+0.05}$ & $13.6_{-0.4}^{+0.4}$  & $383.2/419$\\\hline

161117A&&&&\\\hline
& $118-123$ &$  0.77_{-0.08}^{+0.08}$ & $1132.5_{-77.8}^{+73.9}$  & $295.9/358$\\\hline
& $123-130$ &$  0.85_{-0.07}^{+0.07}$ & $806.3_{-51.4}^{+49.0}$ & $320.9/371$\\\hline
& $130-138$ &$  1.02_{-0.06}^{+0.06}$ & $597.6_{-30.5}^{+29.1}$ & $320.6/440$\\\hline
& $138-147$ &$  1.16_{-0.06}^{+0.06}$ & $429.2_{-21.6}^{+20.9}$ & $343.5/427$\\\hline
& $147-161$ &$  1.44_{-0.05}^{+0.05}$ & $267.9_{-11.5}^{+11.1}$ & $378.2/422$\\\hline
& $161-182$ &$  1.71_{-0.04}^{+0.04}$ & $130.7_{-4.4}^{+4.2}$ & $399.9/458$\\\hline
& $182-219$ &$  1.91_{-0.09}^{+0.09}$ & $80.9_{-5.4}^{+5.0}$  & $190.8/202$\\\hline
& $219-285$ &$  2.17_{-0.04}^{+0.04}$ & $15.0_{-0.4}^{+0.4}$  & $390.1/465$\\\hline
& $285-421$ &$  2.39_{-0.06}^{+0.06}$ & $3.4_{-0.1}^{+0.1}$ & $290.3/335$\\
\hline
\end{tabular}
\caption{{\bf Results of time resolved spectral analysis for the first sample of GRBs-continued.} For each bin we report the time window, the photon index $\alpha$, the un-absorbed flux $F_{(0.5-10)\text{ keV}}$ and the statistics over the degrees of freedom (dof). The uncertainties are reported with $1\sigma$ level of confidence and they are calculated via spectral fitting in XSPEC.}
\end{table}

\begin{table}
\centering
\begin{tabular}{|c|c|c|c|c|}
\hline
GRB & time (s) & $\alpha$ & $F_{(0.5-10)\text{ keV}}$ $(10^{-10}\text{erg cm}^{-2}\text{s}^{-1})$ & cstat/dof\\\hline
170906A&&&&\\\hline
& $88-93$ &$  0.48_{-0.08}^{+0.08}$ & $1936.6_{-132.6}^{+124.7}$  & $264.0/331$\\\hline
& $93-98$ &$  0.82_{-0.08}^{+0.08}$ & $1277.8_{-83.0}^{+78.1}$  & $272.8/330$\\\hline
& $98-105$  &$  1.03_{-0.08}^{+0.08}$ & $729.9_{-48.2}^{+45.3}$ & $235.2/309$\\\hline
& $105-110$ &$  1.27_{-0.08}^{+0.08}$ & $601.6_{-37.7}^{+35.4}$ & $246.0/299$\\\hline
& $110-116$ &$  1.48_{-0.08}^{+0.08}$ & $325.3_{-19.5}^{+18.4}$ & $223.0/284$\\\hline
& $116-123$ &$  1.6_{-0.07}^{+0.06}$  & $329.1_{-15.7}^{+15.0}$ & $304.6/337$\\\hline
& $123-129$ &$  1.81_{-0.08}^{+0.08}$ & $240.6_{-13.2}^{+12.6}$ & $239.4/256$\\\hline
& $129-139$ &$  1.85_{-0.07}^{+0.07}$ & $178.0_{-8.8}^{+8.4}$ & $239.7/292$\\\hline
& $139-152$ &$  1.91_{-0.04}^{+0.04}$ & $76.9_{-2.1}^{+2.1}$  & $414.8/440$\\\hline
& $152-174$ &$  2.07_{-0.04}^{+0.04}$ & $45.5_{-1.2}^{+1.2}$  & $365.0/443$\\\hline
& $174-215$ &$  2.13_{-0.04}^{+0.04}$ & $22.7_{-0.6}^{+0.6}$  & $393.1/421$\\
\hline
180325A&&&&\\\hline
& $85-90$ &$  0.76_{-0.07}^{+0.07}$ & $193.7_{-13.2}^{+12.4}$ & $262.1/384$\\\hline
& $90-96$ &$  0.87_{-0.07}^{+0.07}$ & $146.7_{-10.1}^{+9.5}$  & $289.9/349$\\\hline
& $96-106$  &$  1.16_{-0.07}^{+0.07}$ & $78.1_{-5.3}^{+5.1}$  & $237.4/325$\\\hline
& $106-124$ &$  1.84_{-0.08}^{+0.08}$ & $22.9_{-1.4}^{+1.3}$  & $267.5/274$\\\hline
& $124-168$ &$  2.03_{-0.08}^{+0.08}$ & $11.3_{-0.7}^{+0.6}$  & $229.5/268$\\
\hline
190604B&&&&\\\hline
& $216-222$ &$  0.84_{-0.09}^{+0.09}$ & $1460.4_{-107.2}^{+100.1}$  & $247.5/297$\\\hline
& $222-228$ &$  1.01_{-0.09}^{+0.09}$ & $1144.3_{-89.8}^{+84.0}$  & $219.1/252$\\\hline
& $228-234$ &$  1.4_{-0.09}^{+0.09}$  & $619.1_{-43.1}^{+40.2}$ & $251.5/255$\\\hline
& $234-243$ &$  1.41_{-0.07}^{+0.07}$ & $477.2_{-28.1}^{+26.5}$ & $278.4/301$\\\hline
& $243-254$ &$  1.78_{-0.06}^{+0.06}$ & $201.5_{-8.4}^{+8.0}$ & $327.9/346$\\\hline
& $254-270$ &$  2.08_{-0.06}^{+0.06}$ & $87.8_{-3.1}^{+3.0}$  & $305.8/331$\\\hline

\end{tabular}
\caption{{\bf Results of time resolved spectral analysis for the first sample of GRBs-continued.} For each bin we report the time window, the photon index $\alpha$, the un-absorbed flux $F_{(0.5-10)\text{ keV}}$ and the statistics over the degrees of freedom (dof). The uncertainties are reported with $1\sigma$ level of confidence and they are calculated via spectral fitting in XSPEC.}
\end{table}

\begin{table}
\centering
\begin{tabular}{|c|c|c|c|c|}
\hline
GRB & time (s) & $\alpha$ & $F_{(0.5-10)\text{ keV}}$ $(10^{-10}\text{erg cm}^{-2}\text{s}^{-1})$ & cstat/dof\\\hline
060729&&&&\\\hline
& $130-135$ &$  1.95_{-0.07}^{+0.06}$ & $481.9_{-24.4}^{+23.0}$ & $266.9/265$\\\hline
& $135-141$ &$  2.23_{-0.07}^{+0.08}$ & $323.7_{-16.0}^{+15.8}$ & $252.8/221$\\\hline
& $141-147$ &$  2.37_{-0.08}^{+0.08}$ & $213.9_{-10.6}^{+10.2}$ & $220.7/202$\\\hline
& $147-152$ &$  2.63_{-0.08}^{+0.07}$ & $131.9_{-5.7}^{+5.5}$ & $234.4/197$\\\hline
& $152-159$ &$  2.74_{-0.08}^{+0.08}$ & $77.0_{-3.2}^{+3.1}$  & $232.0/203$\\\hline
& $159-167$ &$  2.82_{-0.07}^{+0.07}$ & $57.1_{-1.9}^{+1.9}$  & $193.1/246$\\
\hline
060904A&&&&\\\hline
& $72-77$ &$  1.09_{-0.08}^{+0.08}$ & $257.2_{-18.2}^{+17.2}$ & $245.7/294$\\\hline
& $77-82$ &$  1.19_{-0.09}^{+0.08}$ & $220.5_{-16.7}^{+15.8}$ & $233.5/256$\\\hline
& $82-88$ &$  1.45_{-0.09}^{+0.09}$ & $112.9_{-8.7}^{+8.1}$ & $172.7/239$\\\hline
& $88-97$ &$  1.55_{-0.05}^{+0.05}$ & $71.0_{-3.1}^{+3.0}$  & $387.7/401$\\\hline
& $97-109$  &$  1.69_{-0.05}^{+0.05}$ & $51.2_{-2.1}^{+2.0}$  & $369.1/378$\\\hline
& $109-128$ &$  1.98_{-0.06}^{+0.06}$ & $27.6_{-1.1}^{+1.1}$  & $320.2/345$\\\hline
& $128-168$ &$  2.37_{-0.06}^{+0.06}$ & $11.1_{-0.4}^{+0.4}$  & $249.3/298$\\\hline
101023A&&&&\\\hline
& $88-94$ &$  1.29_{-0.08}^{+0.08}$ & $294.0_{-19.9}^{+18.7}$ & $252.9/326$\\\hline
& $94-99$ &$  1.35_{-0.07}^{+0.07}$ & $227.6_{-13.8}^{+13.1}$ & $289.7/331$\\\hline
& $99-105$  &$  1.41_{-0.08}^{+0.08}$ & $178.4_{-11.6}^{+11.0}$ & $244.6/304$\\\hline
& $105-110$ &$  1.47_{-0.07}^{+0.07}$ & $165.7_{-9.7}^{+9.2}$ & $298.3/319$\\\hline
& $110-118$ &$  1.58_{-0.08}^{+0.08}$ & $109.9_{-6.8}^{+6.4}$ & $255.5/309$\\\hline
& $118-127$ &$  1.81_{-0.08}^{+0.08}$ & $85.0_{-5.0}^{+4.7}$  & $280.8/291$\\\hline
& $127-143$ &$  1.92_{-0.05}^{+0.05}$ & $50.8_{-1.6}^{+1.6}$  & $417.0/429$\\\hline
& $143-168$ &$  2.21_{-0.05}^{+0.05}$ & $28.6_{-0.8}^{+0.8}$  & $341.2/399$\\\hline
& $168-193$ &$  2.23_{-0.07}^{+0.07}$ & $15.3_{-0.6}^{+0.6}$  & $271.6/311$\\\hline

\end{tabular}
\caption{{\bf Results of time resolved spectral analysis for the second sample of GRBs.} For each bin we report the time window, the photon index $\alpha$, the un-absorbed flux $F_{(0.5-10)\text{ keV}}$ and the statistics over the degrees of freedom (dof). The uncertainties are reported with $1\sigma$ level of confidence and they are calculated via spectral fitting in XSPEC.}
\end{table}

\begin{table}
\centering
\begin{tabular}{|c|c|c|c|c|}
\hline
GRB & time (s) & $\alpha$ & $F_{(0.5-10)\text{ keV}}$ $(10^{-10}\text{erg cm}^{-2}\text{s}^{-1})$ & cstat/dof\\\hline
120922A&&&&\\\hline
& $125-130$ &$  1.09_{-0.07}^{+0.07}$ & $374.2_{-24.2}^{+22.8}$ & $297.7/333$\\\hline
& $130-136$ &$  1.19_{-0.07}^{+0.07}$ & $348.1_{-22.6}^{+21.4}$ & $304.4/321$\\\hline
& $136-141$ &$  1.32_{-0.08}^{+0.08}$ & $251.9_{-18.0}^{+16.9}$ & $268.3/281$\\\hline
& $141-146$ &$  1.28_{-0.07}^{+0.07}$ & $292.0_{-18.3}^{+17.4}$ & $280.1/316$\\\hline
& $146-152$ &$  1.46_{-0.08}^{+0.08}$ & $203.0_{-14.1}^{+13.3}$ & $234.1/276$\\\hline
& $152-157$ &$  1.53_{-0.07}^{+0.07}$ & $186.4_{-11.0}^{+10.5}$ & $263.6/294$\\\hline
& $157-162$ &$  1.58_{-0.07}^{+0.07}$ & $168.2_{-9.8}^{+9.4}$ & $274.1/299$\\\hline
& $162-170$ &$  1.64_{-0.07}^{+0.07}$ & $131.0_{-8.0}^{+7.6}$ & $239.0/286$\\\hline
& $170-178$ &$  1.77_{-0.06}^{+0.06}$ & $97.7_{-5.1}^{+4.8}$  & $291.7/294$\\\hline
& $178-188$ &$  1.82_{-0.07}^{+0.07}$ & $80.2_{-4.2}^{+4.0}$  & $250.9/297$\\\hline
& $188-199$ &$  1.91_{-0.06}^{+0.06}$ & $65.7_{-3.2}^{+3.0}$  & $288.6/300$\\\hline
& $199-212$ &$  1.9_{-0.07}^{+0.07}$  & $62.1_{-3.3}^{+3.2}$  & $233.5/282$\\\hline
& $212-228$ &$  1.84_{-0.04}^{+0.04}$ & $50.9_{-1.6}^{+1.5}$  & $390.6/436$\\\hline
& $228-247$ &$  1.9_{-0.04}^{+0.04}$  & $43.6_{-1.3}^{+1.2}$  & $341.0/444$\\\hline
& $247-276$ &$  1.99_{-0.04}^{+0.04}$ & $28.0_{-0.8}^{+0.8}$  & $375.0/428$\\\hline
& $276-316$ &$  2.2_{-0.04}^{+0.04}$  & $19.1_{-0.5}^{+0.5}$  & $320.7/399$\\\hline

150323A&&&&\\\hline
& $150-156$ &$  1.73_{-0.07}^{+0.07}$ & $333.5_{-18.6}^{+17.6}$ & $238.1/282$\\\hline
& $156-161$ &$  1.8_{-0.09}^{+0.09}$  & $211.7_{-13.1}^{+12.4}$ & $202.8/254$\\\hline
& $161-167$ &$  1.92_{-0.08}^{+0.08}$ & $149.6_{-8.3}^{+7.8}$ & $251.2/276$\\\hline
& $167-176$ &$  2.19_{-0.08}^{+0.08}$ & $86.4_{-4.0}^{+3.9}$  & $254.5/264$\\\hline
& $176-190$ &$  2.23_{-0.07}^{+0.07}$ & $66.9_{-2.8}^{+2.7}$  & $242.1/300$\\\hline
& $190-215$ &$  2.53_{-0.05}^{+0.05}$ & $29.0_{-0.7}^{+0.7}$  & $300.3/364$\\\hline
& $215-267$ &$  2.98_{-0.07}^{+0.06}$ & $10.2_{-0.3}^{+0.3}$  & $251.2/298$\\\hline
\end{tabular}
\caption{{\bf Results of time resolved spectral analysis for the second sample of GRBs-continued.} For each bin we report the time window, the photon index $\alpha$, the un-absorbed flux $F_{(0.5-10)\text{ keV}}$ and the statistics over the degrees of freedom (dof). The uncertainties are reported with $1\sigma$ level of confidence and they are calculated via spectral fitting in XSPEC.}
\end{table}

\begin{table}
\centering
\begin{tabular}{|c|c|c|c|c|}
\hline
GRB & time (s) & $\alpha$ & $F_{(0.5-10)\text{ keV}}$ $(10^{-10}\text{erg cm}^{-2}\text{s}^{-1})$ & cstat/dof\\\hline
160119A&&&&\\\hline
& $164-169$ &$  1.25_{-0.07}^{+0.07}$ & $183.1_{-11.4}^{+10.9}$ & $279.0/347$\\\hline
& $169-174$ &$  1.37_{-0.08}^{+0.08}$ & $171.7_{-11.4}^{+10.7}$ & $241.7/303$\\\hline
& $174-180$ &$  1.47_{-0.07}^{+0.07}$ & $134.6_{-8.4}^{+7.9}$ & $219.9/320$\\\hline
& $180-185$ &$  1.54_{-0.07}^{+0.07}$ & $109.8_{-6.6}^{+6.3}$ & $217.4/309$\\\hline
& $185-193$ &$  1.66_{-0.05}^{+0.05}$ & $129.6_{-5.5}^{+5.3}$ & $323.5/389$\\\hline
& $193-200$ &$  1.63_{-0.06}^{+0.06}$ & $126.9_{-5.9}^{+5.6}$ & $341.0/376$\\\hline
& $200-209$ &$  1.8_{-0.06}^{+0.06}$  & $93.4_{-4.1}^{+3.9}$  & $347.9/370$\\\hline
& $209-223$ &$  1.94_{-0.06}^{+0.06}$ & $63.6_{-2.6}^{+2.5}$  & $324.0/342$\\\hline
& $223-242$ &$  2.12_{-0.06}^{+0.06}$ & $37.9_{-1.5}^{+1.5}$  & $292.8/326$\\\hline
& $242-291$ &$  2.43_{-0.07}^{+0.07}$ & $13.8_{-0.5}^{+0.5}$  & $249.1/308$\\\hline

190106A&&&&\\\hline
& $96-101$  &$  1.56_{-0.08}^{+0.08}$ & $221.9_{-14.4}^{+13.4}$ & $247.0/263$\\\hline
& $101-106$ &$  2.02_{-0.09}^{+0.09}$ & $120.6_{-7.9}^{+7.5}$ & $226.2/220$\\\hline
& $106-114$ &$  2.31_{-0.1}^{+0.1}$ & $67.7_{-4.0}^{+3.8}$  & $150.5/198$\\\hline
& $114-130$ &$  2.55_{-0.06}^{+0.06}$ & $33.2_{-1.0}^{+1.0}$  & $271.4/319$\\\hline
& $130-181$ &$  2.69_{-0.07}^{+0.07}$ & $9.6_{-0.3}^{+0.3}$ & $359.4/332$\\\hline
& $181-274$ &$  2.38_{-0.09}^{+0.09}$ & $2.6_{-0.1}^{+0.1}$ & $242.7/268$\\\hline

190219A&&&&\\\hline
& $111-117$ &$  1.99_{-0.08}^{+0.08}$ & $99.8_{-5.6}^{+5.3}$  & $199.8/251$\\\hline
& $117-124$ &$  2.29_{-0.07}^{+0.07}$ & $116.9_{-5.1}^{+5.0}$ & $218.2/267$\\\hline
& $124-134$ &$  2.46_{-0.07}^{+0.07}$ & $82.8_{-3.5}^{+3.4}$  & $238.2/254$\\\hline
& $134-146$ &$  2.51_{-0.09}^{+0.09}$ & $46.0_{-2.3}^{+2.3}$  & $188.4/215$\\\hline
& $146-164$ &$  2.67_{-0.05}^{+0.05}$ & $33.8_{-0.9}^{+0.9}$  & $328.2/316$\\\hline
& $164-208$ &$  3.05_{-0.06}^{+0.06}$ & $13.8_{-0.4}^{+0.4}$  & $295.2/306$\\\hline

\end{tabular}
\caption{{\bf Results of time resolved spectral analysis for the second sample of GRBs-continued.} For each bin we report the time window, the photon index $\alpha$, the un-absorbed flux $F_{(0.5-10)\text{ keV}}$ and the statistics over the degrees of freedom (dof). The uncertainties are reported with $1\sigma$ level of confidence and they are calculated via spectral fitting in XSPEC.}
\end{table}


\begin{thebibliography}{10} 

\bibitem{Shemi1990} Shemi, A. \& Piran, T. \ The appearance of cosmic fireballs. \textit{Asphysic. J. Lett.} \textbf{365}, L55 (1990)

\bibitem{Usov1992} Usov, V.~V.\ Millisecond pulsars with extremely strong magnetic fields as a cosmological source of $\gamma$-ray bursts. \textit{Nature} \textbf{{357}}, 472 (1992)

\bibitem{Pe'er2006} Pe'er, A., M{\'e}sz{\'a}ros, P., \& Rees, M.~J.	\ The observable effects of a photospheric component on GRB and XRF prompt emission spectrum. \textit{Asphysic. J. Lett.} \textbf{642}, 995 (2006)

\bibitem{Rees1994} Rees, M.~J., Meszaros, P.\ Unsteady outflow models for cosmological gamma-ray bursts.\ \textit{Astrophys. J.} \textbf{430}, L93 (1994)

\bibitem{Zhang2011} Zhang, B., Yan, H.\ The internal-collision-induced magnetic reconnection and turbulence (ICMART) model of gamma-ray bursts. \textit{Astrophys. J.} \textbf{726}, 90 (2011)

\bibitem{Nousek2006} Nousek, J.~A. et al.\ Evidence for a canonical gamma-ray burst afterglow light curve in the Swift XRT data.\ \textit{Astrophys. J.} \textbf{642}, 389 (2006)

\bibitem{Zhang2006} Zhang, B. et al.\ Physical processes shaping gamma-ray burst X-Ray afterglow light curves: theoretical implications from the Swift X-Ray telescope observations. \textit{Astrophys. J.} \textbf{642}, 354.

\bibitem{Tagliaferri2005} Tagliaferri, G. et al.\ An unexpectedly rapid decline in the X-ray afterglow emission of long {\ensuremath{\gamma}}-ray bursts. \textit{Nature} \textbf{436}, 985 (2005)

\bibitem{O'Brien2006} O'Brien, P.~T. et al.\ The early X-Ray emission from GRBs. \textit{Astrophys. J.} \textbf{647}, 1213 (2006)

\bibitem{Paczynski1993} Paczynski, B., Rhoads, J.~E.\ Radio transients from gamma-ray bursters. \textit{Astrophys. J.} \textbf{418}, L5 (1993)

\bibitem{Meszaros1997} M{\'e}sz{\'a}ros, P., Rees, M.~J.\ Optical and long-wavelength afterglow from gamma-ray bursts.\ \textit{Astrophys. J.} \textbf{476}, 232 (1997)

\bibitem{Sari1998} Sari, R., Piran, T., Narayan, R.\ Spectra and light curves of gamma-ray burst afterglows. \textit{Astrophys. J.} \textbf{497}, L17 (1998)

\bibitem{Fenimore1996} Fenimore, E.~E., Madras, C.~D., Nayakshin, S.\ Expanding relativistic shells and gamma-ray burst temporal structure. \textit{Astrophys. J.} \textbf{473}, 998 (1996)

\bibitem{Kumar2000} Kumar, P., Panaitescu, A.\ Afterglow emission from naked gamma-ray bursts. \textit{Astrophys. J.} \textbf{541}, L51 (2000)

\bibitem{Liang2006} Liang, E.~W. et al.\ Testing the curvature effect and internal origin of gamma-ray burst prompt emissions and X-Ray flares with Swift data. \textit{Astrophys. J.} \textbf{646}, 351 (2006)


\bibitem{Lin2017} Lin, D.-B., Mu, H.-J., Liang, Y.-F., et al.\ Steep decay phase shaped by the curvature effect. II. Spectral evolution. \textit{Astrophys. J.} \textbf{840}, 118 (2017)


\bibitem{Zhang07} Zhang, B.-B., Liang, E.-W., \& Zhang, B. A Comprehensive Analysis of Swift XRT Data. I. Apparent Spectral Evolution of Gamma-Ray Burst X-Ray Tails. \textit{Astrophys. J.} \textbf{666}, 1002 (2007)

\bibitem{Mangano2011} Mangano, V. \& Sbarufatti, B.\ Modeling the spectral evolution in the decaying tail of gamma-ray bursts observed by Swift,\ \textit{Advances in Space Research}, \textbf{47}, 1367 (2011)


\bibitem{Gehrels2004} Gehrels, N. et al.\ The Swift gamma-ray burst mission. \textit{Astrophys. J.} \textbf{611}, 1005 (2004)


\bibitem{Frontera2000} Frontera, F. et al.\ Prompt and delayed emission properties of gamma-ray bursts observed with BeppoSAX.\ \textit{Astrophys. J. Suppl. Series} \textbf{127}, 59 (2000)

\bibitem{Kaneko2006} Kaneko, Y., Preece, R.~D., Briggs, M.~S., Paciesas, W.~S., Meegan, C.~A., Band, D.~L.\ The complete spectral catalog of bright BATSE gamma-ray bursts. \textit{Astrophys. J. Suppl. Series} \textbf{166}, 298 (2006)

\bibitem{Nava2011} Nava, L., Ghirlanda, G., Ghisellini, G., Celotti, A.\ Spectral properties of 438 GRBs detected by Fermi/GBM. \textit{Astron. Astrophys.} \textbf{530}, A21 (2011)


\bibitem{Band1993} Band, D. et al.\ BATSE observations of gamma-ray burst spectra. I. Spectral diversity. \textit{Astrophys. J.} \textbf{413}, 281 (1993)

\bibitem{Genet2009} Genet, F., Granot, J.\ Realistic analytic model for the prompt and high-latitude emission in GRBs. \textit{Mon. Not. R. Astron. Soc.} \textbf{399}, 1328 (2009)

\bibitem{Uhm2015} Uhm, Z.~L., Zhang, B.\ On the curvature effect of a relativistic spherical shell. \textit{Astrophys. J.} \textbf{808}, 33 (2015)

\bibitem{Narayan2009} Narayan, R., \& Kumar, P.\ A turbulent model of gamma-ray burst variability.\ \textit{Mon. Not. R. Astron. Soc.} \textbf{394}, L117 (2009)

\bibitem{Duran2016} Barniol Duran, R., Leng, M., Giannios, D.\ An anisotropic minijets model for the GRB prompt emission. \textit{Mon. Not. R. Astron. Soc.} \textbf{455}, L6 (2016)

\bibitem{Geng17} Geng, J.-J., Huang, Y.-F., \& Dai, Z.-G. \ Steep decay of GRB X-Ray flares: the results of anisotropic synchrotron radiation. \textit{Astrophys. J. Lett.} \textbf{841}, L15 (2017)


\bibitem{Pe'er2008} Pe'er, A.\ Temporal evolution of thermal emission from relativistically expanding plasma.\ \textit{Astrophys. J.} \textbf{682}, 463 (2008)

\bibitem{Duran2009} Barniol Duran, R., Kumar, P.\ Adiabatic expansion, early X-ray data and the central engine in GRBs.\ \textit{Mon. Not. R. Astron. Soc.} \textbf{395}, 955 (2009)

\bibitem{Meszaros1999} M{\'e}sz{\'a}ros, P., Rees, M.~J.\ GRB 990123: reverse and internal shock flashes and late afterglow behaviour.\ \textit{Mon. Not. R. Astron. Soc.} \textbf{306}, L39 (1999)

\bibitem{Sari1997} Sari, R., Piran, T.\ Variability in gamma-ray bursts: a clue.\ \textit{Astrophys. J.} \textbf{485}, 270 (1997)

\bibitem{Lyutikov2006} Lyutikov, M.\ Did Swift measure gamma-ray burst prompt emission radii?\ \textit{Mon. Not. R. Astron. Soc.} \textbf{369}, L5 (2006)

\bibitem{Lazzati2006} Lazzati, D., Begelman, M.~C.\ Thick fireballs and the steep decay in the early X-Ray afterglow of gamma-ray bursts.\ \textit{Astrophys. J.} \textbf{641}, 972 (2006)

\bibitem{Walker2000} Walker, K.~C., Schaefer, B.~E., Fenimore, E.~E.\ Gamma-ray bursts have millisecond variability.\ \textit{Astrophys. J.} \textbf{537}, 264 (2000)


\bibitem{Zhang2020} Zhang, B.\ Synchrotron radiation in {\ensuremath{\gamma}}-ray bursts prompt emission.\ \textit{Nat. Astron.} \textbf{4}, 210 (2020)

\bibitem{Kumar08} Kumar, P., \& McMahon, E.\ A general scheme for modelling $\gamma$-ray burst prompt emission
.\ \textit{Mon. Not. R. Astron. Soc.} \textbf{384}, 33 (2008)

\bibitem{Ben13} Beniamini, P., \& Piran, T.\ Constraints on the synchrotron emission mechanism in gamma-ray bursts.\ \textit{Astrophys. J.} \textbf{769}, 69 (2013)

\bibitem{Ben18} Beniamini, P., Barniol Duran, R., \& Giannios, D.\ Marginally fast cooling synchrotron models for prompt GRBs.\ \textit{Mon. Not. R. Astron. Soc.} \textbf{476}, 1785 (2018)


\bibitem{Ghisellini2020} Ghisellini, G. et al.\ Proton-synchrotron as the radiation mechanism of the prompt emission of gamma-ray bursts?\ \textit{Astron. Astrophys.} \textbf{636}, A82 (2020)

\bibitem{Evans2009} Evans, P.~A. et al.\ Methods and results of an automatic analysis of a complete sample of Swift-XRT observations of GRBs.\ \textit{Mon. Not. R. Astron. Soc.} \textbf{397}, 1177 (2009)

\bibitem{Lien2016} Lien, A. et al.\ The third Swift Burst Alert Telescope gamma-ray burst catalog.\ \textit{Astrophys. J.} \textbf{829}, 7 (2016)

\bibitem{Xspec}Arnaud, K.A. XSPEC: The first ten years. \ \textit{Astronomical Data Analysis Software and Systems V}, eds. G. Jacoby and J. Barnes, p17,\ ASP Conf. Series volume \textbf{101} (1996)

\bibitem{2000ApJ...542..914W} Wilms, J., Allen, A., \& McCray, R.\ On the absorption of X-Rays in the interstellar medium.\ \textit{Astrophys. J.} \textbf{542}, 914 (2000)

\bibitem{Kalberla2005} Kalberla, P.~M.~W. et al.\ The Leiden/Argentine/Bonn (LAB) Survey of Galactic HI. Final data release of the combined LDS and IAR surveys with improved stray-radiation corrections.\ \textit{Astron. Astrophys.} \textbf{440}, 775 (2005)

\bibitem{BK2007} Butler, N.~R., Kocevski, D.\ X-Ray Hardness Evolution in GRB Afterglows and Flares: Late-Time GRB Activity without N$_{H}$ Variations.\ \textit{Astrophys. J.} \textbf{663}, 407 (2007)

\bibitem{Mu16} Mu, H.-J., Lin, D.-B., Xi, S.-Q., et al.\ The history of GRB outflows: ejection Lorentz factor and radiation location of X-Ray flares.\  \textit{Astrophys. J.} \textbf{831}, 111 (2016)


\bibitem{Perna2002} Perna, R., Lazzati, D.\ Time-dependent photoionization in a dusty medium. I. code description and general results.\ \textit{Astrophys. J.} \textbf{580}, 261 (2002)

\bibitem{Oganesyan2020} Oganesyan, G., Ascenzi, S., Branchesi, M., Salafia, O.~S., Dall'Osso, S., Ghirlanda, G.\ Structured jets and X-Ray plateaus in gamma-ray burst phenomena.\ \textit{Astrophys. J.} \textbf{893}, 88 (2020)

\bibitem{Rybicky} Rybicki, G.~B., \& Lightman, A.~P.\ Riadiative Processes in Astrophysics.\ A Wiley-Interscience Publication (1979)

\bibitem{Oga17} Oganesyan, G., Nava, L., Ghirlanda, G., et al.\ Detection of low-energy breaks in gamma-ray burst prompt emission spectra.\ \textit{Astrophys. J.} \textbf{846}, 137 (2017)

\bibitem{Oga18} Oganesyan, G., Nava, L., Ghirlanda, G., et al.\ Characterization of gamma-ray burst prompt emission spectra down to soft X-rays.\ \textit{Astron. Astrophys.} \textbf{616}, A138 (2018)

\bibitem{Oga19} Oganesyan, G., Nava, L., Ghirlanda, G., et al.\ Prompt optical emission as a signature of synchrotron radiation in gamma-ray bursts.\  \textit{Astron. Astrophys.} \textbf{628}, A59 (2019)

\bibitem{Rav18} Ravasio, M.~E., Oganesyan, G., Ghirlanda, G., et al.\ Consistency with synchrotron emission in the bright GRB 160625B observed by Fermi.\ \textit{Astron. Astrophys.} \textbf{613}, A16 (2018)

\bibitem{Rav19} Ravasio, M.~E., Ghirlanda, G., Nava, L., et al. \ Evidence of two spectral breaks in the prompt emission of gamma-ray bursts.\ \textit{Astron. Astrophys.} \textbf{625}, A60 (2019)

\bibitem{Bur2020} Burgess, J.~M., B{\'e}gu{\'e}, D., Greiner, J., et al. Gamma-ray bursts as cool synchrotron sources.\ \textit{Nat. Astron.} \textbf{4}, 174 (2020)

\bibitem{Pan19} Panaitescu, A.\ Adiabatic and radiative cooling of relativistic electrons applied to synchrotron spectra and light curves of gamma-ray burst pulses.\ \textit{Astrophys. J.} \textbf{886}, 106 (2019)

\bibitem{Foreman-Mackey} Foreman-Mackey, D., Hogg, D.~W., Lang, D., et al.\ emcee: The MCMC Hammer.\ \textit{Publications of the Astronomical Society of the Pacific}, \textbf{125}, 306 (2013)\\

\end{thebibliography}

\clearpage

\begin{thebibliography}{10} 

\bibitem{Fenimore1996} Fenimore, E.~E., Madras, C.~D., Nayakshin, S.\ Expanding relativistic shells and gamma-ray burst temporal structure.\ \textit{Astrophys. J.} \textbf{473}, 998 (1996)

\bibitem{Dermer2004} Dermer, C.~D.\ Curvature effects in gamma-ray burst colliding shells.\ \textit{Astrophys. J.} \textbf{614}, 284 (2004)

\bibitem{Genet2009} Genet, F., Granot, J.\ Realistic analytic model for the prompt and high-latitude emission in GRBs.\ \textit{Mon. Not. R. Astron. Soc.}  \textbf{399}, 1328 (2009)

\bibitem{Salafia2016} Salafia, O.~S., Ghisellini, G., Pescalli, A., Ghirlanda, G., Nappo, F.\ Light curves and spectra from off-axis gamma-ray bursts.\ \textit{Mon. Not. R. Astron. Soc.} \textbf{461}, 3607 (2016)

\bibitem{Uhm2015} Uhm, Z.~L., Zhang, B.\ On the curvature effect of a relativistic spherical shell.\ \textit{Astrophys. J.} \textbf{808}, 33 (2015)


\bibitem{Uhm2016} Uhm, Z.~L., Zhang, B.\ Evidence of bulk acceleration of the GRB X-ray flare emission region.\ \textit{Astrophys. J.} \textbf{824}, L16 (2016)

\bibitem{Uhm2016b} Uhm, Z.~L., Zhang, B. \ Toward an understanding of GRB prompt emission mechanism. I. The origin of spectral lags.\ \textit{Astrophys. J.} \textbf{825}, 97 (2016)

\bibitem{Uhm2018} Uhm, Z.~L., Zhang, B., Racusin, J.\ Toward an understanding of GRB prompt emission mechanism. II. Patterns of peak energy evolution and their connection to spectral lags.\ \textit{Astrophys. J.} \textbf{869}, 100 (2018)

\bibitem{Lyutikov2003} Lyutikov, M., Blandford, R.\ Gamma bay bursts as electromagnetic outflows.\ arXiv e-prints astro-ph/0312347 (2003)

\bibitem{Narayan2009} Narayan, R., \& Kumar, P.\ A turbulent model of gamma-ray burst variability.\ \textit{Mon. Not. R. Astron. Soc.} \textbf{394}, L117 (2009)

\bibitem{Duran2016} Barniol Duran, R., Leng, M., Giannios, D.\ An anisotropic minijets model for the GRB prompt emission.\ \textit{Mon. Not. R. Astron. Soc.}  \textbf{455}, L6 (2016)

\bibitem{Geng17} Geng, J.-J., Huang, Y.-F., \& Dai, Z.-G. \ Steep eecay of GRB X-ray flares: the results of anisotropic synchrotron radiation.\ \textit{Astrophys. J. Lett.} \textbf{841}, L15 (2017)

\bibitem{Rees1994} Rees, M.~J., Meszaros, P.\ Unsteady outflow models for cosmological gamma-ray bursts.\ \textit{Astrophys. J.} \textbf{430}, L93 (1994)


\bibitem{Drenkhahn2002} Drenkhahn, G., Spruit, H.~C.\ Efficient acceleration and radiation in Poynting flux powered GRB outflows.\ \textit{Astron. Astrophys.} \textbf{391}, 1141 (2002)
\bibitem{Pe'er2008} Pe'er, A.\ Temporal evolution of thermal emission from relativistically expanding plasma.\ \textit{Astrophys. J.} \textbf{682}, 463 (2008)

\bibitem{Piran1999} Piran, T.\ Gamma-ray bursts and the fireball model.\ \textit{Phys. Rep.} \textbf{314}, 575 (1999)

\bibitem{Kumar2008} Kumar, P., Narayan, R., \& Johnson, J.~L.\ Properties of gamma-ray burst progenitor stars.\ \textit{Science} \textbf{321}, 376 (2008)

\bibitem{Hascoet2012} Hasco{\"e}t, R., Daigne, F., Mochkovitch, R.\ Accounting for the XRT early steep decay in models of the prompt gamma-ray burst emission.\ \textit{Astron. Astrophys.} \textbf{542}, L29 (2012)

\bibitem{Beloborodov2011} Beloborodov, A.~M.\ Radiative transfer in ultrarelativistic outflows.\ \textit{Astrophys. J.} \textbf{737}, 68 (2011)

\bibitem{Asano2009} Asano, K., Terasawa, T.\ Slow heating model of gamma-ray burst: photon spectrum and delayed emission.\ \textit{Astrophys. J.} \textbf{705}, 1714 (2009)

\bibitem{Asano2015} Asano, K. \& Terasawa, T.\ Stochastic acceleration model of gamma-ray burst with decaying turbulence. \textit{Mon. Not. R. Astron. Soc.} \textbf{454}, 2242 (2015)

\bibitem{Pan19} Panaitescu, A.\ Adiabatic and radiative cooling of relativistic electrons applied to synchrotron spectra and light curves of gamma-ray burst pulses.\ \textit{Astrophys. J.} \textbf{886}, 106 (2019)

\end{thebibliography}
\end{document}